\documentclass{vldb}

\newif\iffull
\fulltrue

\usepackage{algorithm}
\usepackage[noend]{algorithmic}
\usepackage{alltt}
\usepackage{amsfonts}
\usepackage{amsmath}
\usepackage{balance}
\usepackage{bbding}
\usepackage{color}
\usepackage{colortbl}
\usepackage{enumitem}
\usepackage{float}
\usepackage{graphicx}
\usepackage{listings}
\usepackage{multirow}
\usepackage{soul}
\usepackage{subfigure}
\usepackage{tabularx}
\usepackage{url}
\usepackage{verbatim}

\usepackage{arydshln} 

\newcommand{\kai}[1]{\textcolor{black}{#1}}

\newcommand{\sys}{\texttt{Tempura}}
\newcommand{\xmodel}{TIP model}
\newcommand{\op}[1]{\textit{#1}}
\newcommand{\pl}{\mathbb{P}}
\newcommand{\s}[2]{{{#1}_{#2}}}
\newcommand{\dt}[3]{{\Delta {#1}_{#2}^{#3}}}

\newcommand{\td}[1]{\text{t-dom}(#1)}
\newcommand{\et}[1]{\tau(#1)}
\newcommand{\ivm}{\texttt{IM-1}}
\newcommand{\scm}{\texttt{IM-2}}
\newcommand{\hov}{\texttt{HOV}}
\newcommand{\ojv}{\texttt{OJV}}
\newcommand{\impl}[1]{\texttt{#1}}
\newcommand{\tablename}[1]{\textsf{#1}}

\newcommand{\boldstart}[1]{\textbf{#1}}

\newcommand{\eat}[1]{}
\newcommand{\company}{{CorpX}}

\newtheorem{theorem}{Theorem}
\newtheorem{example}[theorem]{Example}
\newtheorem{definition}[theorem]{Definition}

\def\leftouterjoin{\bowtie^{lo}}

\def\leftantijoin{\bowtie^{la}}
\def\leftsemijoin{\bowtie^{ls}}

\vldbTitle{Tempura: A General Cost-Based Optimizer Framework for Incremental Data Processing}
\vldbAuthors{Zuozhi Wang, Kai Zeng, Botong Huang, Wei Chen, Xiaozong Cui, Bo Wang, Ji Liu, Liya Fan, Dachuan Qu, Zhenyu Ho, Tao Guan, Chen Li, Jingren Zhou}
\vldbDOI{https://doi.org/10.14778/xxxxxxx.xxxxxxx}
\vldbVolume{12}
\vldbNumber{xxx}
\vldbYear{2019}

\begin{document}

\iffull
\title{Tempura: A General Cost-Based Optimizer Framework for Incremental Data Processing (Extended Version)}
\fi


%
%
%
%

\numberofauthors{1} 

\author{
%
%
\alignauthor
Zuozhi Wang$^\dag$, Kai Zeng$^\ddag$, Botong Huang$^\ddag$, Wei Chen$^\ddag$, Xiaozong Cui$^\ddag$, Bo Wang$^\ddag$,\\Ji Liu$^\ddag$, Liya Fan$^\ddag$, Dachuan Qu$^\ddag$, Zhenyu Hou$^\ddag$, Tao Guan$^\ddag$, Chen Li$^\dag$, Jingren Zhou$^\ddag$
\\
\affaddr{$^\dag$University of California, Irvine\hspace*{.7cm}$^\ddag$Alibaba Group}
\email{$^\dag$\{zuozhiw, chenli\}@ics.uci.edu, $^\ddag$\{zengkai.zk, botong.huang, wickeychen.cw, xiaozong.cxz, yanyu.wb, niki.lj, liya.fly, dachuan.qdc, zhenyuhou.hzy, tony.guan, jingren.zhou\}@alibaba-inc.com}
}

\maketitle

\begin{abstract}
Incremental processing is widely-adopted in many applications,
ranging from incremental view maintenance, stream computing,
to recently emerging progressive data warehouse and intermittent query processing.
Despite many algorithms developed on this topic, none of them can produce an incremental plan that always achieves the best performance, since the optimal plan is data dependent.  In this paper, we develop a novel cost-based optimizer framework, called \sys{}, for optimizing incremental data processing.
\kai{We propose an incremental query planning model called TIP based on the concept of time-varying relations,}
which can formally model incremental processing in its most general form. We give a full specification of \sys{}, which can not only unify various existing techniques to generate an optimal incremental plan, but also allow the developer to add their rewrite rules.  We study how to explore the plan space and search for an optimal incremental plan.
We conduct a thorough experimental evaluation of \sys{} in various incremental processing scenarios to show its effectiveness and efficiency.
\end{abstract}


\section{Introduction}
\label{sec:intro}

Incremental processing is widely used
in data computation, where the input data to a query is available
gradually, and the query computation is triggered
multiple times each processing a delta of the input data.
Incremental processing is central to database views with incremental view maintenance (IVM)~\cite{bag-ivm-1,bag-ivm-2,outerjoin-view,dbtoaster} and stream processing~\cite{cql,abadi2005design,ghanem2010supporting, motwani2003query, thakkar2011smm}.
It has been adopted in various application domains
such as active databases~\cite{activedb}, resumable query execution~\cite{resumable}, approximate query processing~\cite{progressive-analysis,iolap,drum}, etc.
New advancements in big data systems
make data ingestion more real-time and analysis increasingly time sensitive,
which further boost the adoption
of the incremental processing model.
Here are a few examples of emerging applications.

\textbf{Progressive Data Warehouse~\cite{grosbeak}.}
  Enterprise data warehouses usually have a large amount of
  automated routine analysis jobs, which usually have a stringent
  schedule and deadline determined by various business logic.  For example, at Alibaba, there is a need to schedule a daily report query after 12 am when the previous
  day's data has been fully collected and deliver the results by 6 am sharp before the bill-settlement time.
  Till now, its automated routine analysis jobs are still predominately handled using batch processing, causing dreadful ``rush hour'' scheduling patterns.
  This approach puts pressure on resources during traffic hours,
  and leaves the resources over-provisioned and wasted during the off-traffic
  hours.
  Incremental processing can answer routine analysis queries
  progressively as data gets ingested, resulting in a more flexible resource usage pattern, which can effectively smoothen the resource skew.


\textbf{Intermittent Query Processing~\cite{intermittent}.}
  Many modern applications require querying an incomplete dataset with the remaining data arriving in an intermittent yet predictable way.
  Intermittent query processing can leverage incremental processing
  to balance latency for maintaining standing queries and resource consumption by exploiting knowledge of data-arrival patterns.
  For instance, when querying dirty data, the data is usually
   first cleaned and then fed into a
  database. The data cleaning step can quickly spill the clean part
  of the data but needs to conduct a time-consuming cleaning
  processing on the dirty part. Intermittent query processing can use
  incremental processing to quickly deliver informative but partial
  results on the clean data to the user, before delivering the full
  results after processing all the cleaned data.

A key problem behind these applications
is that given a query, how to generate an efficient incremental-computation plan.
Previous studies focused on various aspects of the problem,
e.g., incremental computation algorithms for
a specific setting such as~\cite{bag-ivm-1,dbtoaster,outerjoin-view},
or algorithms to determine which intermediate states to materialize given
an incremental plan~\cite{partial-online,icedb,intermittent}.
The following example based on two commonly used algorithms shows that none of them can generate an incremental-computation plan
that is always optimal, since the optimal plan is {\em data dependent}.

\begin{example}[Reporting consolidated revenue]
\begin{alltt}
\footnotesize
\tablename{summary} =
  with \tablename{sales}_\tablename{status} as (
    SELECT \tablename{sales}.\textit{o_id}, \textit{category}, \textit{price}, \textit{cost}
    FROM \tablename{sales} LEFT OUTER JOIN \tablename{returns}
      ON \tablename{sales}.\textit{o_id} = \tablename{returns}.\textit{o_id} )
  SELECT \textit{category},
    SUM(IF(\textit{cost} IS NULL, \textit{price}, -\textit{cost})) AS \textit{gross}
  FROM \tablename{sales}_\tablename{status}
  GROUP BY \textit{category}
\end{alltt}
\label{ex:pipeline}
\end{example}

In the aforementioned progressive data warehouse scenario, consider a routine
analysis job in Example~\ref{ex:pipeline} that reports the gross revenue after
consolidating the sales orders with the returned ones.
We want to incrementally compute the job
as data gets ingested, to utilize the cheaper free resources occasionally available in the cluster.
Thus, we want to find an incremental plan with
the optimal resource usage pattern, i.e.,
carrying out as much early computation as possible
using cheaper free resources to keep the overall resource bill low. This query can be incrementally computed in different ways as
the data in tables~\tablename{sales} and~\tablename{returns} becomes available gradually.
For instance, consider two basic methods used in IVM and stream computing.
(1) A typical view maintenance approach (denoted as \ivm{})
treats \tablename{summary} as views\kai{~\cite{bag-ivm-1,semi-outer-join,bag-ivm-2,iolap}}.
It always maintains \tablename{summary} as if it is directly computed
from the data of \tablename{sales} and \tablename{returns} seen so far.
Therefore, even if a \tablename{sales} order will be returned
in the future, its revenue is counted into the gross revenue temporarily.
(2)
A typical stream-computing method (denoted as \scm{})
avoids such retraction\kai{~\cite{spark-blog,zaniolo,stream-semantics,continuous-append-only}}. It holds back \tablename{sales} orders that do not join
with any \tablename{returns} orders until all data is available.
Clearly, if returned orders are rare, \ivm{} can maximize
the amount of early computation and thus deliver better resource-usage plans.
Otherwise, if returned orders are often, \scm{} can
avoid unnecessary re-computation caused by retraction and thus be better.
(See \S\ref{sec:challenge} for a detailed discussion.) This analysis shows that different data statistics can lead to different preferred methods.


Since the optimal plan for a query \kai{given a user-specified optimization goal} is data dependent, a natural question is how to develop a
principled cost-based optimization framework
to support efficient incremental processing.  To our best knowledge and also to our surprise, there is no such a framework in the literature.  In particular, existing solutions
still rely on users to empirically choose from individual incremental techniques, and
it is not easy to combine the advantages of different techniques
and find the plan that is truly cost optimal.
When developing this framework,
we face more challenges compared to traditional query optimization~\cite{volcano,orca} (See \S\ref{sec:challenge}):
(1)
Incremental query planning needs to do tradeoff analysis on more dimensions than traditional query planning,
such as different incremental computation methods, data arrival patterns, which states to materialize, etc.
(2) The plans for different incremental runs are correlated and may affect each other's optimal choices. Incremental query planning needs to jointly consider the runs across the entire timeline.

In this paper we propose a unified cost-based query optimization framework, which allows users to express and integrate various incremental computation techniques
and provides a turn-key solution to decide optimal incremental execution plans
subject to various objectives. We make the following contributions.

\begin{itemize}[nosep,leftmargin=*]
    \item \kai{We propose a new theory called the {\em \xmodel{}} on top of
    time-varying relation (TVR) that formulates incremental processing using TVR, and defines a formal algebra for TVRs (\S\ref{sec:model}).
    In the \xmodel{}, we also provide a rewrite-rule framework
    to describe different incremental computation techniques, and unify them to explore in a single search space for an optimal incremental plan (\S\ref{sec:tvr-rules}).} This framework allows these techniques to work cooperatively, and enables cost-based search among possible plans.

    \item We build a Cascade-style optimizer named \sys{}.
    It supports cost-based optimization for incremental query planning based on the \xmodel{}. We discuss how to explore the plan space (\S\ref{sec:plan-space-exploration}) and search for an optimal incremental plan  in \sys{} (\S\ref{sec:select-a-plan}).


    \item We conduct a thorough experimental evaluation of the \sys{} optimizer in various application scenarios. The results show the effectiveness and efficiency of \sys{} (\S\ref{sec:experiments}).
\end{itemize}


\section{Problem Formulation}
\label{sec:opt-def}

In this section we formally define the problem of cost-based
optimization for incremental computation. We elaborate on the running example to show that
execution plans generated by different algorithms have different costs. We then illustrate the challenges to solve the problem.

\subsection{Incremental Query Planning}
\label{sec:def}


Despite the different requirements in various applications,
a key problem of cost-based incremental query planning (IQP)
can be modeled uniformly as a quadruple
$(\vec{T}, \vec{D}, \vec{Q}, \tilde{\mathfrak{c}}),$
where:
\begin{itemize}[nosep,leftmargin=*]
  \item \kai{$\vec{T} = [t_1, \ldots, t_k]$ is a vector of time points
    when we can carry out incremental computation.
    Each $t_i$ can be either a concrete physical time, or a discretized logical time.}

  \item \kai{$\vec{D} = [D_1, \cdots, D_k]$ is a vector of data,
      where $D_i$ represents the input data available at time
      $t_i$, e.g., the delta data newly available at $t_i$,
      and/or all the data accumulated up to $t_i$.
      For a future time point $t_i$,
      $D_i$ can be expected data to be available at that time.}

  \item \kai{$\vec{Q} = [Q_1, \ldots, Q_k]$ is a vector of queries. $Q_i$
    defines the expected results that are supposed to be delivered by
    the incremental computation carried out at $t_i$.
    If there is no
    required output at $t_i$, then $Q_i$ is a special empty query $\emptyset$.}

  \item $\tilde{\mathfrak{c}}$ is a cost function that we want to minimize.
\end{itemize}
The goal is to generate an {\em incremental plan} $\pl = [P_1, \ldots, P_k]$
where $P_i$ defines the task (a physical plan) to execute at time $t_i$,
such that (1) $\forall 1\leq i \leq k$, $P_i$ can deliver the results defined by $Q_i$,
and (2) the cost $\tilde{\mathfrak{c}}(\pl)$ is minimized.
\kai{Next we use a few example IQP scenarios to demonstrate how they can be modeled
using the above definition.}

\noindent
\kai{\textbf{Incremental View Maintenance} (\texttt{IVM-PD}).
Consider the problem of incrementally maintaining a view defined by query $Q$.
Instead of using any concrete physical time,
we can use two logical time points $\vec{T} = [t_i, t_{i+1}]$
to represent a general incremental update at $t_{i+1}$ of the result computed at $t_i$.
We assume that the data available at $t_i$ is the data accumulated up to $t_i$,
whereas at $t_{i+1}$ the new delta data (insertions/deletions/updates) between $t_i$ and $t_{i+1}$ is available, denoted by $\vec{D} = [D, \Delta D]$.
At both $t_{i}$ and $t_{i+1}$ we want to keep the view up to date,
i.e., $\vec{Q}$ is defined as $Q_i = Q(D), Q_{i+1} = Q(D + \Delta D)$.
As the main goal is to find the most efficient incremental plan,
we set $\tilde{\mathfrak{c}}$ to be the cost of $P_{i+1}$,
i.e., the execution cost at $t_{i+1}$. (For a formal definition
see $\tilde{c}_v$ in \S\ref{sec:single-opt}.)
Note that if $Q$ involves multiple tables and we want to use different incremental plans
for updates on different tables, we can optimize multiple IQP problems by setting $\Delta D$ to the delta data on only one of the tables at a time.}


\noindent
\kai{\textbf{Progressive Data Warehouse} (\texttt{PDW-PD}).
We model this scenario by choosing $\vec{T}$ as physical time points
of the planned incremental runs.
Note that we only require the incremental plan to deliver the results
defined by the original analysis job $Q$ at the last run, that is,
at the scheduled deadline of the job, without requiring output during the early runs.
Thus, $\vec{Q} = [\emptyset, \cdots, \emptyset, Q]$.
We set $\tilde{\mathfrak{c}}$ as a weighted sum of the costs of all plans in $\pl{}$ (see $\tilde{c}_w(O)$ in \S\ref{sec:single-opt}).}

\kai{A detailed discussion,
such as how $\vec{D}$ is decided for logical times or physical times in the future,
and what if the chosen $\vec{T}$ is subject to change, will be presented in \S\ref{sec:discussion}.}

\subsection{Plan Space and Search Challenges}
\label{sec:challenge}

\begin{figure}[tbh!]
\centering
\tiny
\begin{tabular}{cc}
\subfigure[]{
  \begin{tabular}{|c|c|cc}
  \multicolumn{3}{c}{\tablename{sales}} \\
  \cline{1-3}
  \textit{o\_id} & \textit{cat} & \multicolumn{1}{c|}{\textit{price}} & \\
  \cline{1-3}
  $o_1$ & $c_1$ & \multicolumn{1}{c|}{$100$} & $t_1$ \\
  \cline{1-3}
  $o_2$ & $c_2$ & \multicolumn{1}{c|}{$150$} & $t_1$ \\
  \cline{1-3}
  $o_3$ & $c_1$ & \multicolumn{1}{c|}{$120$} & $t_1$ \\
  \cline{1-3}
  $o_4$ & $c_1$ & \multicolumn{1}{c|}{$170$} & $t_1$ \\
  \cline{1-3}
  $o_5$ & $c_2$ & \multicolumn{1}{c|}{$300$} & $t_2$ \\
  \cline{1-3}
  $o_6$ & $c_1$ & \multicolumn{1}{c|}{$150$} & $t_2$ \\
  \cline{1-3}
  $o_7$ & $c_2$ & \multicolumn{1}{c|}{$220$} & $t_2$ \\
  \cline{1-3}
  \multicolumn{2}{c}{\tablename{returns}} \\
  \cline{1-2}
  \textit{o\_id} & \textit{cost} & \\
  \cline{1-2}
  $o_1$ & $10$ & $t_1$ \\
  \cline{1-2}
  $o_2$ & $20$ & $t_2$ \\
  \cline{1-2}
  $o_6$ & $15$ & $t_2$ \\
  \cline{1-2}
  \end{tabular}
  \label{fig:sales-and-returns}
}
&
\subfigure[]{
  \begin{tabular}{|c|c|c|c|}
  \multicolumn{4}{c}{\tablename{sales\_status}} \\
  \hline
  \textit{o\_id} & \textit{cat} & \textit{price} & \textit{cost} \\
  \hline
  $o_1$ & $c_1$ & $100$ & $10$ \\
  \hline
  $o_2$ & $c_2$ & $150$ & $20$ \\
  \hline
  $o_3$ & $c_1$ & $120$ & null \\
  \hline
  $o_4$ & $c_1$ & $170$ & null \\
  \hline
  $o_5$ & $c_2$ & $300$ & null \\
  \hline
  $o_6$ & $c_1$ & $150$ & $15$ \\
  \hline
  $o_7$ & $c_2$ & $220$ & null \\
  \hline
  \multicolumn{2}{c}{\tablename{summary}} \\
  \cline{1-2}
  \textit{cat} & \textit{gross} \\
  \cline{1-2}
  $c_1$ & $265$ \\
  \cline{1-2}
  $c_2$ & $500$ \\
  \cline{1-2}
  \end{tabular}
  \label{fig:sales-status-and-summary-full}
}
\\
\subfigure[]{
  \begin{tabular}{|c|c|c|c|c}
  \multicolumn{4}{c}{\tablename{sale\_status} at $t_1$} \\
  \cline{1-4}
  \textit{o\_id} & \textit{cat} & \textit{price} & \textit{cost} \\
  \cline{1-4}
  $o_1$ & $c_1$ & $100$ & $10$ \\
  \cline{1-4}
  $o_2$ & $c_2$ & $150$ & null \\
  \cline{1-4}
  $o_3$ & $c_1$ & $120$ & null \\
  \cline{1-4}
  $o_4$ & $c_1$ & $170$ & null \\
  \cline{1-4}
  \multicolumn{5}{c}{Changes to \tablename{sale\_status} at $t_2$} \\
  \hline
  \textit{o\_id} & \textit{cat} & \textit{price} & \textit{cost} & \multicolumn{1}{c|}{\#} \\
  \hline
  \rowcolor[gray]{.7}
  $o_2$ &$c_2$ &$150$ &null & \multicolumn{1}{c|}{$-1$} \\
  \hline
  $o_2$ & $c_2$ & $150$ & $20$ & \multicolumn{1}{c|}{$+1$} \\
  \hline
  $o_5$ & $c_2$ & $300$ & null & \multicolumn{1}{c|}{$+1$} \\
  \hline
  $o_6$ & $c_1$ & $150$ & $15$ & \multicolumn{1}{c|}{$+1$} \\
  \hline
  $o_7$ & $c_2$ & $220$ & null & \multicolumn{1}{c|}{$+1$} \\
  \hline
  \end{tabular}
  \label{fig:view}
}
&
\subfigure[]{
  \begin{tabular}{|c|c|c|c|c}
  \multicolumn{4}{c}{\tablename{sale\_status} at $t_1$} \\
  \cline{1-4}
  \textit{o\_id} & \textit{cat} & \textit{price} & \textit{cost} \\
  \cline{1-4}
  $o_1$ & $c_1$ & $100$ & $10$ \\
  \cline{1-4}
  \multicolumn{4}{c}{Changes to \tablename{sale\_status} at $t_2$} \\
  \hline
  \textit{o\_id} & \textit{cat} & \textit{price} & \textit{cost} & \multicolumn{1}{c|}{\#} \\
  \hline
  $o_2$ & $c_2$ & $150$ & $20$ & \multicolumn{1}{c|}{$+1$} \\
  \hline
  $o_3$ & $c_1$ & $120$ & null & \multicolumn{1}{c|}{$+1$} \\
  \hline
  $o_4$ & $c_1$ & $170$ & null & \multicolumn{1}{c|}{$+1$} \\
  \hline
  $o_5$ & $c_2$ & $300$ & null & \multicolumn{1}{c|}{$+1$} \\
  \hline
  $o_6$ & $c_1$ & $150$ & $15$ & \multicolumn{1}{c|}{$+1$} \\
  \hline
  $o_7$ & $c_2$ & $220$ & null & \multicolumn{1}{c|}{$+1$} \\
  \hline
  \end{tabular}
  \label{fig:stream}
}
\end{tabular}
\caption{(a) Data arrival patterns of \tablename{sales} and \tablename{returns}, (b) results of query \tablename{sales\_status} and \tablename{summary} at $t_2$, (c) incremental results
of \tablename{sales\_status} produced by view maintenance at $t_1$ and $t_2$, and (d) incremental
results of \tablename{sales\_status} produced by stream computing at $t_1,t_2$.}
\end{figure}

We elaborate different plans to answer the query in Example~\ref{ex:pipeline}
using the \texttt{PDW-PD} definition above.
Suppose the query \tablename{summary} is originally scheduled at $t_2$,
but the progressive data warehouse decides to schedule an early
execution at $t_1$ on partial inputs.
Assume the records visible at $t_1$ and $t_2$ in \tablename{sales} and \tablename{returns} are those in Fig.~\ref{fig:sales-and-returns}.
In this IQP problem, we have
$\vec{T} = [t_1, t_2]$ and $\vec{Q} = [\emptyset, q]$, where $q$ is the \tablename{summary} query,
$\vec{D}$ is shown in Fig.~\ref{fig:sales-and-returns}, and $\tilde{\mathfrak{c}}$
is the cost function that takes the weighted sum of the resources used at $t_1$ and $t_2$.
Many existing incremental techniques (e.g., view maintenance, stream computing, mini-batch execution, and so on~\cite{bag-ivm-1,bag-ivm-2,dbtoaster,cql}) can be applied to generate a plan.
Consider two commonly used methods \ivm{} and \scm{}.

\boldstart{Method \ivm{}} treats \tablename{sales\_status} and
  \tablename{summary} as views, and uses incremental computation
  to keep the views always up to date with respect to the data seen so far.
  The incremental computation is done on the delta input. For example,
  the delta input to \tablename{sales} at $t_2$ includes tuples $\{o_5,o_6,o_7\}$.
  Fig.~\ref{fig:view} depicts \tablename{sales\_status}'s
  incremental outputs at $t_1$ and $t_2$, respectively,
  where $\#=+/-1$ denote insertion or deletion respectively.
  Note that a \tablename{returns} record (e.g., $o_2$ at $t_2$) can arrive much later
  than its corresponding \tablename{sales} record (e.g., the shaded $o_2$ at $t_1$).
  Therefore, a \tablename{sales} record may be output early as it cannot join with a \tablename{returns} record
  at $t_1$, but retracted later at $t_2$ when the \tablename{returns} record arrives,
  such as the shaded tuple $o_2$ in Fig.~\ref{fig:view}.

\boldstart{Method \scm{}} can avoid such retraction
  during incremental computation. Specifically, in the outer join of \tablename{sales\_status},
  tuples in \tablename{sales} that do not join with tuples from \tablename{returns} for now
  (e.g., $o_2$, $o_3$, and $o_4$) may join in the future, and thus will be held back at $t_1$.
  Essentially the outer join is computed as an inner join at $t_1$.
  The incremental outputs of \tablename{sales\_status} are shown
  in Fig.~\ref{fig:stream}.

In addition to these two, there are many other methods as well.
Generating one plan with a high performance is non-trivial due to the following reasons. {\em (1) The optimal incremental plan is data dependent,
and should be determined in a cost-based way.}
In the running example, \ivm{} computes $9$ tuples
($5$ tuples in the outer join and $4$ tuples in the aggregate) at $t_1$,
and $10$ tuples at $t_2$. Suppose the cost per unit at $t_1$ is 0.2 (due to fewer queries at that time), and the cost per unit at $t_2$ is 1.  Then its total cost is $9 \times 0.2 + 10 \times 1 = 11.8$.  \scm{} computes $6$ tuples at $t_1$, and $11$ tuples at $t_2$, with a total cost of $6 \times 0.2 + 11 \times 1 = 12.2$.
\ivm{} is more efficient, since it can do more early computation
in the outer join, and more early outputs further enable \tablename{summary}
to do more early computation.
On the contrary, if retraction is often, say, with one more tuple $o_4$ at $t_2$,
then \scm{} will become more efficient, as it costs $12.2$ versus the cost $13.8$ of \ivm{}.
The reason is that retraction wastes early computation and causes more re-computation overhead.  Notice that the performance difference of these two approaches can be arbitrarily large.

{\em (2) The entire space of possible plan alternatives is very large.}
Different parts within a query
can choose different incremental methods.
Even if early computing the entire query does not pay off,
we can still incrementally execute a subpart of the query.
For instance, for the left outer join in \tablename{sales\_status},
we can incrementally shuffle the input data once it is ingested
without waiting for the last time.
IQP needs to search the entire plan space
ranging from the traditional batch plan at one end
to a fully-incrementalized plan at the other.

\emph{(3) Complex
temporal dependencies between different incremental runs can also impact the plan decision.}
For instance, during the continuous ingestion of data,
query \tablename{sales\_status} may prefer a broadcast join at $t_1$
when the \tablename{returns} table is small,
but a shuffled hash join at $t_2$ when the \tablename{returns} table gets bigger.
But such a decision may not be optimal, as
shuffled hash join needs data to be distributed according to the join key,
which broadcast join does not provide.
Thus, different join implementations between $t_1$ and $t_2$ incur reshuffling overhead.
IQP needs to jointly consider \kai{all incremental runs} across the entire timeline.

Such complex reasoning is challenging, if not impossible,
even for very experienced experts.
To solve this problem, we offer a cost-based solution to systematically
search the entire plan space to generate an optimal plan.  Our solution can
unify different incremental computation techniques in a single plan.






\section{The TIP Model}
\label{sec:model}



The core of incremental computation is to deal with relations changing over time,
and understand how the computation on these relations can be expanded
along the time dimension.
\kai{In this section, we introduce a formal theory based on the concept of {\em time-varying relation} (TVR)~\cite{cql,one-query,two-sides}, called the \emph{TVR-based Incremental query Planning (TIP) Model}}.
The model naturally extends the relational model by considering the temporal aspect to formally describe incremental execution.
It also includes various data-manipulation operations on TVR's,
as well as rewrite rules of TVR's
in order for a query optimizer to define and
explore a search space to generate an efficient incremental query plan.
\kai{
To the best of our knowledge, the proposed \xmodel{} is the first one that
not only unifies different incremental computation methods,
but also can be used to develop a principled cost-based optimization framework for incremental execution.} We focus on definitions and algebra of TVR's in this section,
and dwell on TVR rewrite rules in \S\ref{sec:tvr-rules}.

\subsection{Time-Varying Relations}
\label{sec:concept}


\begin{figure*}[tbh!]
\centering
\includegraphics[width=.95\textwidth]{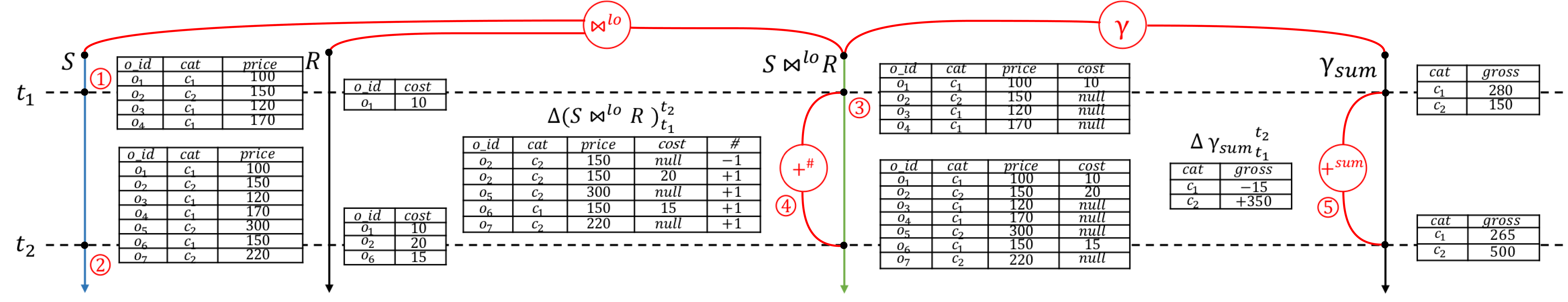}
\caption{Example TVR's and their relationships.}
\label{fig:tvr-basic}
\end{figure*}


\begin{definition}
A {\em time-varying relation (TVR) $R$} is a mapping from a time domain $\mathcal{T}$ to a bag of tuples
belonging to a schema.
\end{definition}
A \emph{snapshot} of $R$ at a time $t$, denoted $\s{R}{t}$, is the instance of $R$ at time $t$.
For example, due to continuous ingestion,
table \tablename{sales} ($S$) in Example~\ref{ex:pipeline} is a TVR,
depicted as the blue line in Fig.~\ref{fig:tvr-basic}.
On the line, tables \textcircled{1} and \textcircled{2} show the snapshots of $S$ at $t_1$ and $t_2$, i.e., $\s{S}{t_1}$ and $\s{S}{t_2}$, respectively.  Traditional data warehouses run queries  on relations at a specific time, while incremental execution runs queries on TVR's.
\begin{definition}[Querying TVR]
\kai{Given a TVR $R$ on time domain $\mathcal{T}$, applying a query $Q$ on $R$ defines another TVR
$Q(R)$ on $\mathcal{T}$, where
$\s{[Q(R)]}{t} = Q(\s{R}{t}), \forall t \in \mathcal{T}$.}
\label{def:query-tvr}
\end{definition}

In other words, the snapshot of $Q(R)$ at $t$ is the same as
applying $Q$ as a query on the snapshot of $R$ at $t$.
For instance, in Fig.~\ref{fig:tvr-basic},
joining two TVR's \tablename{sales}  ($S$) and \tablename{returns} ($R$)
yields a TVR $(S\leftouterjoin R)$, depicted as the green line.
Snapshot $\s{(S\leftouterjoin R)}{t_1}$ is shown as table \textcircled{3},
which is equal to joining $\s{S}{t_1}$ and $\s{R}{t_1}$.
We denote left outer-join as $\leftouterjoin$,
left anti-join as $\leftantijoin$, left semi-join as $\leftsemijoin$, and aggregate as $\gamma$.
For brevity, we use ``$Q$'' to refer to the TVR ``$Q(R)$'' when there is no ambiguity.

\subsection{Basic Operations on TVR's}

Besides as a sequence of snapshots, a TVR can be encoded from a delta perspective
using the changes between two snapshots.
We denote the difference between two snapshots of TVR $R$ at $t, t' \in T$ ($t < t'$)
as the \emph{delta} of $R$ from $t$ to $t'$, denoted $\dt{R}{t}{t'}$, which defines a second-order TVR.
\begin{definition}[TVR difference]
$\dt{R}{t}{t'}$ defines a mapping from a time interval to a bag of tuples
belonging to the same schema, such that there is a merge operator ``$+$'' satisfying
$\s{R}{t} + \dt{R}{t}{t'} = \s{R}{t'}.$
\end{definition}
Table \textcircled{4} in Fig.~\ref{fig:tvr-basic} shows $\dt{(S\leftouterjoin R)}{t_1}{t_2}$, which is the delta of snapshots $\s{(S\leftouterjoin R)}{t_1}$ and $\s{(S\leftouterjoin R)}{t_2}$.
Here multiplicities $\#=+1/-1$ represent insertion and deletion of the corresponding tuple, respectively.
The merge operator $+$ is defined as additive union on relations
with bag semantics,
which adds up the multiplicities of tuples in bags.

\kai{Interestingly, a TVR can have different snapshot/delta views.}
For instance, the delta $\dt{\gamma(S\leftouterjoin R)}{t_1}{t_2}$
can be defined differently as Table \textcircled{5} in Fig.~\ref{fig:tvr-basic}.
Here the merge operator $+$ directly sums up the partial \texttt{SUM} values
(the \textit{gross} attribute) per \textit{category}.
For \textit{category} $c_1$, summing up the partial \texttt{SUM}'s
in $\s{\gamma(S\leftouterjoin R)}{t_1}$ and $\dt{\gamma(S\leftouterjoin R)}{t_1}{t_2}$
yields the value in $\s{\gamma(S\leftouterjoin R)}{t_2}$, i.e., $280 + (-15) = 265$.
To differentiate these two merge operators, we denote the merge operator for
$S\leftouterjoin R$ as $+^{\#}$, and the merge operator for
$\gamma(S\leftouterjoin R)$ as $+^{sum}$.

This observation shows that
the way to define TVR deltas and the merge operator $+$ is not unique.
In general, as studied in previous research~\cite{partial,iolap},
the difference between two snapshots $\s{R}{t}$ and $\s{R}{t'}$
can have two types:

\kai{(1) \emph{Multiplicity Perspective}. $\s{R}{t}$ and $\s{R}{t'}$ may have different multiplicities of tuples. $\s{R}{t}$ may have less or more tuples than $\s{R}{t'}$.
In this case, the merge operator (e.g., $+^{\#}$) combines the same tuples by adding up their multiplicities.}

\kai{(2) \emph{Attribute Perspective}. $\s{R}{t}$ may have different attribute values
in some tuples compared to $\s{R}{t'}$. In this case, the merge operator (e.g., $+^{sum}$) groups tuples with the
same primary key, and combines the delta updates on the changed attributes into one value.
Aggregation operators usually produce this type of snapshots and deltas.
Formally, distributed aggregation in data-parallel computing platforms
is often modeled using four methods~\cite{dist-aggr}:
\iffull
\begin{enumerate}[nosep, leftmargin=*]
\item \texttt{Initialize}: It is called once before any data is supplied with a given key
to initialize the aggregate state.
\item \texttt{Iterate}: It is called every time a tuple is provided with a matching key
to combine the tuple into the aggregate state.
\item \texttt{Merge}: It is called every time when combining two aggregate states
with the same key into a single aggregate state.
\item \texttt{Final}: It is called at the end on the final aggregate state to produce a result.
\end{enumerate}
\fi
For an aggregate function $\gamma$, the snapshots/deltas are the aggregate states
computed using \texttt{Initialize} and \texttt{Iterate} on partial data;
the merge operator $+^\gamma$ is defined using \texttt{Merge};
and at this end, the attribute-perspective snapshot is converted by \texttt{Final} to
produce the multiplicity-perspective snapshot, i.e., the final result.\footnote{\kai{
Note that \texttt{Final} also needs to filter out empty groups with zero contributing tuples.
We omit this detail due to the limited space.}}
For instance, for the aggregate function \texttt{AVG},
its snapshot/delta
is an aggregate state consisting of a running \texttt{SUM} and \texttt{COUNT} on the partial data;
the merge operator $+^{\texttt{AVG}}$ sums up the running \texttt{SUM} and \texttt{COUNT},
and at the end, the running \texttt{SUM} is divided by the running \texttt{COUNT} to get the final average.
Note that for aggregates such as \texttt{MEDIAN} whose state needs to be the full set of tuples, \texttt{Iterate} and \texttt{Merge} degenerate
to no-op.}

Furthermore, for some merge operator $+$,
there is an inverse operator $-$, such that $\s{R}{t'} - \s{R}{t} = \dt{R}{t}{t'}$.
For instance, the inverse operator $-^{sum}$ for $+^{sum}$ is defined as taking the difference
of \texttt{SUM} values per \textit{category} between two snapshots.


\section{TVR Rewrite Rules}
\label{sec:tvr-rules}

Rewrite rules expressing relational algebra equivalence
are the key mechanism that enables traditional query optimizers
to explore the entire plan space.
As TVR snapshots and deltas are simply static relations,
traditional rewrite rules still hold within a single snapshot/delta.
However, these rewrite rules are not enough for incremental
query planning, due to their inability to express algebra equivalence
between TVR concepts. 

To capture this more general form of equivalence,
in this section, we introduce \emph{TVR rewrite rules} \kai{in the \xmodel{}, focusing on logical plans}.
We propose a trichotomy of TVR rewrite rules,
namely \emph{TVR-generating rules}, \emph{intra-TVR rules},
and \emph{inter-TVR rules}, and show how to model
existing incremental techniques using these three types of rules.
This modeling enables us to unify existing incremental
techniques and leverage them uniformly when exploring the plan space;
it also allows IQP to evolve
by adding new TVR rewrite rules.


\subsection{TVR-Generating and Intra-TVR Rules}
\label{sec:tvr-rule-1}

Most existing work on incremental computation revolves around
the notion of delta query that can be described as
Eq.~\ref{eq:delta} below.
\begin{equation}
Q(\s{R}{t'}) = Q(\s{R}{t} + \dt{R}{t}{t'}) = Q(\s{R}{t}) + \mathfrak{d}Q(\s{R}{t}, \dt{R}{t}{t'}).
\label{eq:delta}
\end{equation}
The idea is intuitive:
when an input delta $\dt{R}{t}{t'}$ arrives,
instead of recomputing the query on the new input snapshot $\s{R}{t'}$,
one can directly compute a delta update to the previous query result
$Q(\s{R}{t})$ using a new delta query $\mathfrak{d}Q$.
Essentially, Eq.~\ref{eq:delta} contains two key parts---the delta
query $\mathfrak{d}Q$ and the merge operator $+$,
which correspond to the first two types of TVR rewrite rules, namely
\emph{TVR-generating rules} and \emph{intra-TVR rules}, respectively.

\noindent
\textbf{TVR-Generating Rules}.
Formally, TVR-generating rules define for each relational
operator on a TVR, how to compute its deltas from the
snapshots and deltas of its input TVRs.
In other words, TVR-generating rules define $\mathfrak{d}Q$ for each
relational operator $Q$ such that
$\dt{Q}{t}{t'} = \mathfrak{d}Q(\s{R}{t}, \dt{R}{t}{t'})$.
Many previous studies on deriving delta queries under different semantics~\cite{set-ivm-1,set-ivm-2,bag-ivm-1,semi-outer-join,bag-ivm-2}
fall into this category.
As an example, Fig.~\ref{fig:gen-intra-tvr} shows the TVR-generating rules used by \ivm{} in Example~\ref{ex:pipeline}.
The rules for left outer-join (Rule $(1)$)\kai{\footnote{\kai{
For brevity, some padding of null to match outer join's schema is omitted in
Fig.~\ref{fig:gen-intra-tvr} and Fig.~\ref{fig:inter-tvr-outer}.
This padding can simply be implemented using a project operator.}}}
and aggregate (Rule $(2)$) are
from~\cite{semi-outer-join} and~\cite{bag-ivm-2}, respectively.
For simplicity, we separate the inserted/deleted part
in a TVR delta, and denote them by superscripting $\Delta$ with $+/-$.
The blue lines in Fig.~\ref{fig:gen-intra-tvr} demonstrate these TVR-generating rules in a plan space.



\noindent
\textbf{Intra-TVR Rules}.
Intra-TVR rules define the conversion between the snapshots and deltas
of a single TVR.
As in Eq.~\ref{eq:delta},
the merge operator $+$ defines how to merge $Q$'s snapshot $\s{Q}{t}$
and delta $\dt{Q}{t}{t'}$ into a new snapshot $\s{Q}{t'}$.
Other examples of intra-TVR rules include
rules that merge deltas into a new delta,
e.g., 
for a TVR $R$,
$\dt{R}{t}{t'} + \dt{R}{t'}{t''}=\dt{R}{t}{t''}$,
or rules that take the difference between snapshots/deltas if the merge operator $+$ has an inverse operator $-$,
e.g., $\s{R}{t'} - \s{R}{t} = \dt{R}{t}{t'}$.
The red lines in Fig.~\ref{fig:gen-intra-tvr} demonstrate the intra-TVR
rules used by \ivm{} in Example~\ref{ex:pipeline}.
Note that when merging the snapshot/delta of $S \leftouterjoin R$
(subquery \tablename{sales\_status}),
we use $+^\#$ 
(Rule $(3)$),
whereas when merging the snapshot/delta of $\gamma(S \leftouterjoin R)$
(query \tablename{summary}),
we use $+^{sum}$
((Rule $(4)$).

\subsection{Inter-TVR Rules}
\label{sec:inter-tvr}

\begin{figure*}[tb!]
\centering
\begin{tabular}{cc}
\subfigure[]{
	\includegraphics[height=0.2\textwidth]{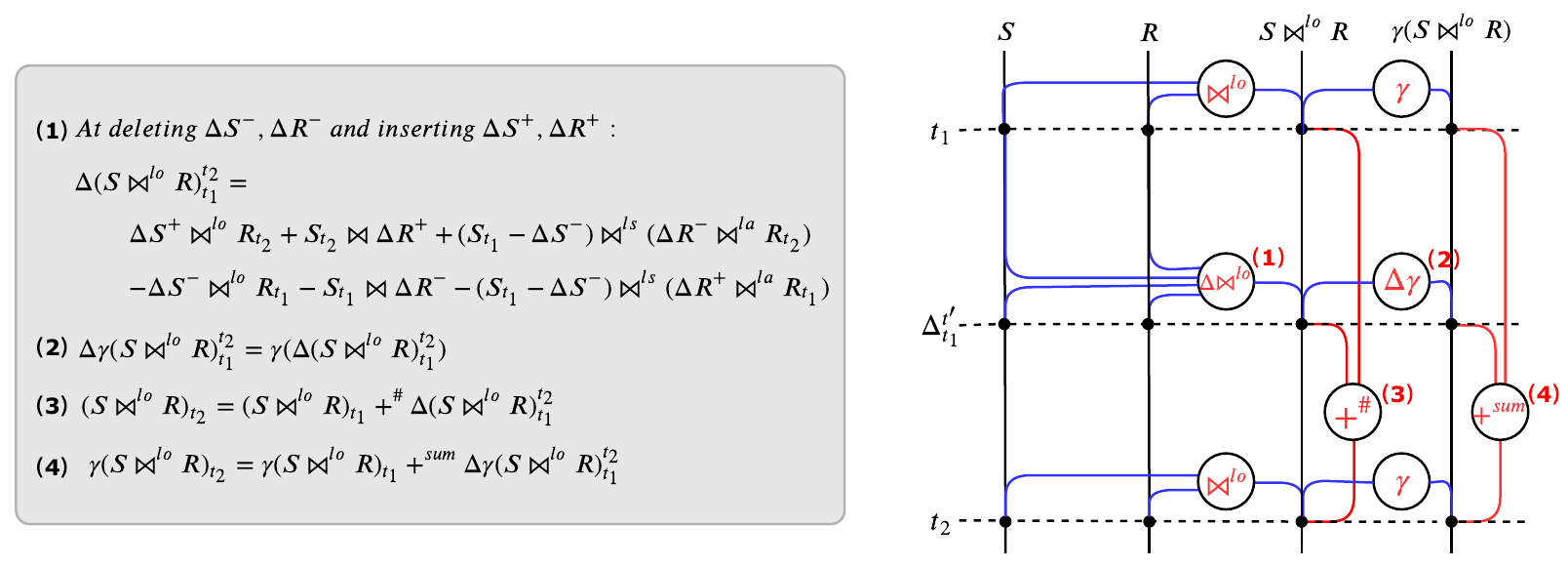}
	\label{fig:gen-intra-tvr}
}
&
\subfigure[]{
	\includegraphics[height=0.2\textwidth]{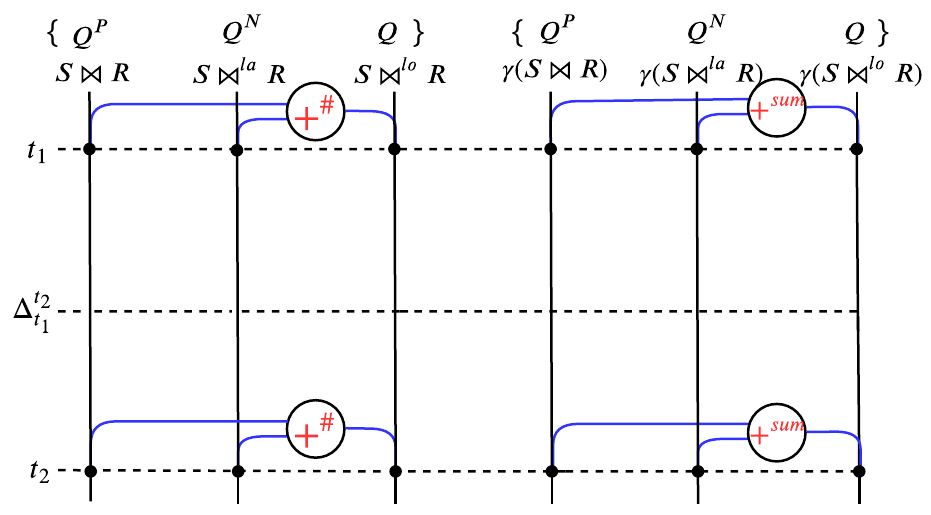}
	\label{fig:inter-tvr-stream}
}
\\
\subfigure[]{
	\includegraphics[height=0.2\textwidth]{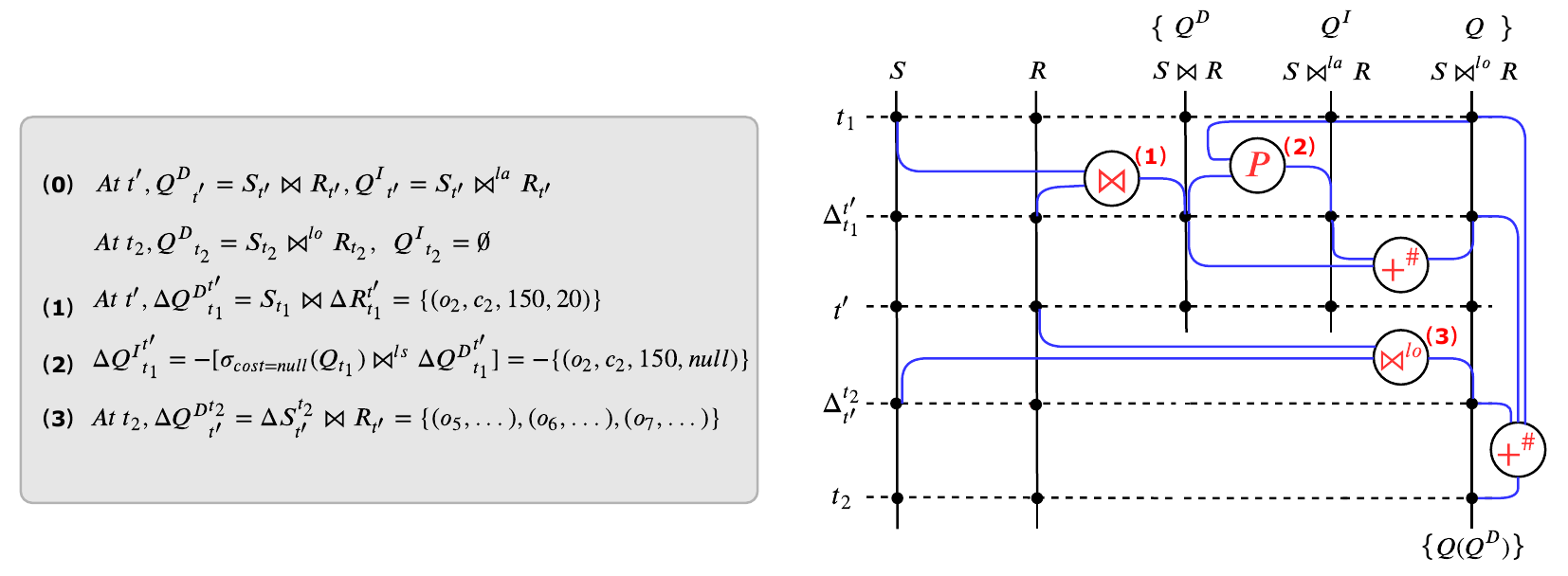}
	\label{fig:inter-tvr-outer}
}
&
\subfigure[]{
	\includegraphics[height=0.2\textwidth]{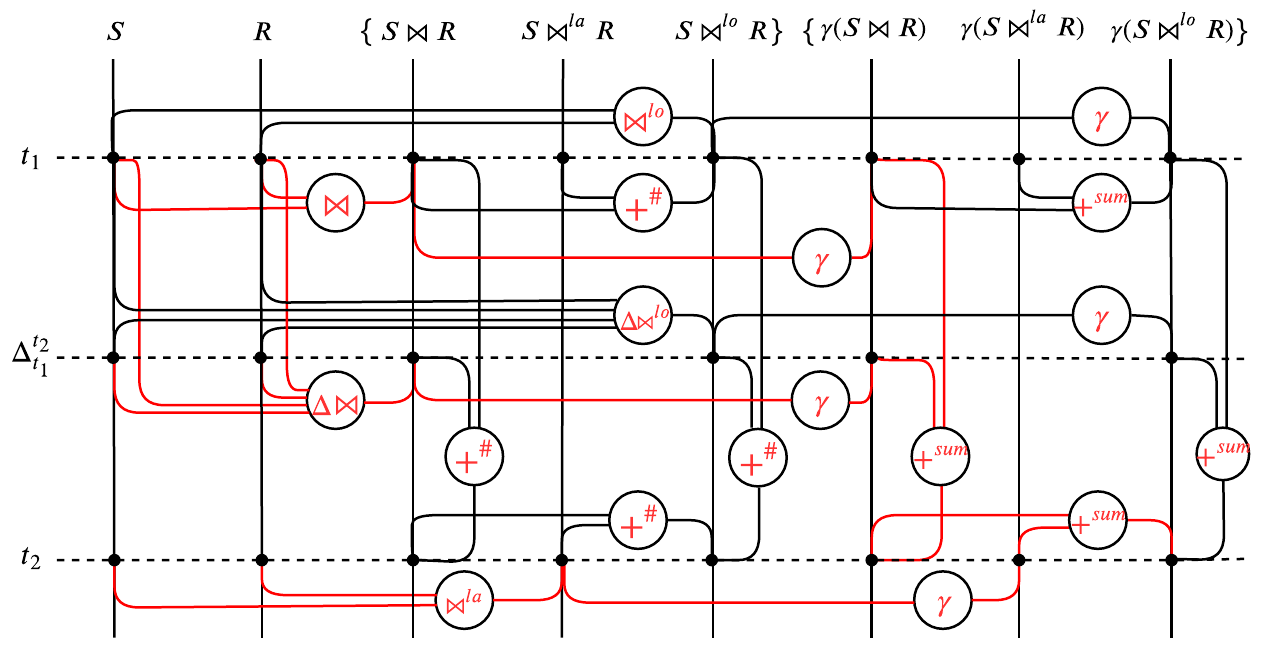}
	\label{fig:tvr-transform}
}
\end{tabular}
\caption{(a) Examples of TVR-generating and intra-TVR rules, (b) examples of inter-TVR equivalence rules in stream computing,
(c) examples of inter-TVR equivalence rules in outer-join view maintenance, and (d) the incremental plan space of Example~\ref{ex:pipeline}.}
\end{figure*}

There are incremental techniques
that cannot be modeled using the two aforementioned types rules alone.
The \scm{} approach in Example~\ref{ex:pipeline} is such an example.
Different from \ivm{}, approach \scm{} does not directly deliver
the snapshot of $S \leftouterjoin R$ at $t_1$. Instead, it delivers only the subset of $S \leftouterjoin R$
that is guaranteed not to be retracted in the future,
essentially the results of $S \bowtie R$.
At $t_2$ when the data is known to be complete,
\scm{} computes the rest part of $S \leftouterjoin R$,
essentially $S \leftantijoin R$, then pads with nulls to match the output schema.

This observation shows another family of incremental techniques:
without computing $Q$ directly,
one can incrementally compute a set of different queries
$\{Q'_1,\cdots,Q'_k\}$, and then apply another query $P$ on their results to get that of $Q$, formally described as Eq.~\ref{eq:decomposition}.
The intuition is that $\{Q'_1,\cdots,Q'_k\}$ may be more amenable
to incremental computation and thus may be more efficient than directly incrementally computing $Q$:
\begin{equation}
Q(R) = P(Q'_1(R),\cdots,Q'_k(R)).
\label{eq:decomposition}
\end{equation}

The family of techniques that Eq.~\ref{eq:decomposition} can describe
are very general. They all rely on certain rewrite rules describing
the equivalence between snapshots/deltas of multiple TVRs.
We summarize this family of techniques into a third type of rules namely
\emph{inter-TVR rules}. Below we demonstrate using a couple of existing
incremental techniques how they can be modeled by inter-TVR rules.

\emph{(1) The \scm{} approach}:
Let us revisit \scm{} using the terminology of inter-TVR rules.
Formally, \scm{} decomposes $Q=S \leftouterjoin R$ into
two parts, $Q^P$ and $Q^N$, defined below:
\begin{align}
\s{Q^P}{t}=\s{S}{t} \bowtie \s{R}{t} &,& \s{Q^N}{t}=\s{S}{t} \leftantijoin \s{R}{t} &,& \s{Q}{t} = \s{Q^P}{t}+^{\#}\s{Q^N}{t}
\label{eq:scm}
\end{align}
where $Q^P$ is a positive part that will not retract any tuple
if both $S$ and $R$ are append-only,
whereas $Q^N$ represents a part that could cause retractions
at insertions to $S$ and $R$.
The inter-TVR rule in Eq.~\ref{eq:scm} states that
any snapshot of $Q$ can be decomposed into snapshots of
$Q^P$ and $Q^N$ at the same time.
Similar decomposition holds for the aggregate $\gamma$ in
\textit{summary} too,
just with a different merge operator $+^{sum}$.
Fig.~\ref{fig:inter-tvr-stream} depicts these rules in a plan space.
As it is much easier to incrementally compute inner join
than left outer join,
$Q^P$ can be incrementally computed using the
TVR rewrite rules in \S\ref{sec:tvr-rule-1}
in a more efficient manner than $Q$,
whereas $Q^N$ cannot be easily incrementalized,
and is not computed until the completion time.

 \emph{(2) Outer-join view maintenance}: \cite{outerjoin-view}
	proposed a method to incrementally maintain outer-join views.
	Its main idea can be summarized using two types of inter-TVR rules:\\
	\begin{subequations}
	\centering
	\begin{tabularx}{0.48\textwidth}{XX}
	\begin{equation}
	\s{Q}{t} = \s{Q^D}{t} +^{\#} \s{Q^I}{t} +^{\#} \s{Q^U}{t}
	\label{eq:ojv-1}
	\end{equation}
	&
	\begin{equation}
	\dt{Q^I}{t}{t'} = P(\dt{Q^D}{t}{t'}, \s{Q}{t})
	\label{eq:ojv-2}
	\end{equation}
	\end{tabularx}
	\end{subequations}
	The first type of rules described by Eq.~\ref{eq:ojv-1}
	decompose a query into three parts
	given an update to a single input table:
	a directly affected part $Q^D$, an indirectly affected part $Q^I$,
	and an unaffected part $Q^U$, where $Q^D$, $Q^I$, and $Q^U$ are defined formally using the join-disjunctive normal form of $Q$.
	Due to space limitation we refer the readers to~\cite{outerjoin-view} for formal details.
	Intuitively, an insertion (deletion) into the input table will cause
	insertions (deletions) to $Q^D$ and deletions (insertions) to $Q^I$,
	but leave $Q^U$ unaffected.
	Eq.~\ref{eq:ojv-2} describes the second type of rules
	that give a way to directly compute the deltas of $Q^I$ from
	the delta of $Q^D$ and the previous snapshot of $Q$.
	At updates, one can use the TVR-generating rules to compute the delta of $Q^D$, and the inter-TVR rules in Eq.~\ref{eq:ojv-2} to get delta of $Q^I$,
	and these two deltas can be merged to incrementally compute $Q$,
	i.e., $\dt{Q}{t}{t'} = \dt{Q^D}{t}{t'} +^{\#} \dt{Q^I}{t}{t'}$.

	Take query \tablename{sales\_status} as an example.
	Fig.~\ref{fig:inter-tvr-outer} shows the corresponding inter-TVR rules.
	As the algorithm in~\cite{outerjoin-view} considers updating one input table at a time,
	we insert a virtual time point $t'$ between $t_1$ and $t_2$,
	assuming $R$ and $S$ are updated separately at $t'$ and $t_2$.
	Rule $(0)$ shows the decomposition of \tablename{sales\_status}
	at $t'$ and $t_2$ following the inter-TVR rule in Eq.~\ref{eq:ojv-1}.
	By applying the TVR-generating rules, $Q^D$ can be incrementally
	computed as rules $(1)$ and $(3)$;
	whereas $Q^I$ can be incrementally computed following the inter-TVR
	rule in Eq.~\ref{eq:ojv-2}, as shown in rule $(2)$.
	Combining them yields the delta of $Q$ as in Table~\textcircled{4} in Fig.~\ref{fig:tvr-basic}.

\iffull

 \emph{(3) Higher-order view maintenance}: \cite{dbtoaster,batch-dbtoaster} proposed a higher-order view-maintenance algorithm,
 which can also be expressed by inter-TVR rules.
 The main idea is to treat the deltas of a query $Q$ as another TVR,
 and continue applying TVR rewrite rules to incrementally compute it.
 Formally, considering a query $Q$ and updates to one of its inputs $R$,
 the algorithm can be summarized as the following inter-TVR rule.
 \begin{equation}
 \dt{Q}{t}{t'} = \mathfrak{d}Q(\s{R}{t},\dt{R}{t}{t'}) = P(\s{M}{t}, \dt{R}{t}{t'})
 \label{eq:dbtoaster}
 \end{equation}
 \kai{The rule decomposes the delta query into two parts: the delta update $\dt{R}{t}{t'}$,
 and an update-independent subquery $M$ that does not
 involve $\dt{R}{t}{t'}$.
 The two parts are combined using a query $P$ to get the delta of $Q$.
 If $M$ is a query involving input relations other than $R$,
  it can be further decomposed again with respect to updates to each of its input relations
  according to Eq.~\ref{eq:dbtoaster}, until it becomes a constant.
  We refer the readers to~\cite{dbtoaster} for a detailed algorithm.}
 \kai{Take the \tablename{summary} query and updates to \tablename{sales} ($S$) as an example (we denote \tablename{returns} as R).
  Applying Eq.~\ref{eq:dbtoaster}, we can decompose it as
  \begin{align*}
  \dt{Q}{t}{t'} &= \gamma_{\textit{category};\texttt{SUM}(r)}(\dt{S}{t}{t'} \leftouterjoin \s{M}{t}), \\
 \text{where~~~~} M_t &= \gamma_{\textit{o\_id};\textit{total}=\texttt{SUM}(\textit{cost})} (\textit{R}_t) , \\
  r &= \texttt{IF}(\textit{total} \texttt{ IS NULL}, \textit{price}, -\textit{total}).
  \end{align*}}

 $M$ essentially preprocesses \tablename{returns} by computing the total cost per \textit{o\_id},\kai{\footnote{\kai{
 Here we do not assume \textit{o\_id} as the primary key of \tablename{returns}. Say \tablename{returns} could contain multiple records for a returned order due to different costs
   such as shipping cost, product damage, inventory carrying cost, etc.}}}
  and $P$ computes the gross revenue per category by summing up the precomputed total cost in
  $M$ or the prices of the new orders added to $S$.
  \kai{Then $M$ is materialized as a higher-order view and can be
    further incrementally maintained with respect to updates to \tablename{returns} by repeatedly
    applying the inter-TVR rule to generate higher-order views.}

\fi

\subsection{Putting Everything Together}

The above concepts and observations lay a theoretical foundation
for our IQP framework.
Different TVR rules can be extended individually and can work with
each other automatically. For example, TVR-generating rules
can be applied on any TVR created by inter-TVR rules.
By jointly applying TVR rewrite rules and traditional rewrite rules,
we can explore a plan space much larger than
that of any individual existing incremental method.
For instance, if we overlay Fig.~\ref{fig:gen-intra-tvr} and~\ref{fig:inter-tvr-stream},
we can achieve the plan space as shown in Fig.~\ref{fig:tvr-transform}.
Any tree rooted at $\s{\gamma(S\leftouterjoin R)}{t_2}$
is a valid incremental plan for Example~\ref{ex:pipeline}, e.g., the red lines indicate the plan used by \scm{}.

In the next two sections, we discuss how to build an optimizer framework based on the \xmodel{}, including plan-space exploration (\S\ref{sec:plan-space-exploration}) and selecting an optimal incremental plan (\S\ref{sec:select-a-plan}).


\section{Plan-Space Exploration}
\label{sec:plan-space-exploration}

In this section we study how~\sys{} explores the space of incremental plans.
Existing query optimizers do the exploration only for a specific time.
To support query optimization in incremental processing, we need to explore
a much bigger plan space by considering not only relations at different times,
but also transformations between them.
We study how to extend existing query optimizers
to support cost-based optimization for incremental processing based on the \xmodel{}.
As an example, we consider one of the state-of-the-art solutions,
the Cascades-style cost-based optimizer~\cite{cascades,volcano}.
We illustrate how to incorporate the \xmodel{} into such an optimization framework
to develop the corresponding optimizer framework called~\sys{}.



\sys{} consists of two main modules. (1) \emph{Memo}: it keeps track of the explored plan space, i.e., all plan alternatives generated,
in a succinct data structure, typically represented as an AND/OR tree, for detecting redundant derivations and fast retrieval.
(2) \emph{Rule engine}: it manages all the transformation rules, which specify algebraic equivalence laws
and suitable physical implementations of logical operators, and
monitors new plan alternatives generated in the memo.
Whenever there are changes, the rule engine fires applicable transformation rules
on the newly-generated plans to add more plan alternatives to the memo.  The memo and rule engine of a traditional Cascades optimizer lack
the capability to support IQP.
We will focus on the key adaptations we made on the two modules to incorporate the \xmodel{}.

\begin{figure*}[htb!]
\centering
\begin{tabular}{cc}
\subfigure[An example memo of subquery \tablename{sales\_status}]{
    \includegraphics[width=0.47\textwidth]{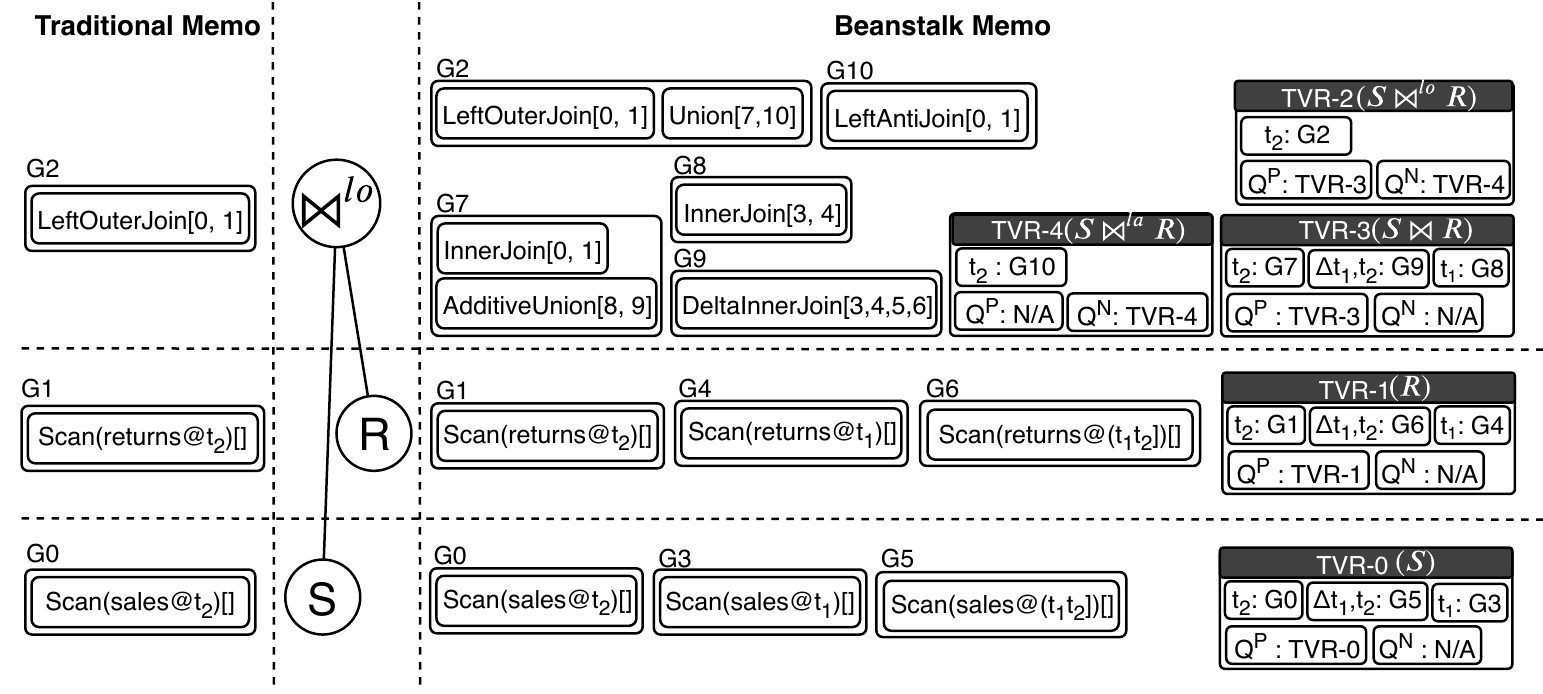}
    \label{fig:memo}
}
&
\subfigure[A step-wise illustration of the growth of the memo]{
  \includegraphics[width=0.47\textwidth]{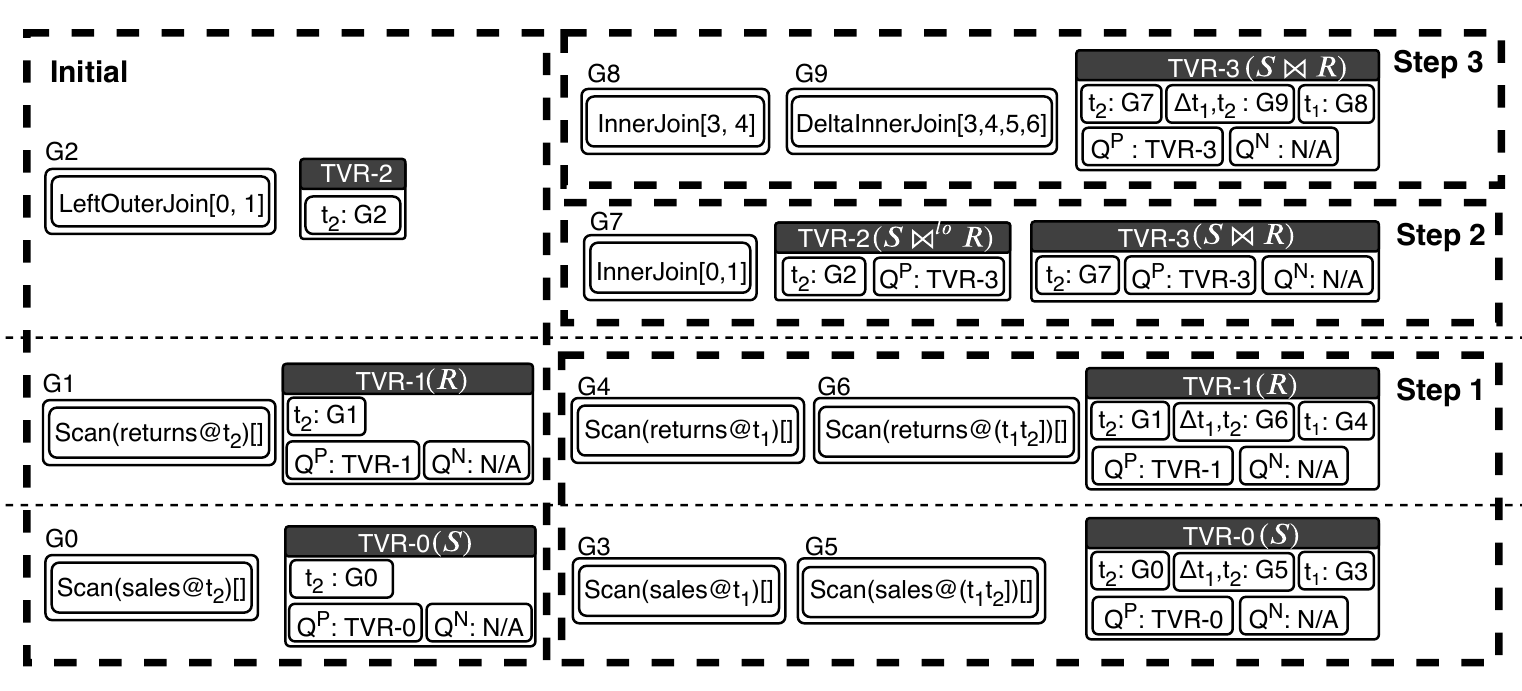}
  \label{fig:memo-breakdown}
}
\\
\subfigure[A TVR-generating rule pattern]{
  \includegraphics[width=0.47\textwidth]{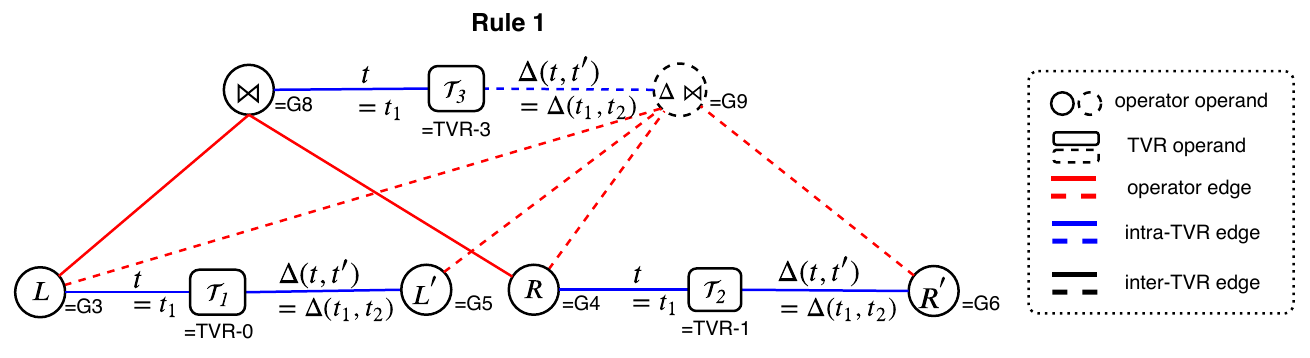}
  \label{fig:tvr-generating-rule}
}
&
\subfigure[An inter-TVR rule pattern]{
  \includegraphics[width=0.47\textwidth]{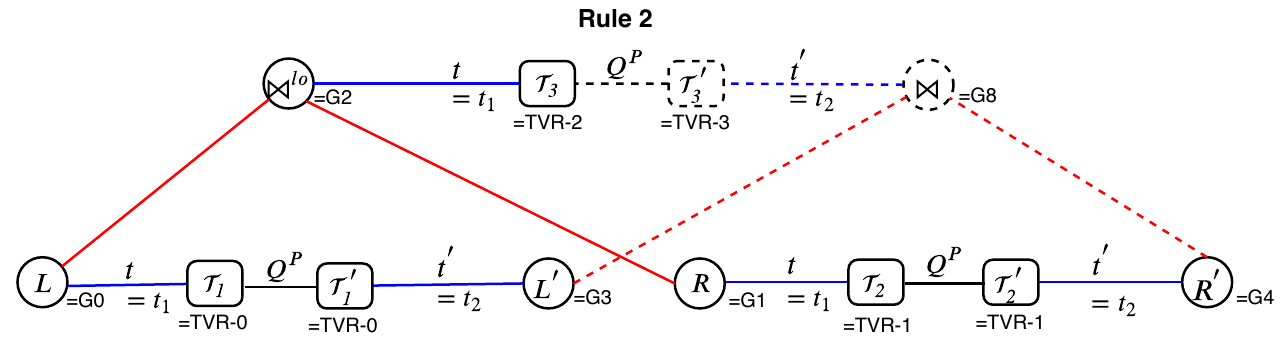}
  \label{fig:inter-tvr-rule}
}
\end{tabular}
\caption{Examples of the memo and TVR rewrite-rule patterns in \sys{}.}
\end{figure*}

\subsection{Memo: Capturing TVR Relationships}

The memo in the traditional Cascades-style optimizer only captures two levels of equivalence relationship:
\emph{logical equivalence} and \emph{physical equivalence}.
A logical equivalence class groups operators that generate the same result set;
within each logical equivalence class,
operators are further grouped into physical equivalence classes by their physical properties such as sort order, distribution, etc.
The ``Traditional Memo'' part in Fig.~\ref{fig:memo} depicts the traditional memo
of the \tablename{sales\_status} query.
For brevity, we omit the physical equivalence classes.
For instance, \op{LeftOuterJoin}[0,1] has Groups G0 and G1 as children,
and it corresponds to the plan tree rooted at $\leftouterjoin$.
G2 represents all plans logically equivalent to \op{LeftOuterJoin}[0,1].

However, the above two equivalences are not enough to capture the
rich relationships along the time dimension and between different incremental computation methods in the \xmodel{}.
For example, the relationship between snapshots and deltas of a TVR
cannot be modeled using the logical equivalence due to the following reasons.
Two snapshots at different times produce different relations, and
the snapshots and deltas do not even have the same schema (deltas
have an extra $\#$ column).
To solve this problem, on top of logical/physical equivalence classes,
we explicitly introduce TVR nodes into the memo, and keep track of the following relationships, shown as
the ``\sys{} Memo'' part in Fig.~\ref{fig:memo}.
\begin{itemize}[nosep,leftmargin=*]
	\item \textbf{Intra-TVR relationship} specifies the snapshot/delta relationship between logical equivalence classes of operators and the corresponding TVR's.
  For example,
  the traditional memo only models scanning the full content of $S$, i.e., $\s{S}{t_2}$,
  represented by G0,
  while the \sys{} memo models two more scans: scanning the partial content of $S$ available at $t_1$ ($\s{S}{t_1}$),
  and scanning the delta input of $S$ newly available at $t_2$ ($\dt{S}{t_1}{t_2}$).
  These two new scans are represented by G3 and G5, and the memo uses an explicit
  TVR-0 to keep track of these intra-TVR relationships.

	\item \textbf{Inter-TVR relationship} specifies the user-defined relationship between TVR's introduced by inter-TVR equivalence rules.
  For example, the \scm{} approach decomposes TVR-2 ($S \leftouterjoin R$)
  into two parts $Q^P$ and $Q^N$ as discussed in \S\ref{sec:model},
  represented by TVR-3 and TVR-4, respectively.
  It is worth noting that the above relationships are transitive.
  For instance, as G7 is the snapshot at $t_2$ of TVR-3 and TVR-3 is the $Q^P$ part of TVR-2,
  it is the snapshot at $t_2$ of the $Q^P$ part of TVR-2.
\end{itemize}

\subsection{Rule Engine: Enabling TVR Rewritings}

As the memo of \sys{} strictly subsumes a traditional Cascades memo,
traditional rewrite rules can be adopted and work without modifications.
Besides, the rule engine of \sys{} supports TVR rewrite rules.
\sys{} allows optimizer developers to define TVR rewrite rules by
specifying a graph pattern on both relational operators and TVR nodes in the memo.
A TVR rewrite rule pattern consists of two types of nodes and three types of edges:
(1) \emph{operator operands} that match relational operators;
(2) \emph{TVR operands} that match TVR nodes;
(3) \emph{operator edges} between operator operands that specify traditional parent-child relationship of operators;
(4) \emph{intra-TVR edges} between operator operands and TVR operands that specify intra-TVR relationships; and
(5) \emph{inter-TVR edges} between TVR operands that specify inter-TVR relationships.
All nodes and intra/inter-TVR edges can have predicates.
Once fired, TVR rewrite rules can register new TVR nodes and intra/inter-TVR relationships.

Fig.~\ref{fig:tvr-generating-rule}-\ref{fig:inter-tvr-rule} depict two TVR rewrite rules,
where solid nodes and edges specify the patterns to match, and dotted ones are newly registered
by the rules.
In the figures, we also show an example match of these rules
when applied on the memo in Fig.~\ref{fig:memo}:
\begin{itemize}[nosep,leftmargin=*]
  \item \textbf{Rule 1} is the TVR-generating rule for computing the delta of an inner join.
  It matches a snapshot of an \op{InnerJoin}, whose children $L$ ($R$)
  have a delta sibling $L'$ ($R'$) respectively.
  The rule will generate a \op{DeltaInnerJoin} taking
  $L$, $R$, $L'$, and $R'$ as inputs,
  and register it as a delta sibling of the original \op{InnerJoin}.

  \item \textbf{Rule 2} is an inter-TVR rule of \scm{}.
  The rule matches a snapshot of a \op{LeftOuterJoin},
  whose children $L$, $R$ each have a $Q^P$ snapshot sibling $L'$, $R'$.
  The rule will generate an \op{InnerJoin} of $L'$ and $R'$,
  and register it as the $Q^P$ snapshot sibling of the original \op{LeftOuterJoin}.
\end{itemize}

Fig.~\ref{fig:memo-breakdown} demonstrates the growth of a memo in \sys{}.
For each step, we only draw the updated part due to space limitation.
The memo starts with G0 to G2 and their corresponding TVR-0 to TVR-2.
In step 1, we first populate
the snapshots and deltas of the \op{scan} operators, e.g.,
G3 to G6, and register the intra-TVR relationship in TVR-0 and TVR-1.
We also populate their $Q^P$ and $Q^N$ inter-TVR relationships as in \scm{}
(for base tables these relationships are trivial).
In step 2, rule 2 fires on the tree rooted at \op{LeftOuterJoin[0,1]} in G2 as
in Fig.~\ref{fig:inter-tvr-rule}.
In step 3, rule 1 fires on the tree rooted at \op{InnerJoin[0,1]} in G7 as
in Fig.~\ref{fig:tvr-generating-rule}.
Similarly by applying other TVR rewrite rules,
we can eventually populate the memo in Fig.~\ref{fig:memo}.

\iffull

\subsection{Implementation}

\kai{Next we explain the implementation details of \sys{},
based on the terminology of Apache Calcite 1.17.0~\cite{calcite-website}.}

\kai{
Apache Calcite uses \impl{RelSet} and \impl{RelSubset} to represent
the logical and physical equivalence classes, respectively,
and \impl{Trait} to represent the physical properties of a physical equivalence class.
On top of that, \sys{} introduces \impl{TvrMetaSet}
to represent the TVRs, as well as \impl{IntraTvrTrait} and \impl{InterTvrTrait} to represent
the intra-TVR and inter-TVR relationship respectively.
Each intra-TVR/inter-TVR relationship is recorded in the involved TVRs and operators.
E.g., an intra-TVR relationship is modeled as a triple $\langle \impl{TvrMetaSet}, \impl{IntraTvrTrait}, \impl{RelSet} \rangle$ and stored in the corresponding \impl{TvrMetaSet} and
\impl{RelSet}.
We allow users to define their own \impl{IntraTvrTrait}'s and \impl{InterTvrTrait}'s,
and also implement several commonly-used ones as in \S\ref{sec:model}.
For instance, the attribute-perspective \impl{IntraTvrTrait}
for the group-by aggregate operator consists of the group-by keys (the primary keys
of the aggregate results), and the merge operator $+^\gamma$ for each aggregate column $\gamma$.
The \impl{InterTvrTrait} for the \scm{} approach comprises of an indicator
whether it is the $Q^P$ or $Q^N$ part.}

\kai{
To facilitate fast rule triggering, \sys{} indexes the rule patterns
by their containing operator operands and intra/inter-TVR edges.
Similar to Calcite, \sys{} monitors new structural changes
in the memo.
Whenever new operators or intra/inter-TVR relationships are registered in the memo,
\sys{} only fires the rule patterns that contain operator operands
or intra/inter-TVR edges that match the changes.
Note that we do not index the TVR operands and the operator edges,
as they do not have predicates to facilitate filtering.}

\kai{Initially, the rule engine starts with registering
the original logical plan,
and associates each \impl{RelSet} to a \impl{TvrMetaSet} with a default \impl{IntraTvrTrait}.
When a traditional rule is fired on a set of operators, if the \impl{RelSet} of every matched
operator is already connected to a \impl{TvrMetaSet} via the default \impl{IntraTvrTrait},
then besides registering the new operators generated by the rule, \sys{}
also creates a new \impl{TvrMetaSet} for each new operator and
connects them with the default \impl{IntraTvrTrait}.
In general, all structural changes in the memo will cause rules
to match and fire,
and further generate new nodes and edges.
\sys{} does not distinguish TVR rules from traditional rules in terms of rule firing.
All rule matches are stored in the same queue, and the firing order is determined by the customizable scoring function.
We used the default Calcite scoring function, which takes into consideration the rule importance and the location of the matched relation operators in the memo. We adjusted the scoring function for TVR rules by giving them a boosting factor, because TVR rules are transformations on logical plans and we want them to be fired with higher priorities.
Similar to Calcite, \sys{} deduplicates the TVRs besides the operators
when registering them in the memo.
Two TVRs are considered equivalent if they are both connected to a \impl{RelSet}
with the same default \impl{IntraTvrTrait}.}

\fi

\subsection{Speeding Up Exploration Process}
\label{sec:speed-up-opt}

In this section, we discuss optimizations to speed up the exploration process, which is needed since IQP explores a much bigger plan space than traditional query planning.


\noindent
\textbf{Translational symmetry of TVR's.}
The structures in the \sys{} memo usually have translation symmetry along the timeline,  because the same rule generates similar patterns
when applied on snapshots or deltas of the same set of TVR's.
For instance, in Fig.~\ref{fig:inter-tvr-rule}, if we let $t'=t_1$ instead,
$L'$ ($R'$) will match G0 (G1) instead of G3 (G4),
and we will generate the \op{InnerJoin} in G7 instead of G8.
In other words, \op{InnerJoin}[0,1] in G7 and \op{InnerJoin}[3,4] are translation symmetric,
modulo the fact that G0, G1, and G7 (G3, G4, and G8) are all snapshot $t_1$ ($t_2$) of the corresponding TVR's respectively.

By leveraging this symmetry, instead of repeatedly firing TVR rewrite rules
on snapshots/deltas of the same set of TVR's, we can apply the rules on just
one snapshot/delta, and copy the structures to other snapshots/deltas.
\kai{This helps eliminate the expensive repetitive matching process of
the same rule patterns on the memo.}
The improved process is as follows:
\begin{enumerate}[nosep,leftmargin=*]
\item \kai{We seed the TVR's of the leaf operators (usually \op{Scan})
with only one snapshot plus a consecutive delta, and fire all the rewrite rules to populate the memo.}

\item We seed the TVR's leaf operators with all snapshots and deltas,
and copy the memo from step 1 by substituting its leaf operators with their snapshot/delta
siblings in the corresponding TVR's.

\item We continue optimizing the copied memo, as we can further apply time-specific optimization, e.g., pruning empty relations if a delta at a specific time is empty.
\end{enumerate}

\noindent
\textbf{Pruning non-promising alternatives.}
There are multiple ways to compute a TVR's snapshot or delta,
within which certain ways are usually more costly than others.
We can prune the non-promising alternatives.
For instance, to compute a delta, one can take the difference
of two snapshots, or use TVR-generating rules to directly compute
from deltas of the inputs. Based on the experience of previous research on incremental
computation~\cite{ring}, we know that the plans generated by TVR-generating rules are usually more efficient.
Therefore, for operators that are known to be easily incrementally maintained, such as filter and project, we assign a lower importance to intra-TVR rules for generating deltas
to defer their firing.
Once we find a delta that can be generated through TVR-generating rules,
we skip the corresponding intra-TVR rules altogether.
\kai{To implement this optimization, we can give this subset of intra-TVR rules a lower priority than all other rules,
and thus other TVR rewrite rules and traditional rewrite rules will always
be ranked higher.
Each intra-TVR rule also has an extra skipping condition, which is tested to see whether the target delta is already generated
before firing the rule. If so, the rule is skipped.}

\noindent
\textbf{Guided exploration.}
Inside a TVR, snapshots and deltas  consecutive in time
can be merged together, leading to combinatorial explosion of rule applications.
However, the merge order of these snapshots and deltas usually do not
affect the cost of the final plan.
Thus, we limit the exploration to a left-deep merge order.
\kai{Specifically, we disable merging of consecutive deltas,
but instead only let rules match patterns that merge a snapshot
with its immediately consecutive delta.
In this way, we always use a left-deep merge order.}




\section{Selecting an Optimal Plan}
\label{sec:select-a-plan}

In this section we discuss how~\sys{} selects an optimal incremental plan in the explored space.
The problem
has two distinctions from existing query optimization: (1) costing the plan space and searching through the space need to consider the temporal execution of a plan; and (2) the optimal plan needs to decide which states to materialize
  to maximize the sharing opportunities between different time points within a query. 


\subsection{Time-Point Annotations of Operators}
\label{sec:costing}

\begin{figure}
\centering
\begin{tabular}{c}
\subfigure[]{
  \includegraphics[width=0.43\textwidth]{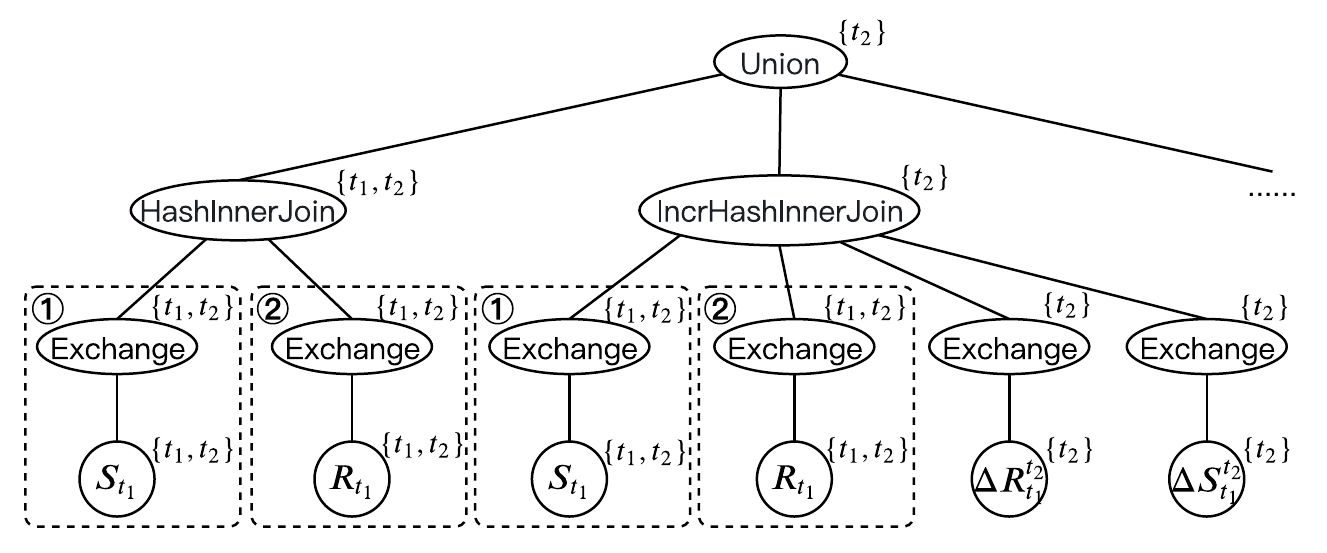}
  \label{fig:temporal-space}
}
\\
\subfigure[]{
  \includegraphics[width=0.43\textwidth]{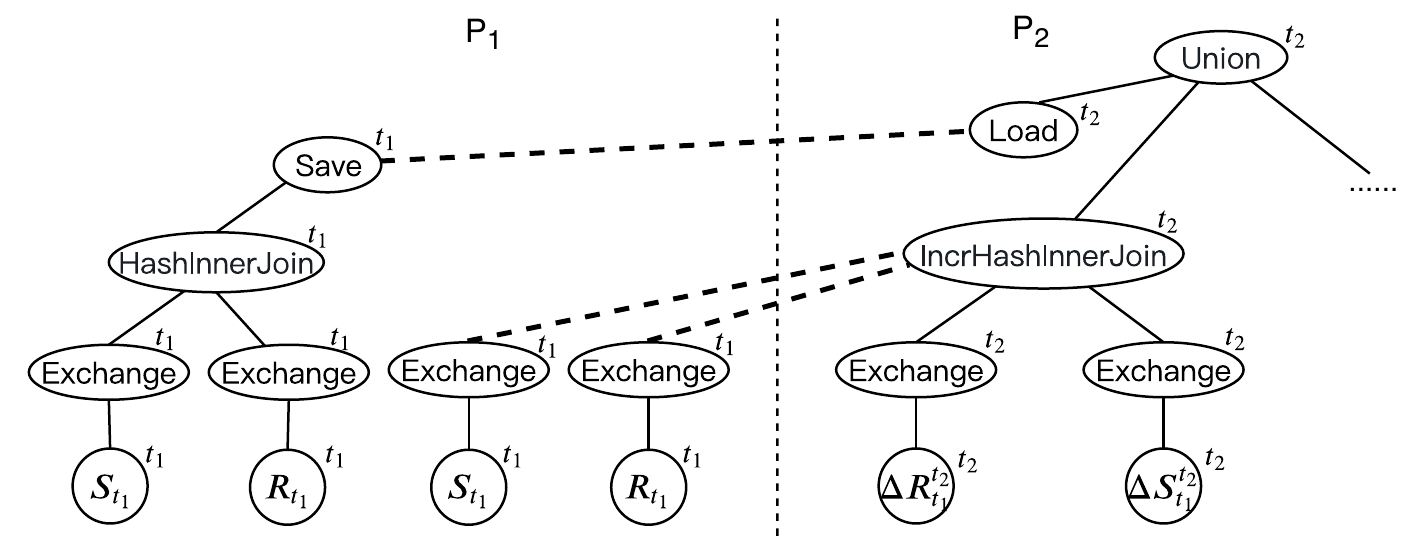}
  \label{fig:temporal-ex}
}
\end{tabular}
\caption{Examples of (a) the temporal plan space, and (b) an temporal assignment of the physical plan for subquery \tablename{sales\_status}.}
\end{figure}

Costing the plan alternatives properly is crucial for correct optimization.
However, as the temporal dimension is involved in query planning, costing is not trivial.
Fig.~\ref{fig:temporal-space} depicts one physical plan alternative derived from the plan
rooted at $\s{(S \leftouterjoin R)}{t_2}$ shown in red in Fig.~\ref{fig:tvr-transform}.
This plan only specifies the concrete physical operations
taken on the data, but does not specify when these physical operations
are executed.
Actually, each operator in the plan usually has multiple choices of execution time.
In Fig.~\ref{fig:temporal-space}, the time points annotated
alongside each operator shows the possible temporal domain in which each operator
can be executed.
For instance, snapshots $\s{S}{t_1}$ and $\s{R}{t_1}$ are available at $t_1$,
and thus can execute at any time after that, i.e., $t_1$ or $t_2$.
Deltas $\dt{R}{t_1}{t_2}$ and $\dt{S}{t_1}{t_2}$ are not available until $t_2$,
and thus any operators taking it as input, including the \op{IncrHashInnerJoin},
can only be executed at $t_2$.
The temporal domain of each operator $O$, denoted $\td{O}$, can be defined inductively:
\begin{itemize}[nosep,leftmargin=*]
  \item For a base relation $R$, $\td{R}$ is the set of execution times that are
  no earlier than the time point when $R$ is available.

  \item For an operator $O$ with inputs $I_1,\ldots,I_k$, $\td{R}$ is the intersection
  of its inputs' temporal domains:
  $\td{R}$ $=$ $\cap_{1\leq j\leq k}{\td{I_j}}$.
\end{itemize}

To fully describe a physical plan, one has to assign each operator in the plan an
execution time from the corresponding temporal domain.
We denote a specific execution time of an operator $O$ as $\et{O}$.
We have the following definition of a valid temporal assignment.
\begin{definition}[Valid Temporal Assignment]
An assignment of execution times to a physical plan is valid if and only if
 for each operator $O$, its execution time $\et{O}$ satisfies
$\et{O} \in \td{O}$ and $\et{O} \ge \et{O'}$ for all operators $O'$ in the subtree rooted at $O$.
\end{definition}
Fig.~\ref{fig:temporal-ex} demonstrates a valid temporal assignment
of the physical plan in Fig.~\ref{fig:temporal-space}.
As $\s{S}{t_1}$ and $\s{R}{t_1}$ are already available at $t_1$,
the plan chooses to  compute \op{HashInnerJoin} of $\s{S}{t_1}$ and $\s{R}{t_1}$ at $t_1$,
as well as shuffling $\s{S}{t_1}$ and $\s{R}{t_1}$ in order to prepare for
\op{IncrHashInnerJoin}.
At $t_2$, the plan shuffles the new deltas $\dt{S}{t_1}{t_2}$ and $\dt{R}{t_1}{t_2}$,
finishes \op{IncrHashInnerJoin}, and unions the results with that of \op{HashInnerJoin}
computed at $t_1$.
Note that if an operator $O$ and its input $I$ have different execution times,
then the output of $I$ needs to be saved first at $\et{I}$,
and later loaded and fed into $O$ at $\et{O}$,
e.g., \op{Union} at $t_2$ and \op{HashInnerJoin} at $t_1$.
The cost of \op{Save} and \op{Load} needs to be properly included in the plan cost.
It is worth noting that some operators save and load the output as a by-product,
for which we can spare \op{Save} and \op{Load},
e.g., \op{Exchange} of $\s{S}{t_1},\s{R}{t_1}$ at $t_1$ for \op{IncrHashInnerJoin}.

\subsection{Time-Point-Based Cost Functions}
\label{sec:single-opt}

The cost of an incremental plan
is defined under a specific assignment of execution times.
Therefore, the optimization problem of searching
for the optimal incremental plan is formulated as:
given a plan space, find a physical plan and temporal assignment
that achieve the lowest cost.
In this section, we discuss the cost model and optimization algorithm for this problem
\kai{without considering sharing common sub-plans.
We will discuss the problem of how to decide the optimal sharable sub-plans to materialize
in \S\ref{sec:mqo-sharing}.}

\kai{As an incremental plan can span across multiple time points,
the cost function $\tilde{\mathfrak{c}}$ in an IQP problem (as in \S\ref{sec:def}) is
extended to a function taking into consideration of costs at different times.
For the cost at each time point,
we inherit the general cost model used in traditional query optimizers,
i.e., the cost of a plan is the sum of the costs of all its operators.}
Below we give two examples of $\tilde{\mathfrak{c}}$.
We denote traditional cost functions as $c$, and $c_i$ is the cost at time $t_i$.
As before, $c$ can be a number, e.g., estimated monetized resource cost,
or a structure, e.g., a vector of CPU time and I/O.
\begin{enumerate}[nosep,leftmargin=*]
\item $\tilde{c}_w(O) = \sum_{i = 1..T}{w_i \cdot c_i(O)}$.
That is, the extended cost of an operator is a weighted sum of its cost at each time $t_i$.
For the example setting in \S\ref{sec:challenge}, $w_1 = 0.2$ for $t_1$ and $w_2=1$ for $t_2$.


\item $\tilde{c}_v(O) = [c_1(O),\ldots,c_T(O)]$. That is, the extended cost is
a vector combining costs at different times. \kai{$\tilde{c}_v$ can be compared
entry-wise in a reverse lexical order.
Formally, $\tilde{c}_v(O_1) > \tilde{c}_v(O_2)$ iff $\exists j$ s.t. $c_j(O_1) > c_j(O_2)$
and $c_i(O_1) = c_i(O_2)$ for all $i, j < i \leq T$. }
\end{enumerate}
\kai{
\noindent Consider the plan in Fig.~\ref{fig:temporal-space} as an example.
To get the result of \op{HashInnerJoin} at $t_2$, we have two options:
(i) compute the join at $t_2$; or (ii) as in Fig.~\ref{fig:temporal-ex},
compute the join at $t_1$, save the result, and load it back at $t_2$.
Assume the cost of computing \op{HashInnerJoin}, saving the result, and loading it are $10$, $5$, $4$, respectively.
Then for option (i) $(c_1, c_2) = (0, 10)$,
for option (ii) $(c_1, c_2) = (15, 4)$.
Say that we use $\tilde{c}_w$ as the cost function.
If $w_1 = 0.6$ and $w_2=1$ then option (i) is better,
whereas if $w_1 = 0.2$ and $w_2 = 1$, option (ii) becomes better.
}

\boldstart{Dynamic programming} used predominantly in existing query optimizers~\cite{r-opt,starburst,volcano}
also need to be adapted to handle the cost model extensions.
In these existing query optimizers, the state space of the dynamic programming is
the set of all operators in the plan space, represented as $\{O\}$.
Each operator $O$ records the best cost of all the subtrees rooted at $O$.
We extend the state space by considering all combinations of operators and their execution times, i.e., $\{O\} \times \td{\{O\}}$.
Also instead of recording a single optimum, each operator $O$ records multiple optimums,
one for each execution time $\et{O}$,
which represents the best cost of all the subtrees rooted at $O$
if $O$ is generated at $\tau$.
During optimization, the state-transition function is as
Eq.~\ref{eq:dp}. That is, the best cost of $O$ if executed at $\tau$ is the best cost
of all possible plans of computing $O$ with all possible valid temporal assignments compatible with $\tau$.
\begin{equation}
\tilde{\mathfrak{c}}[O, \tau] = min_{\forall \text{ valid } \tau_j} \bigl(\sum_j\tilde{\mathfrak{c}}[I_j, \tau_j] + c_\tau(O)\bigr).
\label{eq:dp}
\end{equation}
In general, we can apply dynamic programming to the optimization problem
for any cost function satisfying the property of optimal substructure.  We have the following correctness result of the above dynamic programming algorithm.
\begin{theorem}
\kai{The optimization problem under cost functions $\tilde{c}_w$ and $\tilde{c}_v$
without sharing common sub-plans
satisfies the property of optimal substructure, and dynamic programming
is applicable.}
\label{thm:dp}
\end{theorem}



\subsection{Deciding States to Materialize}
\label{sec:mqo-sharing}

The problem of deciding the states to materialize
can be modeled as finding the sharing opportunities in the plan space.
In other words, a shared sub-plan between $P_i$ and $P_j$ in an incremental plan
is essentially an intermediate state that can be saved by $P_i$ and reused by $P_j$.
For example, in Fig.~\ref{fig:temporal-space},
since both \op{HashInnerJoin} and \op{IncrHashInnerJoin} require shuffling $\s{S}{t_1}$ and $\s{R}{t_1}$,
the two relations can be shuffled only once and reused for both joins.
The parts \textcircled{1} and\textcircled{2} circled in dashed lines depict the sharable sub-plans.

\iffull

Finding the optimal common sub-plans to share is a multi-query optimization (MQO) problem,
which has been extensively studied~\cite{mqo,mqo-jr,mqo-pods}.
In this paper, we extend the MQO algorithm in~\cite{mqo-pods},
which proposes a greedy framework on top of Cascade-style optimizers for MQO.
For the sake of completeness, we list the algorithm in
Algo.~\ref{alg:greedy}, by highlighting the extensions for progressive planning.
The algorithm runs in an iterative fashion.
In each iteration, it picks one more candidate from all possible shareable candidates,
which if materialized can minimize the plan cost (line 4),
where $bestPlan(\mathbb{S})$ means the best plan with $\mathbb{S}$ materialized and shared.
The algorithm terminates when all candidates are considered or adding candidates can no longer
improve the plan cost.
As IQP needs to consider the temporal dimension,
the shareable candidates are no longer solely the set of shareable sub-plans,
but pairs of a shareable sub-plan $s$ and a choice of its execution time $\et{s}$.
Pair $\langle s, \et{s} \rangle$ means computing and materializing the sub-plan $s$
at time $\et{s}$, which can only benefit the computation that happens after $\et{s}$.
For instance, considering the physical plan space in Fig.~\ref{fig:temporal-space},
the sharable candidates are $\{\langle \text{\textcircled{1}}, t_1 \rangle, \langle \text{\textcircled{1}}, t_2 \rangle, \langle \text{\textcircled{2}}, t_1 \rangle, \langle \text{\textcircled{2}}, t_2 \rangle\}$.
The optimizations in~\cite{mqo-pods} are still applicable to Algo.~\ref{alg:greedy}.

\begin{algorithm}[th]
\footnotesize
\algsetup{indent=2em}
\begin{algorithmic}[1]
\STATE $\mathbb{S}=\emptyset$
\STATE $\mathbb{C}=$ \textbf{shareable candidate set consiting of all shareable nodes and their potential execution times $\{\langle s, \et{s} \rangle\}$}
\WHILE {$\mathbb{C} \neq \emptyset$}
  \STATE Pick $\langle s, \et{s} \rangle \in \mathbb{C}$ that minimizes
  $\tilde{\mathfrak{c}}(bestPlan(\mathbb{S}'))$
  where $\mathbb{S}' = \{\langle s, \et{s} \rangle \} \cup \mathbb{S}$
  \IF {$\tilde{\mathfrak{c}}(bestPlan(\mathbb{S}')) < \tilde{\mathfrak{c}}(bestPlan(\mathbb{S}'))$}
    \STATE $\mathbb{C} = \mathbb{C} - \{\langle s, \et{s} \rangle\}$
    \STATE $\mathbb{S} = \mathbb{S}'$
  \ELSE
    \STATE $\mathbb{C}=\emptyset$
  \ENDIF
\ENDWHILE
\RETURN $\mathbb{S}$
\end{algorithmic}
\caption{Greedy Algorithm for MQO}
\label{alg:greedy}
\end{algorithm}


As expanded with execution time options, the enumeration space of the shareable candidate set becomes much larger than the original algorithm in~\cite{mqo-pods}.
Interestingly, we find that under certain cost models we can reduce the enumeration space
down to a size comparable to the original algorithm, formally summarized in
Theorem~\ref{thm:store-early}.
This theorem relies on the fact that
materializing a shareable sub-plan at its earliest possible time
subsumes other materialization choices. Due to space limit, we omit the proof.

%
%
%
\begin{theorem}
For a cost function $\tilde{c}_w$ satisfying $w_i < w_j$ if $i < j$,
or a cost function $\tilde{c}_v$ satisfying the property that an
entry $i$ has a lower priority than an entry $j$ if $i < j$ in the lexical order,
we only need to consider the earliest valid execution time for each shareable sub-plan.
That is, for each shareable sub-plan $s$, we only need to consider the shareable candidate $\langle s, min(\td{s}) \rangle$ in Algorithm~\ref{alg:greedy}.
\label{thm:store-early}
\end{theorem}

\fi

\section{Beanstalk in Action}
\label{sec:discussion}

\kai{In this section, we discuss a few important considerations
when applying \sys{} in practice.}

\noindent
\kai{\textbf{Dynamic Re-optimization of incremental plans.}
We have studied the IQP problem assuming a static setting,
i.e., in $(\vec{T}, \vec{D}, \vec{Q}, \tilde{\mathfrak{c}})$
where $\vec{T}$ and $\vec{D}$
are given and fixed.
In many cases, the setting can be much more dynamic
where $\vec{T}$ and $\vec{D}$ are subject to change.
Fortunately, \sys{} can be adapted to a dynamic setting using re-optimization.
Generally, an incremental plan $\pl{} = [P_1, \cdots, P_{i-1}, P_i,$ $\cdots, P_k]$
for $\vec{T} = [t_1, \cdots, t_{i-1}, t_i, \cdots, t_k]$
is only executed up to $t_{i-1}$,
after which $\vec{T}$ and $\vec{D}$ change
to $\vec{T'} = [t_{i'}, \cdots, t_{k'}]$ and $\vec{D'} = [D_{i'}, \cdots, D_{k'}]$.
\sys{} can adapt to this change by re-optimizing the plan
under $\vec{T'}$ and $\vec{D'}$. We want to remark that
during re-optimization, \sys{} can incorporate the materialized states
generated by $P_1, \cdots, P_{i-1}$ as materialized views. In this way \sys{}
can choose to reuse the materialized states instead of blindly recomputing everything.}

\kai{Take the progressive data warehouse scenario as an example.
Since we may not know the exact schedule and data statistics in the future,
we do not optimize for a large number of time points at once.
Instead, one can plan adaptively using re-optimization.
Consider a query originally scheduled at $t^*$.
Say, initially at $t_1$ 
we decide to have one incremental run. 
At this planning stage, we only need to consider a simple schedule $[t_1, t^*]$.
In the resulting plan $[P_1, P^*]$, only $P_1$ is executed,
but co-planing $P_1,P^*$ makes $P_1$ robust for future runs.
At a future time $t_i$, if we decide to have another run,
we can always re-optimize for $[t_i, t^*]$
with accurate statistics for $t_i$ (as data is already available).
We can also take into consideration the materialized states from previous runs
before $t_i$.
Note that at each planning, we only optimize a small IQP problem with
a limited number of time points.
This way, we avoid planning many time points
with uncertain schedule and statistics.
The above methods are used in~\cite{grosbeak}.}

\noindent
\kai{\textbf{Data statistics estimation.}
Incremental processing scenarios usually involve
planning for logical times or physical times in the future,
for which estimating the data statistics becomes very challenging.
IQP scenarios, including both planning for logical times (e.g., \texttt{IVM-PD})
and physical times (e.g., \texttt{PWD-PD}) as described in \S\ref{sec:def},
usually involve recurring queries.
Thus, we can use historical data's arrival patterns to estimate the statistics
of available data at logical time points or physical time points in the future.
Having inaccurate statistics is not a new problem
to query optimization, and many techniques have been proposed~\cite{robust-planning-survey}
to tackle this issue.
Note that we can always re-optimize the plan when we find that the previously estimated
statistics is not accurate.
Also, techniques such as robust planning~\cite{babu-proactive,graefe-robust,optimality-range}
and parametric planning~\cite{ioannidis-parametric} can be adopted to IQP too.
It is out of the scope of this paper.}

\section{Experiments}
\label{sec:experiments}

\begin{figure*}[tbh!]
\centering
\begin{tabular}{ccc}
\subfigure[]{
  \includegraphics[height=0.16\textwidth]{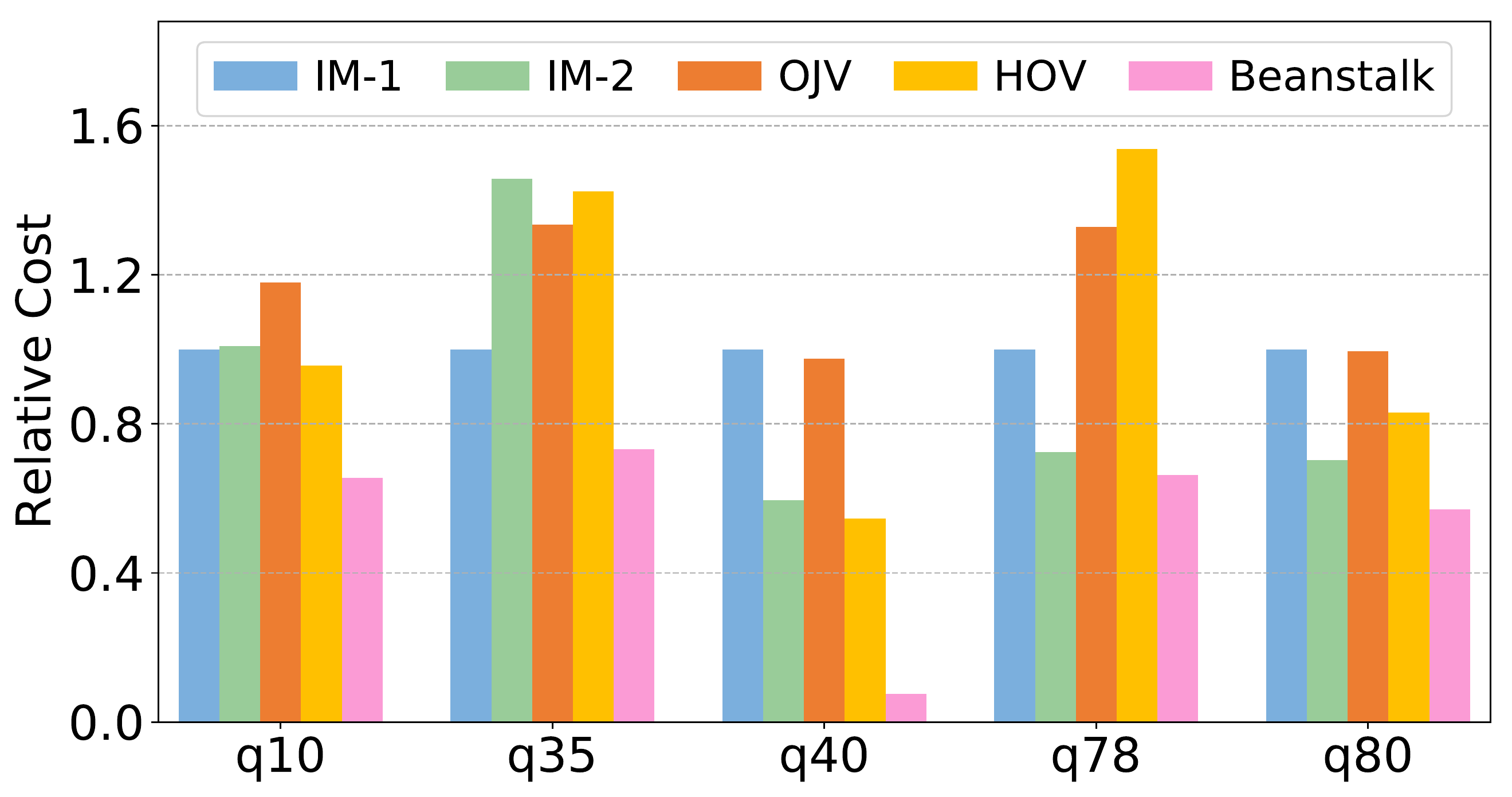}
  \label{fig:ivm-query}
}
&
\subfigure[]{
  \includegraphics[height=0.16\textwidth]{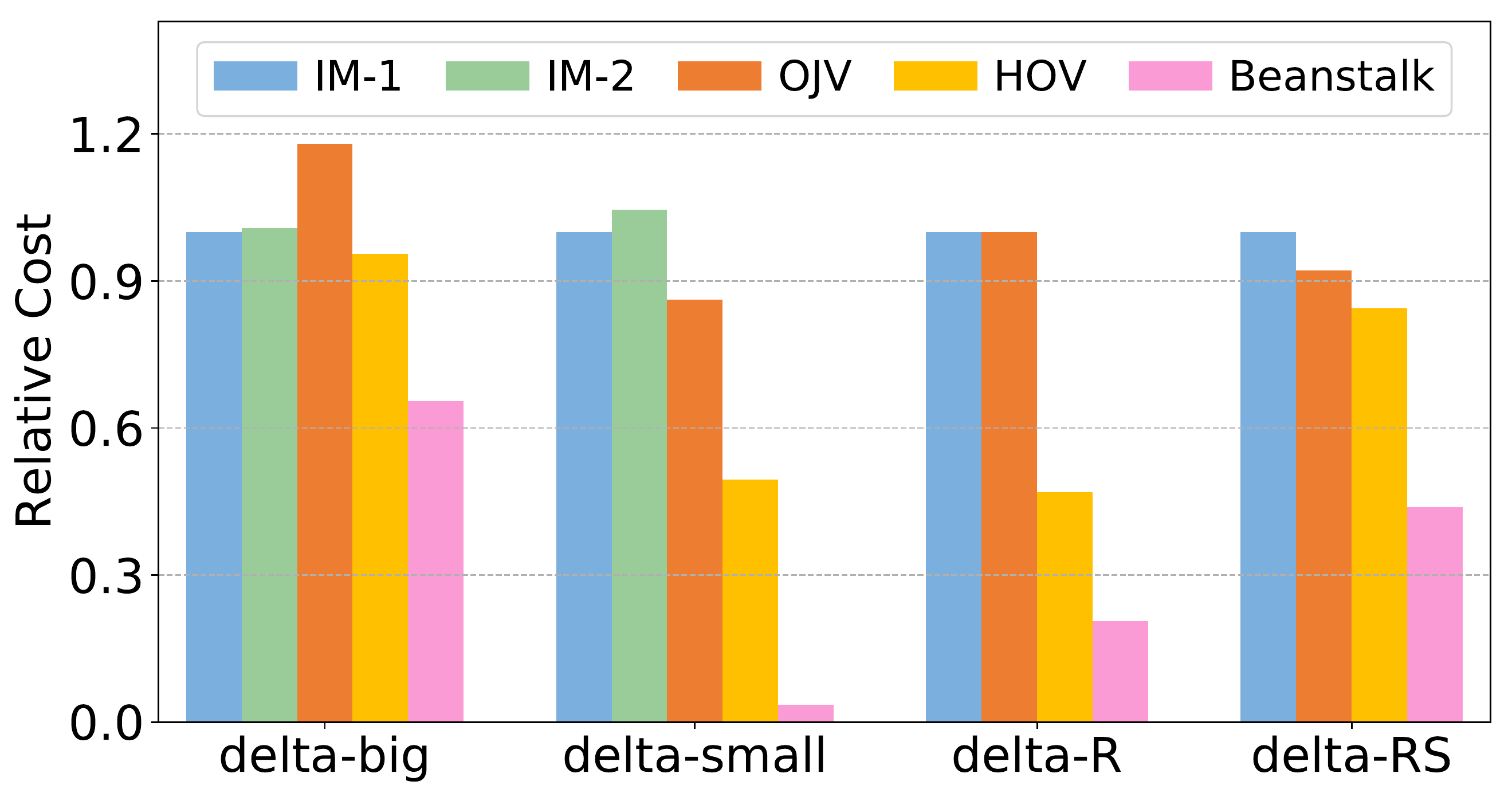}
  \label{fig:ivm-delta}
}
&
\subfigure[]{
  \includegraphics[height=0.16\textwidth]{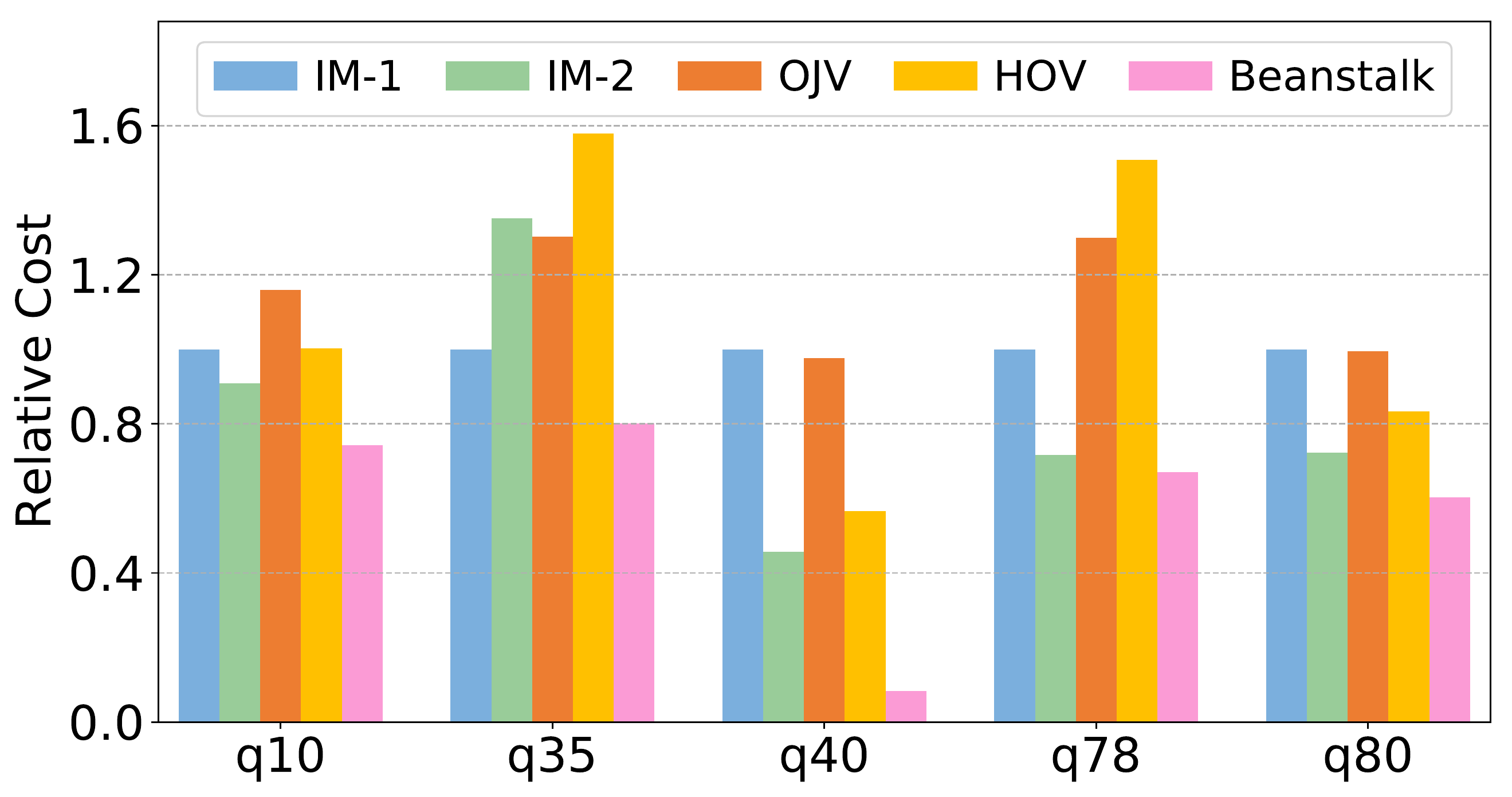}
  \label{fig:pdw-query}
}
\\
\subfigure[]{
  \includegraphics[height=0.16\textwidth]{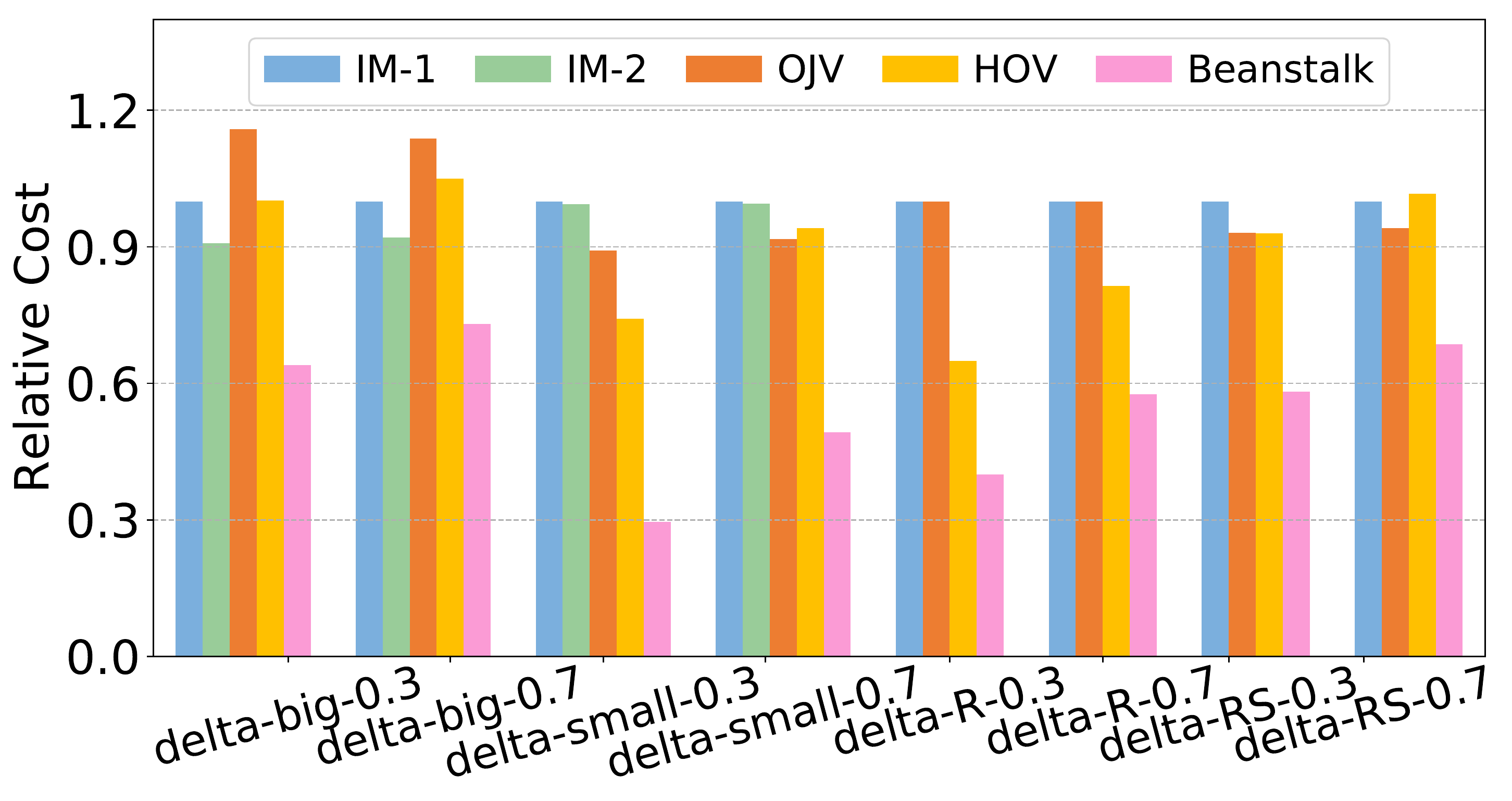}
  \label{fig:pdw-delta}
}
&
\subfigure[]{
  \includegraphics[height=0.16\textwidth]{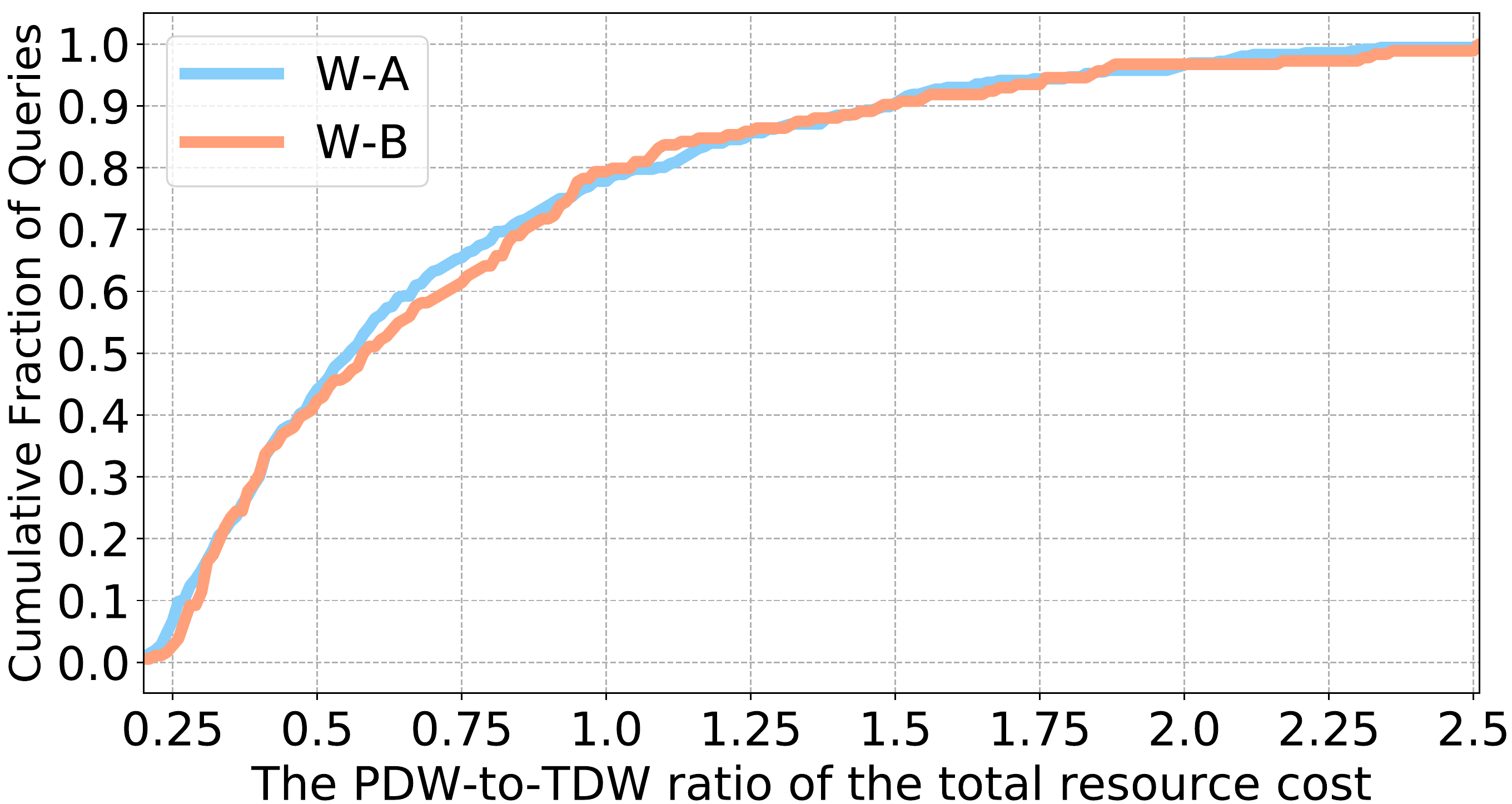}
  \label{fig:total-vs-total}
}
&
\subfigure[]{
  \includegraphics[height=0.16\textwidth]{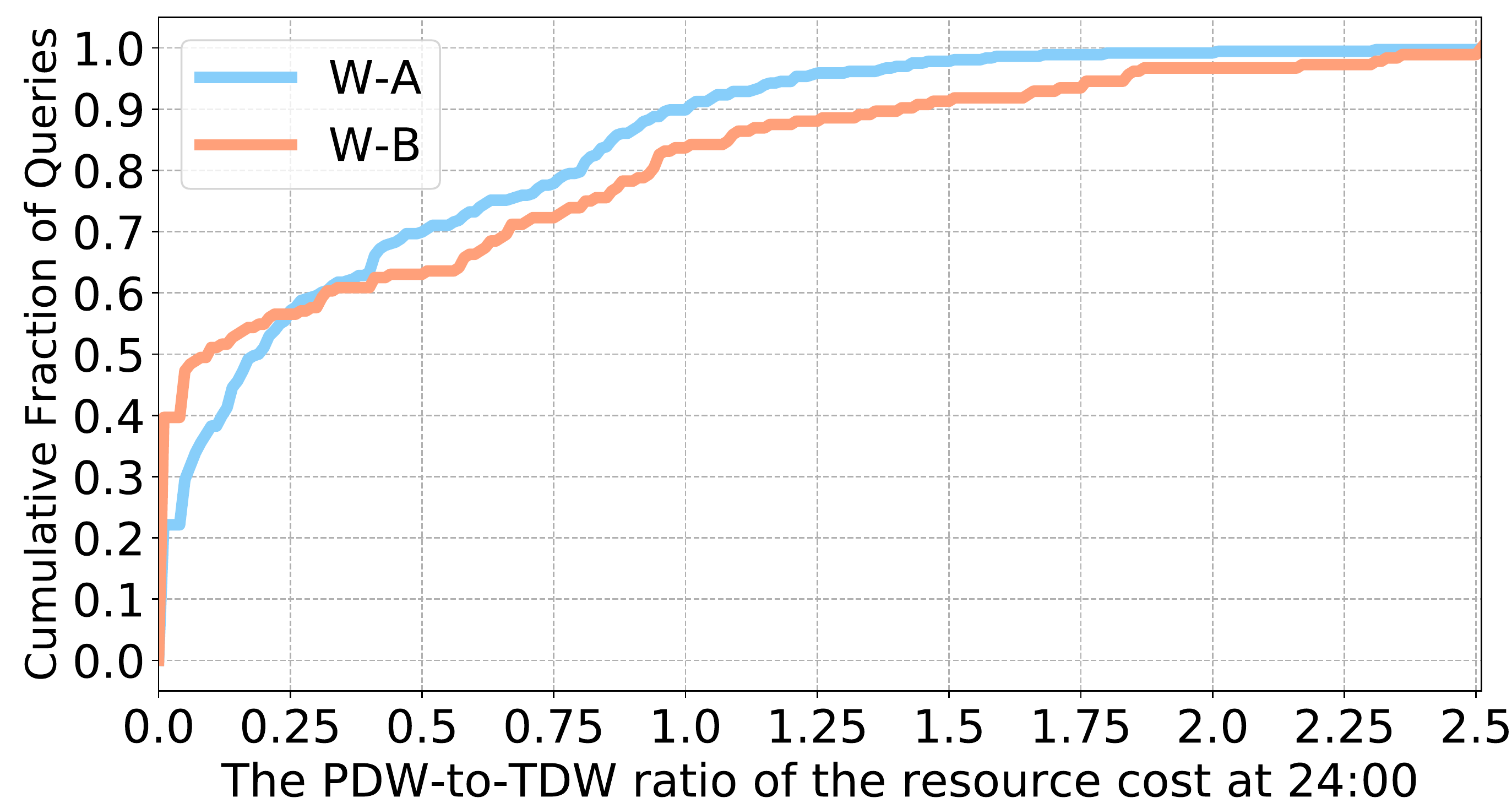}
  \label{fig:last-vs-total}
}
\end{tabular}
\caption{(a)(b) The optimal \kai{estimated} costs of different incremental plans in \texttt{IVM-PD}
for different queries and data-arrival patterns.
(c)(d) The optimal \kai{estimated} costs of different incremental plans in \texttt{PDW-PD} for different queries, data-arrival patterns and cost weights.
(e)(f) The PDW-to-TDW ratio of the \kai{real} total CPU cost and CPU cost at $24$:$00$ for the data warehouse workloads respectively.}
\label{fig:plan-cost}
\end{figure*}

In this section, we report our experimental study on the effectiveness and efficiency of \sys{}
in IQP.

\subsection{Settings}

We used the query optimizer of Alibaba Cloud MaxCompute~\cite{maxcompute}, which was built on Apache Calcite 1.17.0~\cite{calcite-website}, as a traditional optimizer baseline.
We implemented \sys{} on the optimizer of MaxCompute.
We integrated four commonly used incremental methods into \sys{} by expressing them as
TVR-rewrite rules:
(1) \textbf{IM-1}: as described in \S\ref{sec:challenge}.
(2) \textbf{IM-2}: as described in \S\ref{sec:challenge} and \S\ref{sec:inter-tvr}.
(3) \textbf{OJV}: the outer-join view maintenance algorithm in \S\ref{sec:inter-tvr}.
(4) \textbf{HOV}: the higher-order view maintenance algorithm in \S\ref{sec:inter-tvr}.
By default, \sys{} jointly considered all four methods to generate an optimal plan.
In the experiments, we used \sys{} to simulate each method
by turning off the inter-TVR rules of the other methods.

\noindent
\textbf{Incremental Processing Scenarios.}
We used two incremental processing scenarios
to demonstrate \sys{}. Below are the incremental planning problem definitions
for these two scenarios.
\begin{itemize}[nosep,leftmargin=*]
  \item \textbf{Progressive data warehouse.}
  We use the \texttt{PDW-PD} definition as in \S\ref{sec:challenge}.
  Specifically,
  We used $\tilde{c}_w(O)$ (in \S\ref{sec:single-opt}) as the cost function,
  where the cost at each time $c_i$ was a linear combination of the estimated usage
  of CPU, IO, memory, and network transfer, and the weight $w_i \in [0.25,0.3]$ for early runs
  and $w_i = 1$ for the last run. Note that it helped us achieve a balance between re-computation and the size of materialized states by jointly considering CPU and memory in the cost function.
  We chose $w_i \leq 0.3$ for early runs to simulate the tiered rates
  of resources on Amazon AWS~\cite{aws-spot}, where spot instances usually cost less
  than $30\%$ of on-demand instances.

  \item \textbf{Incremental view Maintenance.}
  We use the \texttt{IVM-PD} definition as defined in \S\ref{sec:def}.
\end{itemize}

\noindent
\textbf{Query Workloads.}
We used the TPC-DS benchmark~\cite{tpcds} (\kai{$1TB$}) to study the effectiveness (\S\ref{sec:effect}) and performance (\S\ref{sec:perf}) of \sys{}.
To further demonstrate the effectiveness of the incremental plans produced by \sys{},
in \S\ref{sec:case}, we used two real-world analysis workloads consisting of
recurrent daily jobs from Alibaba's enterprise progressive data warehouse,
denoted as \texttt{W-A} and \texttt{W-B}.
\iffull
Table~\ref{tbl:workloads} shows statistics of the two workloads.
\begin{table}[tbh!]
\tiny
\caption{Statistics of two workloads at Alibaba}
\label{tbl:workloads}
\begin{tabular}{cccccc}
\hline
 & \# Queries & Avg. \# Joins & \# with $\ge 1$ join & \# with $\ge 2$ joins \\
\hline
\texttt{W-A} & $274$ & $1.14$ & $167$ & $83$ \\
\hline
\texttt{W-B} & $554$ & $1.18$ & $357$ & $144$ \\
\hline
\end{tabular}
\end{table}
\fi

\noindent
\kai{\textbf{Running Environment.}
In the experiments,
query optimization was carried out single-threaded on a machine
with an Intel Xeon Platinum $8163$ CPU @ $2.50$GHz and $512$GB memory,
whereas the generated query was executed on a cluster
of thousands of machines shared with other production workloads.}

\subsection{Effectiveness of IQP}
\label{sec:effect}

We first evaluated the effectiveness of IQP
by comparing \sys{} with four individual incremental methods \ivm{}, \scm{}, \ojv{},
and \hov{}, in both the \texttt{PDW-PD} and \texttt{IVM-PD} scenarios.
We controlled and varied two factors in the experiments:
(1) Queries. We chose five representative queries covering complex joins (inner-, left-outer-,~and left-semi-joins) and aggregates.
(2) Data-arrival patterns. We controlled the amount of input data available in the 1st and
2nd incremental runs ($D_1, D_2$) and whether there are retractions in the input data.
Correspondingly, we chose the following four data-arrival patterns,
namely delta-big ($|D_1|/|D_2|=1:1$), delta-small ($|D_1|/|D_2|=4:1$), delta-R ($|D_1|/|D_2|=2:1$ with retractions in the \tablename{sales} tables),
and delta-RS ($|D_1|/|D_2|=2:1$ with retractions in both \tablename{sales} and \tablename{returns} tables).
\kai{As the accuracy of cost estimation is orthogonal to \sys{},
to isolate its interference,
we mainly compared the estimated costs of plans produced by the optimizer,
and reported them in relative scale
(dividing them by the corresponding costs of \ivm{}) for easy comparison.
We reported the real execution costs \iffull as a reference later. \fi}
Due to space limit,
we only report the most significant entries in the cost vector of $\tilde{c}_v$
for \texttt{IVM-PD}.

\noindent
\textbf{IVM-PD}. We first fixed the data-arrival pattern to delta-big and varied the queries.
The optimal-plan costs are reported in Fig.~\ref{fig:ivm-query}.
As shown, different queries prefer different incremental methods.
For example, \ivm{} outperformed both \ojv{} and \hov{} for complex queries such as q35.
This is because \ojv{} computed $Q^I$ by left-semi joining the delta of $Q^D$
with the previous snapshot (\S\ref{sec:inter-tvr}), and a bigger delta incurred
a higher cost of computing $Q^I$.
Whereas for simpler queries such as q80, \ojv{} degenerated to a similar plan as \ivm{},
and thus had similar costs. Note that \hov{} cost much less than both \ojv{} and \ivm{}.
This is because \hov{} maintained extra higher-order views
(e.g., \tablename{catalog\_sales} inner joining with \tablename{warehouse}, \tablename{item} and \tablename{date\_dim}) and thus avoided repeated recomputation of these views as in \ojv{} and \ivm{}.
\sys{} outperformed the individual incremental methods on all queries.
The reason is that \sys{} was able to combine the benefits of different incremental methods.

Next we chose query q10 as a complex query with multiple left outer joins,
and varied the data-arrival patterns.
The results are plotted in Fig.~\ref{fig:ivm-delta}.
Again, the data-arrival patterns affected the preference of incremental methods.
For example, \scm{} could not handle input data with retractions.
Compared to delta-big, \hov{} started to outperform \ivm{} by a large margin
in delta-small, as both of them could use different join orders when applying updates to different
input relations, and a smaller delta help significantly reduce the cost of incrementally
computing $M$ in \hov{} and $Q^D,Q^I$ in \ojv{}.

For both experiments, \sys{} consistently delivered the best plans among all the methods.
For example,
for q40 in Fig.~\ref{fig:ivm-query} and the delta-small case in Fig.~\ref{fig:ivm-delta},
\sys{} delivered a plan $5$-$10$X better than others.
\sys{} leveraged all three of \hov{}, \scm{} and \ivm{} to generate a mixed optimal plan:
\sys{} chose different join orders when applying updates to different input relations,
which was similar to \hov{}. It leveraged the fact that joining the smaller delta
earlier can quickly reduce the output sizes.
Interestingly, when it came to combining higher-order views $M$ and $\Delta R$ as required
by \hov{}, \sys{} used the \scm{} approach,
and applied \ivm{} to incrementally compute the $Q^N$ part in \scm{}.

\noindent
\textbf{PDW-PD}. For the \texttt{PDW-PD} scenario, we conducted the same experiments
by varying the queries and the data-arrival patterns as in \texttt{IVM-PD},
and in addition tried different weights used in the cost functions ($w_1 = 0.3$ vs. $w_1 = 0.7$).
We reported the results in \kai{Figures~\ref{fig:pdw-query} and~\ref{fig:pdw-delta}}. We have similar conclusions as in \texttt{IVM-PD}.
We make two remarks.
(1) Since the \texttt{PDW-PD} setting did not require any outputs at earlier runs,
\sys{} automatically avoided unnecessary computation,
e.g., computing the $Q^N$ part in an \scm{} approach,
which usually cannot be efficiently incrementally maintained.
This result is also shown in the figures,
as the \scm{} approach was more preferred uniformly in \texttt{PDW-PD} than in \texttt{IVM-PD}.
(2) The weights in the cost function can also affect the choice of the optimizer.
For instance, in Fig.~\ref{fig:pdw-delta}, q10 preferred \hov{} to \ojv{} when $w_1=0.3$, but the other way when $w_1 = 0.7$. This result was because as the cost of early execution
became higher, it was less preferable to store many intermediate states as in \hov{}.
\sys{} automatically exploited this fact and adjusted the amount of computation
in each incremental run.
When $w_1$ increased from $0.3$ to $0.7$,
\sys{} moved some early computation from the first incremental run to the second.

\iffull

\begin{figure*}[tbh!]
\centering
\begin{tabular}{ccc}
\subfigure[]{
  \includegraphics[height=0.16\textwidth]{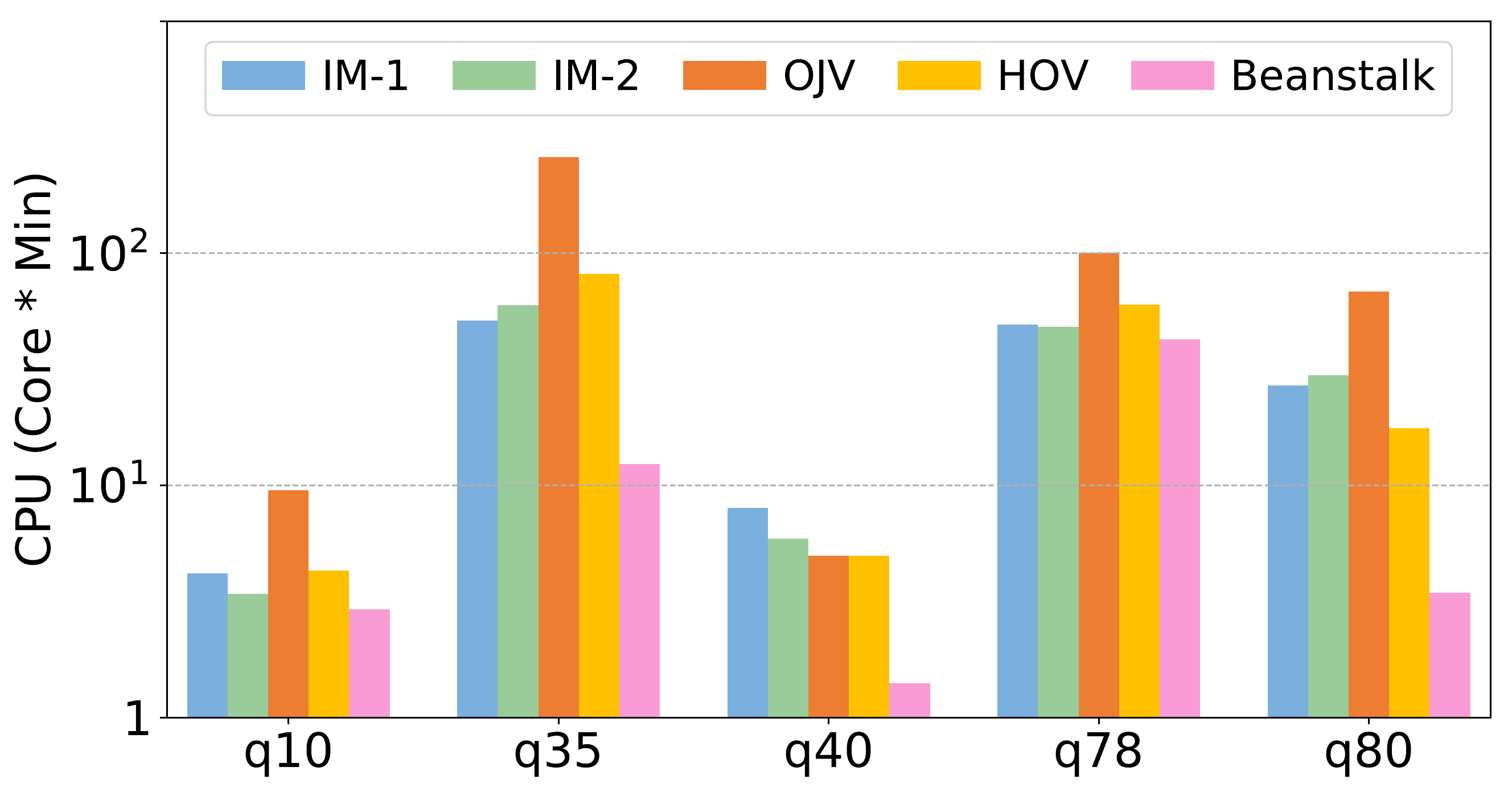}
  \label{fig:ivm-query-cpu}
}
&
\subfigure[]{
  \includegraphics[height=0.16\textwidth]{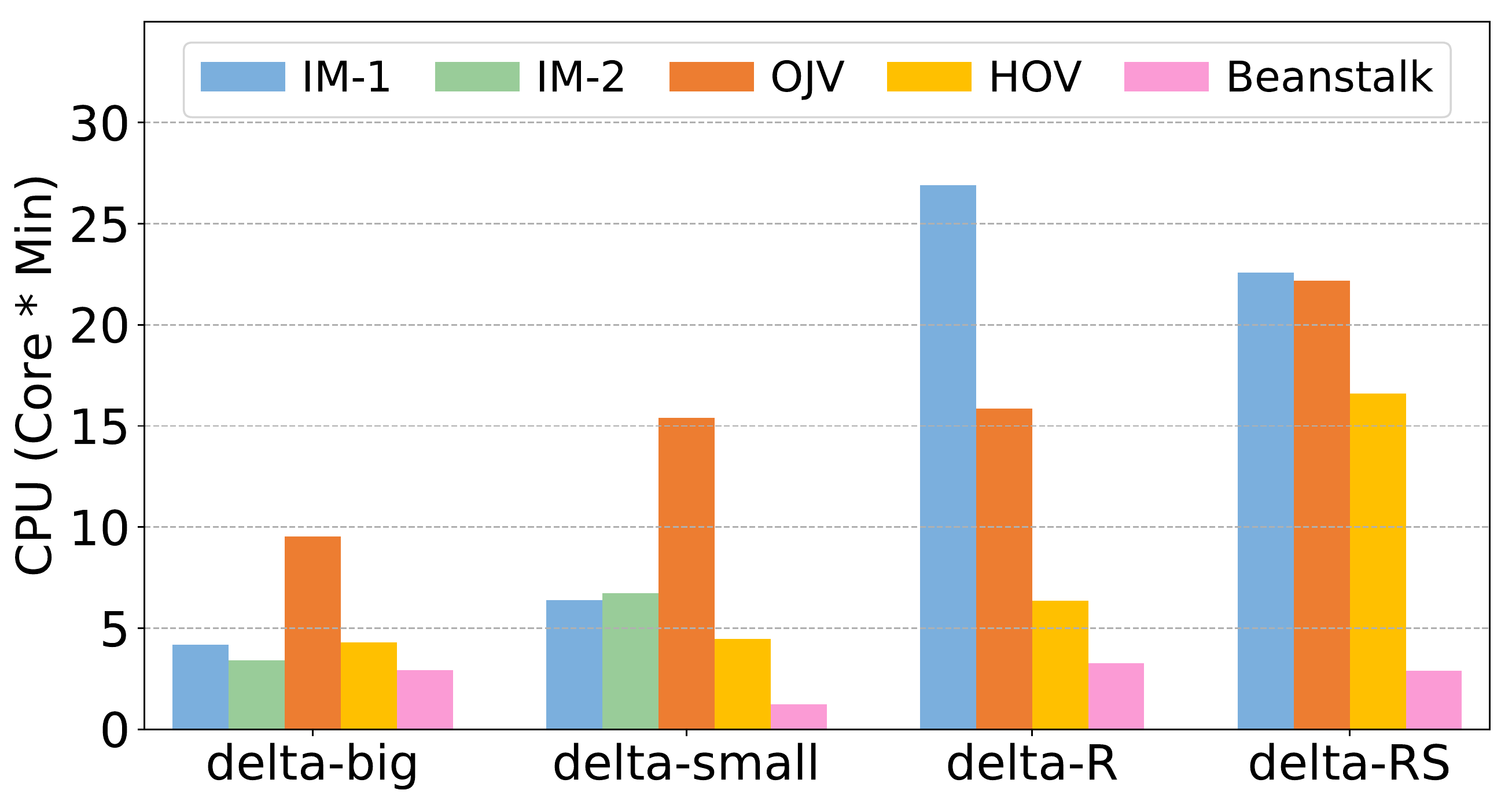}
  \label{fig:ivm-delta-cpu}
}
&
\subfigure[]{
  \includegraphics[height=0.16\textwidth]{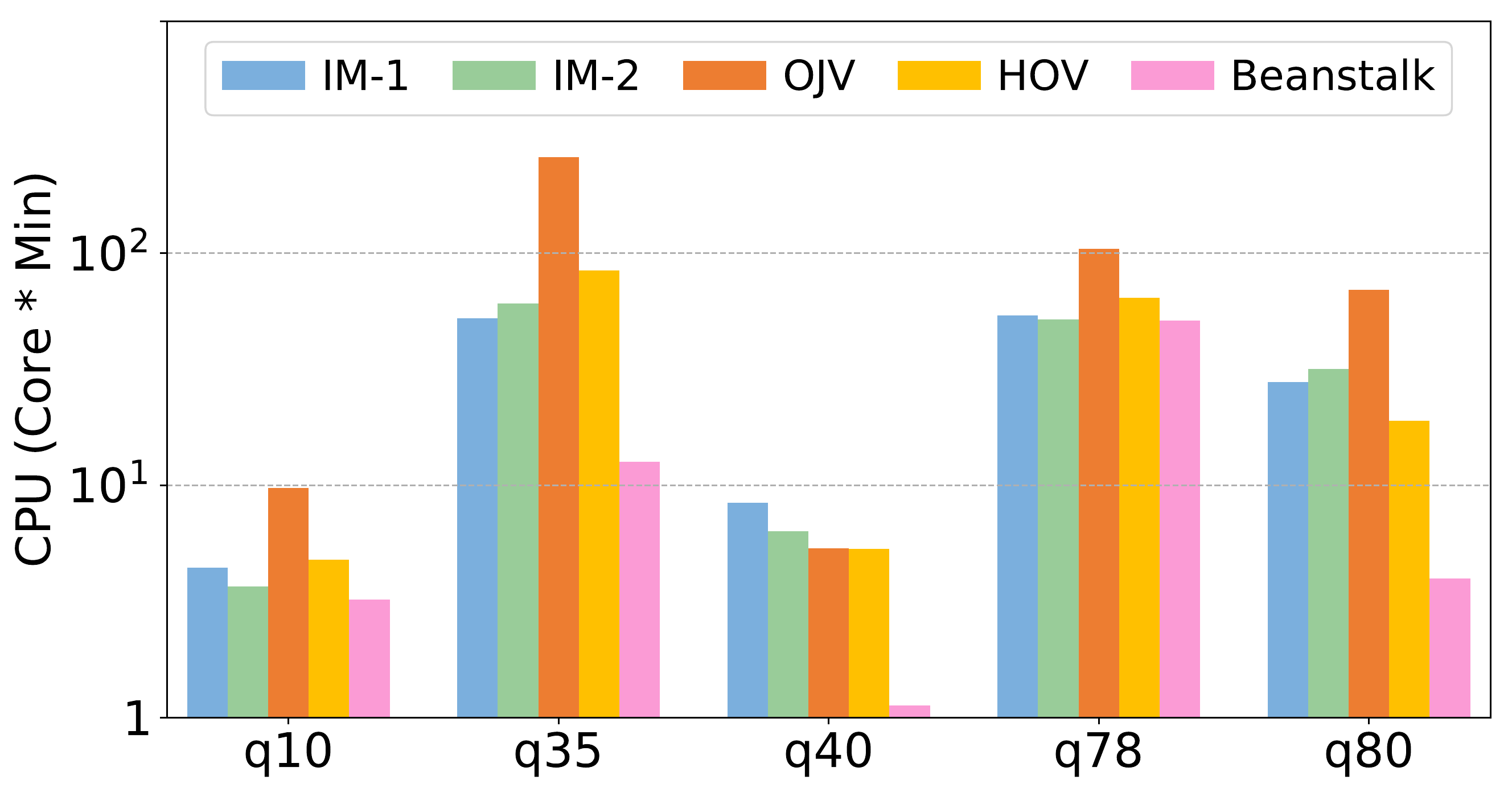}
  \label{fig:pdw-query-cpu}
}
\\
\subfigure[]{
  \includegraphics[height=0.16\textwidth]{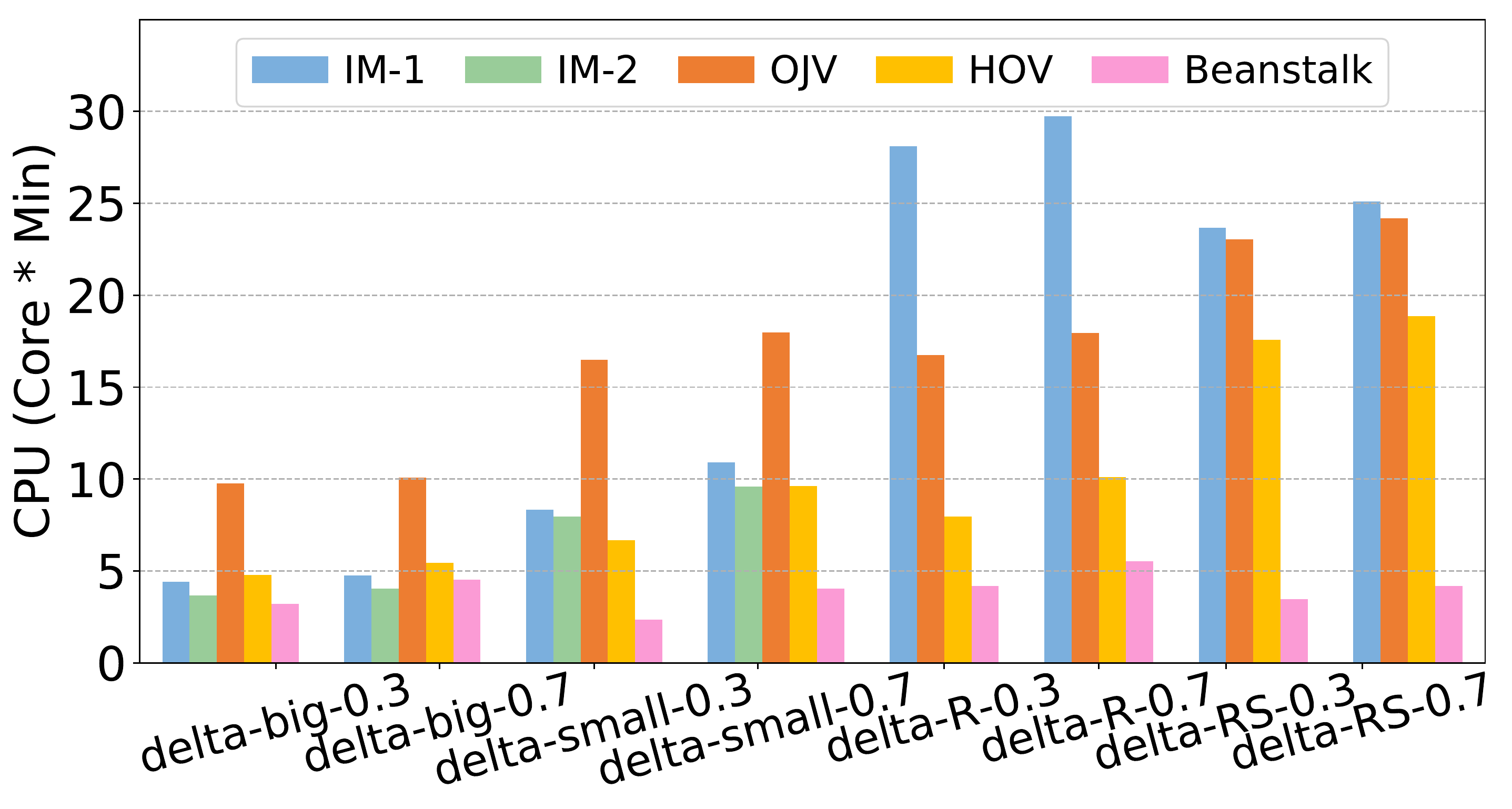}
  \label{fig:pdw-delta-cpu}
}
&
\subfigure[]{
  \includegraphics[height=0.16\textwidth]{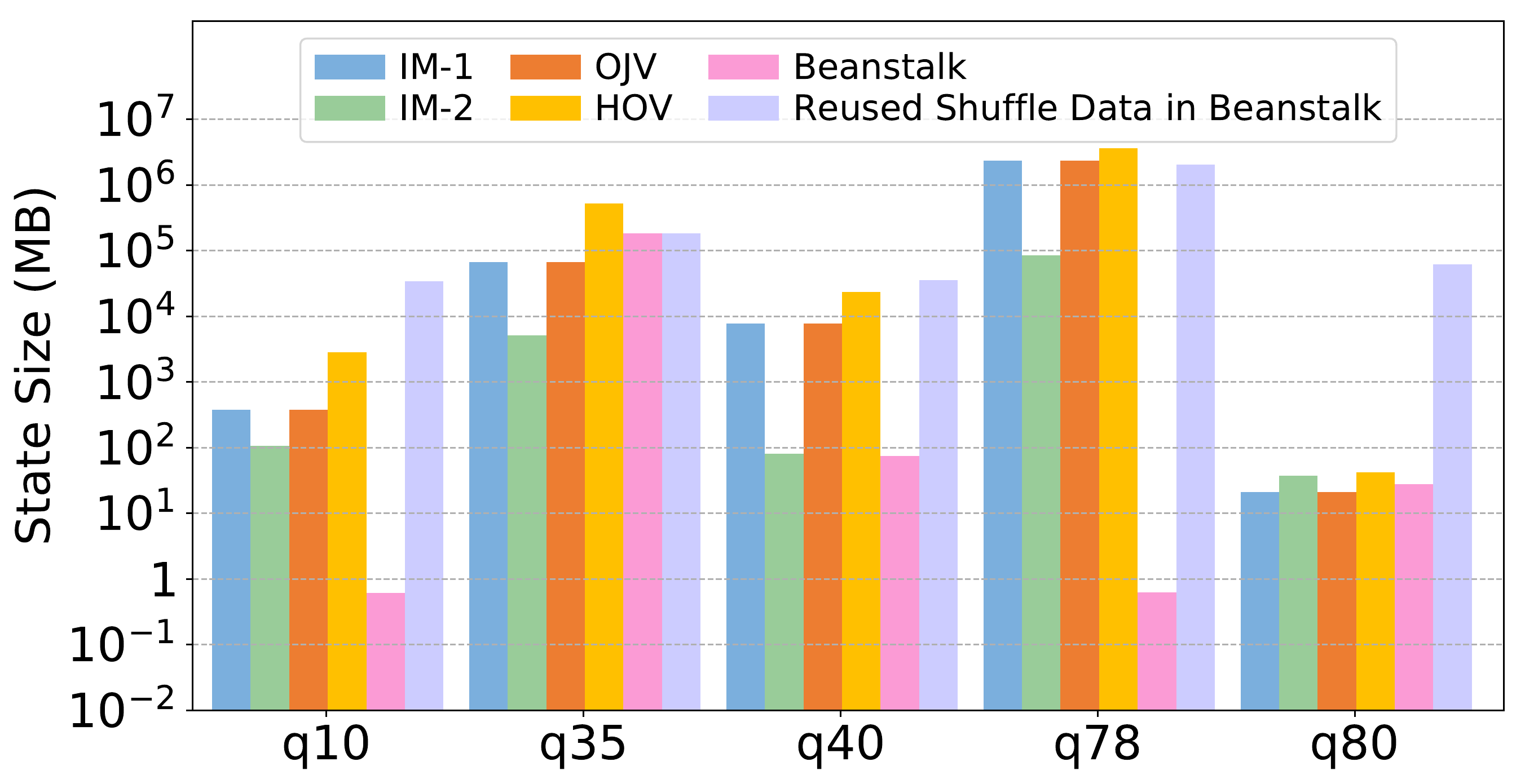}
  \label{fig:ivm-state-size}
}
&
\subfigure[]{
  \includegraphics[height=0.16\textwidth]{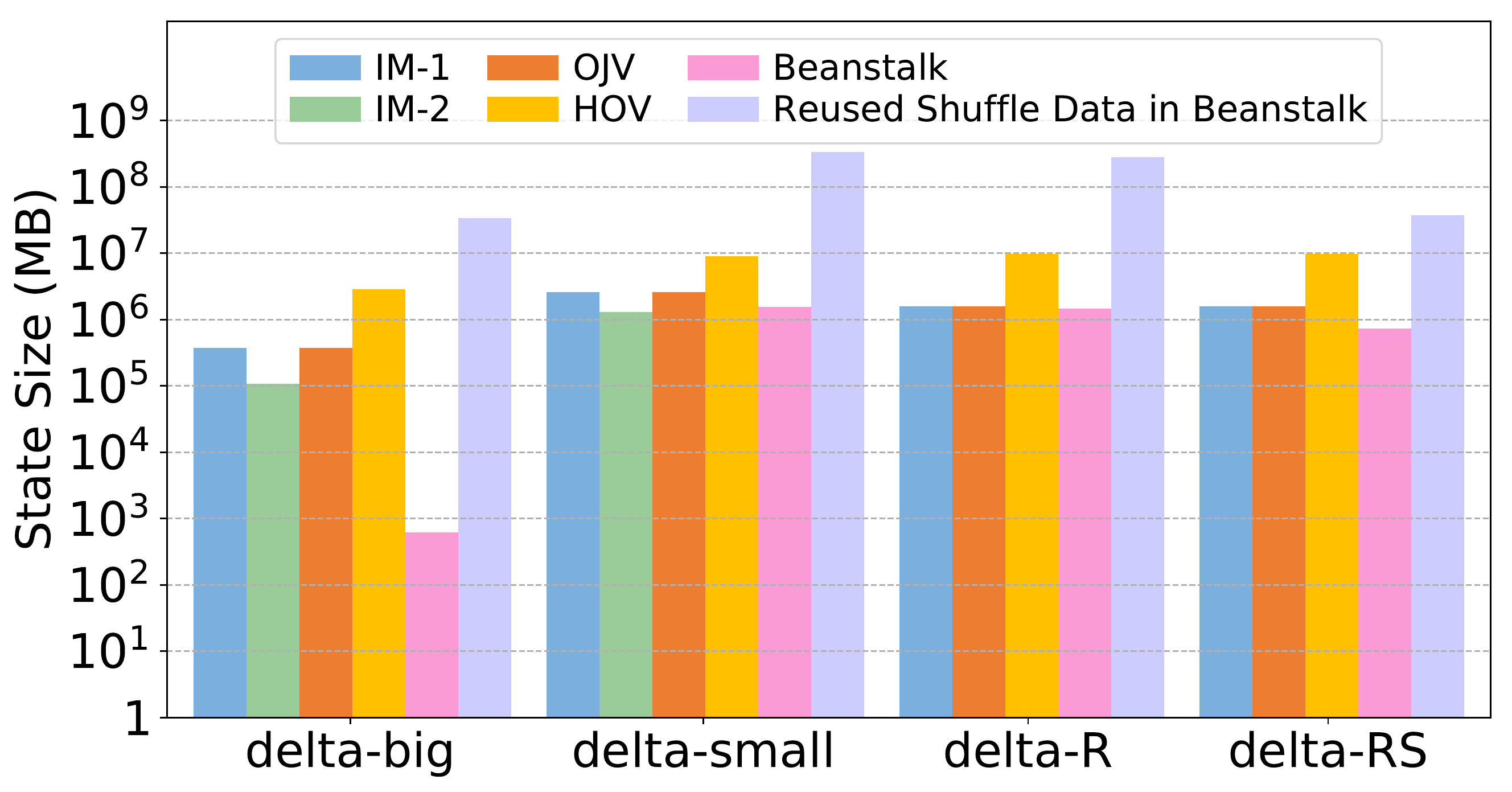}
  \label{fig:ivm-delta-state-size}
}
\\
\subfigure[]{
  \includegraphics[height=0.16\textwidth]{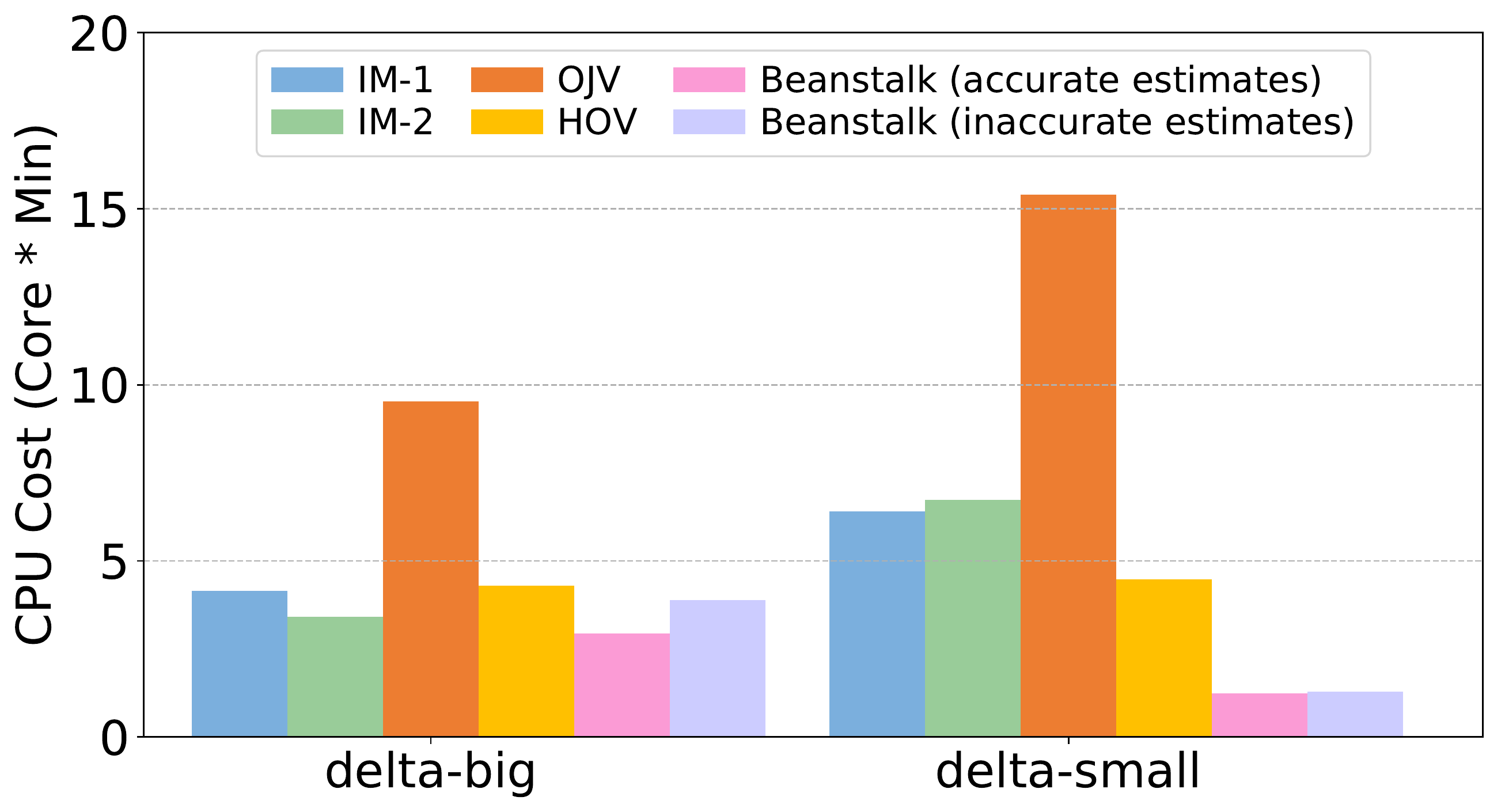}
  \label{fig:sensitivity}
}
&
\subfigure[]{
  \includegraphics[height=0.16\textwidth]{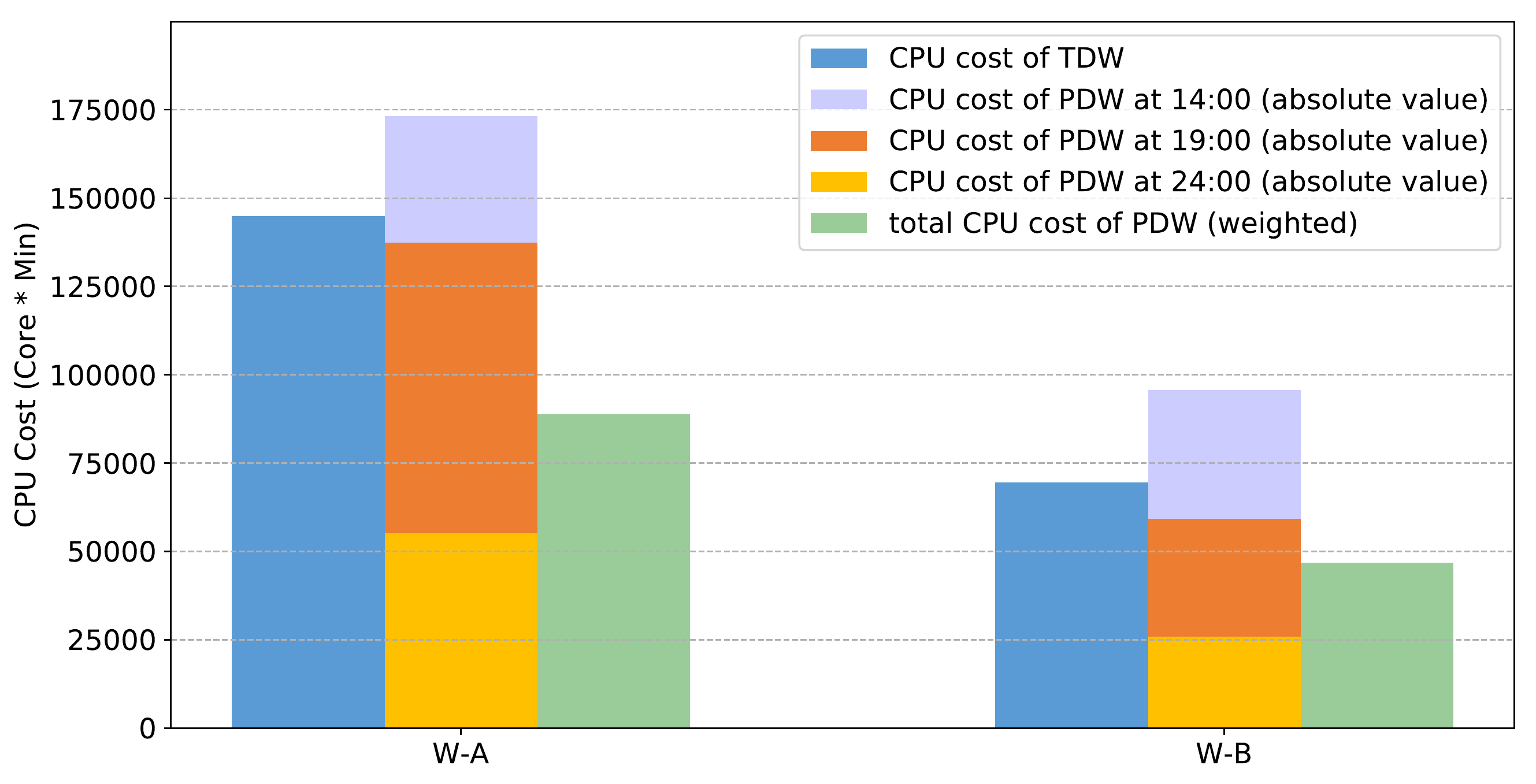}
  \label{fig:cpu-overall}
}
&
\subfigure[]{
  \includegraphics[height=0.16\textwidth]{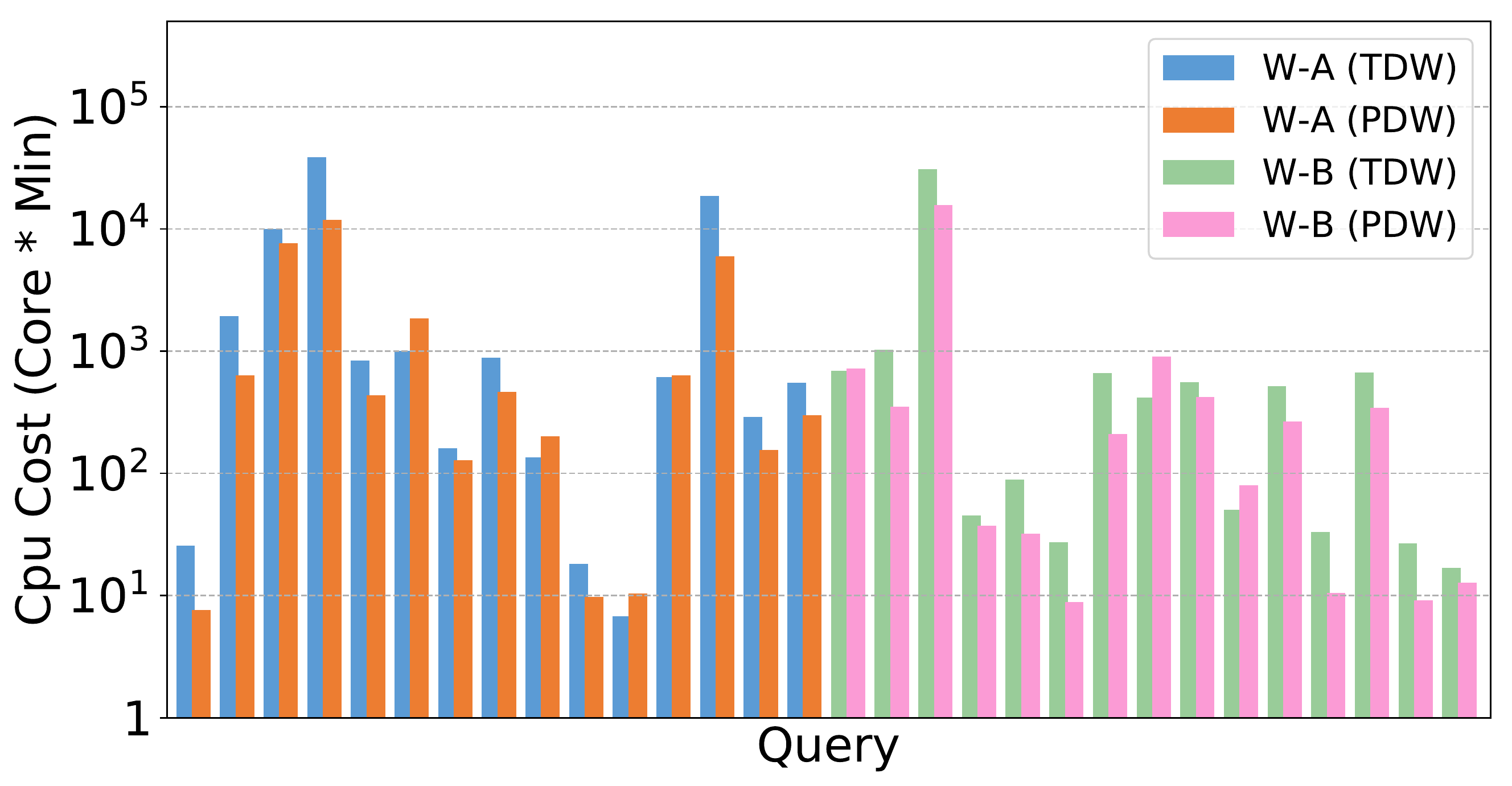}
  \label{fig:cpu-sample}
}
\end{tabular}
\caption{
(a)(b) The optimal real CPU costs of different incremental plans in \texttt{IVM-PD}
for different queries and data-arrival patterns.
(c)(d) The optimal real CPU costs of different incremental plans in \texttt{PDW-PD} for different queries, data-arrival patterns and cost weights.
(e)(f) The state sizes of different incremental plans in \texttt{IVM-PD} for different queries and data-arrival patterns.
(g) The plan quality of \sys{} under inaccurate cardinality estimation.
(h) The comparison between TDW and PDW on the CPU cost breakdowns of all queries in \texttt{W-A} and \texttt{W-B}, and (i) a detailed comparison of $30$ randomly sampled queries in \texttt{W-A} and \texttt{W-B}.}
\label{fig:real-cost}
\end{figure*}

\noindent
\kai{\textbf{Real CPU Costs}.
We reported the real CPU costs (in the unit of CPU$\cdot$minute)
in Fig.~\ref{fig:ivm-query-cpu}-\ref{fig:pdw-delta-cpu}
for the experiments in Fig.~\ref{fig:ivm-query}-\ref{fig:pdw-delta}.
In the \texttt{PDW-PD} experiments (Fig.~\ref{fig:pdw-query-cpu} and~\ref{fig:pdw-delta-cpu}),
the CPU costs were weighted according to the cost function in \texttt{PWD-PD}.
Note that Fig.~\ref{fig:ivm-query-cpu} and~\ref{fig:pdw-query-cpu} are plotted
in log scale due to the huge differences in CPU costs for different queries.
As we can see, the real CPU costs were agreed with the planner's estimation
(Fig.~\ref{fig:ivm-query}-\ref{fig:pdw-delta}) pretty well.
Some of the real costs were different from the estimated ones because of
the inaccuracy of the cost model.
But note that \sys{} consistently delivered the best plans
with the lowest CPU consumption across all experiments.}

\noindent
\kai{\textbf{State Sizes}.
In this set of experiments, we study the storage costs of materialized states
between \sys{} and each individual incremental methods.
We first fixed the data-arrival pattern to delta-big
and tested different queries under \texttt{IVM-PD}
settings respectively.
The results are reported in Fig.~\ref{fig:ivm-state-size}.
As shown, for most queries, the sizes of states materialized by \sys{}
were smaller than or comparable to each individual incremental algorithms.
This is due to the fact that \sys{} is able to
reuse the shuffled data as the states without incurring additional storage overheads
(see \S\ref{sec:costing}).
Thus, we further reported the sizes of the shuffled data reused by \sys{} in the figures.
Next we chose query q10 and varied the data-arrival patterns.
The results are reported in Fig.~\ref{fig:ivm-delta-state-size}.
Again, the storage costs of \sys{} were lower than or comparable to
that of each individual incremental algorithms.}

\noindent
\kai{\textbf{Sensitivity to Inaccurate Estimates}.
Next, we evaluated the sensitivity of \sys{} to inaccurate cardinality estimation.
To set up the experiment, we used q10 in the \texttt{IVM-PD} scenario.
We gave \sys{} the estimation of delta-small when running q10 with input delta-big,
and gave the estimation of delta-big when running q10 with input delta-small.
Fig.~\ref{fig:sensitivity} reported the real CPU costs.
For delta-big, \sys{} with the inaccurate estimation ran slower compared
to \sys{} with accurate estimation.
This is expected because \sys{} chose a plan that is optimal to the inaccurate cost model.
Nevertheless, \sys{} was still faster than \ivm{}, \ojv{}, \hov{}, and comparable to \scm{}.
For delta-small, inaccurate estimation had a small impact on execution time,
and \sys{} was still faster than each individual incremental method.}

\fi

\noindent
\textbf{Remarks}. In conclusion, the optimal incremental plan is affected by different factors
and does need to be searched in a principled cost-based way,
and \sys{} can consistently find better plans than each incremental method alone.

\subsection{Case Study: Progressive Data Warehouse}
\label{sec:case}

To validate the effectiveness of \sys{} in a real application,
we conducted a case study of the \texttt{PDW-PD} scenario
using two real-world analysis workloads \texttt{W-A} and \texttt{W-B} at Alibaba.
We compared the resource usage of executing these workloads in a traditional
data warehouse and a progressive one:
(1) \textbf{Traditional} (TDW), where we ran the analysis workloads at $24$:$00$
according to a schedule using the plans generated by a traditional optimizer;
and (2) \textbf{Progressive} (PDW), where besides $24$:$00$,
we also early executed the analysis workloads at $14$:$00$ and $19$:$00$ with only partial data
using the incremental plans generated by \sys{}.
The two early-execution time points were chosen to simulate
the observed cluster usage pattern (the cluster was often under-utilized at these times),
for which we set the weights of resource cost to $0.25$ and $0.3$, respectively.

Fig.~\ref{fig:total-vs-total} shows the real CPU cost of executing the workloads
(scored using the cost function in the \texttt{PDW-PD} setting), where we plotted the cumulative distribution of
the ratio between the CPU cost in PDW versus that in TDW.
We can see that PDW delivered better CPU cost for $80\%$ of the queries.
For about $60\%$ of the queries,
PDW was able to cut the CPU cost by more than $35\%$.
Remarkably, PDW delivered a total cost reduction of $56.2\%$ and $55.5\%$
for \texttt{W-A} and \texttt{W-B}, respectively.
Note that \sys{} searched plans based on the estimated costs
which could be different from the real execution cost.
As a consequence, for some of the queries (less than $10\%$)
we see more than $50\%$ cost increase.
Accuracy of cost estimation is not within the scope of the paper.
We further reported the PDW-to-TDW ratio of the CPU cost at $24$:$00$
in Fig.~\ref{fig:last-vs-total}, as this ratio indicated the resource reduction during the ``rush hours.''
As shown, for both workloads, PDW reduced the resource usage at peak hours
for over $85\%$ of the queries, and for over $70\%$ of the queries we
can see significant reduction of more than $25\%$.
\iffull

\kai{
We also reported the absolute values of CPU costs of \texttt{W-A} and \texttt{W-B}.
However, as \texttt{W-A} and \texttt{W-B} have $274$ and $554$ queries each,
it is not realistic to show all of them.
Instead we reported the total CPU cost breakdowns for TDW and PDW
in Fig.~\ref{fig:cpu-overall}.
Specifically for PDW, we reported the absolute values of CPU costs at each time,
and the total CPU costs weighted according to the cost function in \texttt{PDW-PD}.
As we can see, \sys{} indeed picked better plans with less resource consumption:
PDW saved $38.7\%$ and $32.6\%$ CPU costs compared to TDW for \texttt{W-A} and \texttt{W-B} respectively.
On the other hand, with incremental computation, PDW brought relatively low overheads
compared to TDW, $19.6\%$ and $37.6\%$ for \texttt{W-A} and \texttt{W-B} respectively.
The PDW overheads are computed by summing up the absolute values of CPU costs at each time,
minus the CPU costs of TDW.
We further randomly selected $15$ queries from \texttt{W-A} and \texttt{W-B} respectively,
and reported their CPU costs in TDW and PDW in Fig.~\ref{fig:cpu-sample}.
Again, for most queries PDW reduced the CPU costs by a significant amount.}
\fi

\subsection{Performance of IQP}
\label{sec:perf}

Next, we evaluated the time performance of \sys{} in IQP.
Compared to traditional query planning, IQP
has two salient characteristics:
(1) In plan-space exploration, IQP explores a larger plan space.
(2) Besides, IQP needs to decide the intermediate states to share,
which are not considered by traditional query planning.
We will present performance results
on these two phases: \emph{Plan-Space Exploration} (PSE)
and \emph{State Materialization Optimization} (SMO).

We used PDW-PD as the IQP problem definition.
Unless otherwise specified, in the problem definition we set $|\vec{T}|=3$.
To help readers better understand how \sys{} performs on different types
of queries, we use the TPC-DS queries,
and present the optimization time of \sys{} on them.
Besides the overall performance study, we also present a detailed study on four aspects of \sys{}'s optimization performance:
\iffull
\begin{table}[tbh!]
\tiny
\caption{Statistics of selected representative queries}
\label{tbl:selected-queries}
\begin{tabular}{ccccccccccc}
\hline
Query & Q22& Q20 & Q43 & Q67 & Q27 & Q99 & Q85 & Q91 & Q5 & Q33 \\
\hline
\# Joins & 2 & 2 & 2 & 3 & 4 & 4 & 6 & 6 & 7 & 9 \\
\hline
\# Agg-\\regates & 1 & 1 & 1 & 1 & 1 & 1 & 1 & 1 & 4 & 4 \\
\hline
\# Sub-\\Queries & 0 & 0 & 0 & 2 & 0 & 0 & 0 & 0 & 7 & 7 \\
\hline
\end{tabular}
\end{table}
\fi
\begin{enumerate}[nosep,leftmargin=*]
  \item \textbf{Query complexity}: How does \sys{} perform when queries become increasingly complex, e.g.,
  with more joins or subqueries?

  \item \textbf{Size of IQP}: How does \sys{} perform when the number of incremental runs (i.e., $|\vec{T}|$) in the IQP problem definition changes?

  \item \textbf{Number of incremental methods}: How does \sys{} perform when users integrate more incremental methods into it?

  \item \textbf{Optimization breakdown}: How effective are the speed-up optimizations
  discussed in \S\ref{sec:speed-up-opt}?
\end{enumerate}
\iffull
To study the above four aspects,
we further selected ten representative TPC-DS queries with different
numbers of joins, aggregates, and subqueries.
The selected queries are shown in Table~\ref{tbl:selected-queries}.
\fi

\begin{figure*}[tb!]
\centering
\begin{tabular}{ccc}
\multicolumn{3}{c}{
\subfigure[]{
  \includegraphics[width=.96\textwidth]{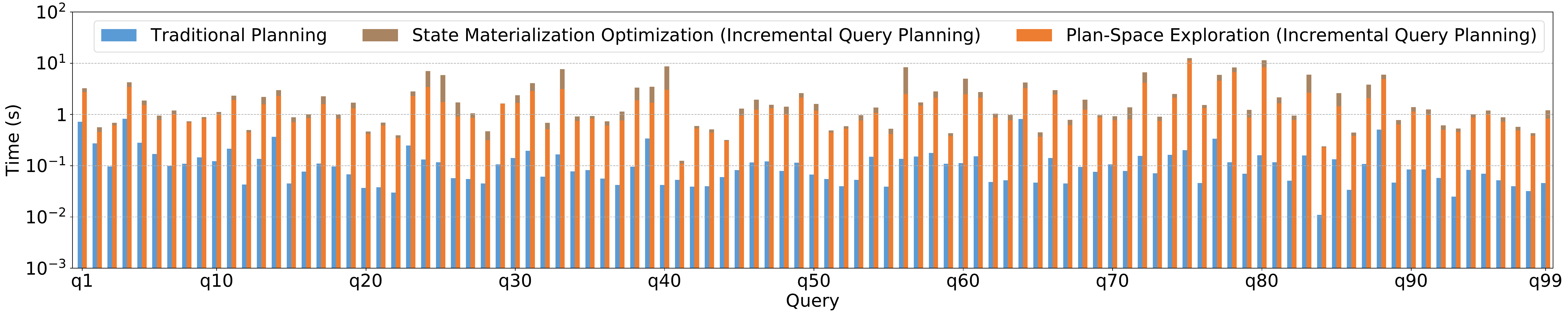}
  \label{fig:opt-perf}
}
}
\\
\subfigure[]{
  \includegraphics[height=0.18\textwidth]{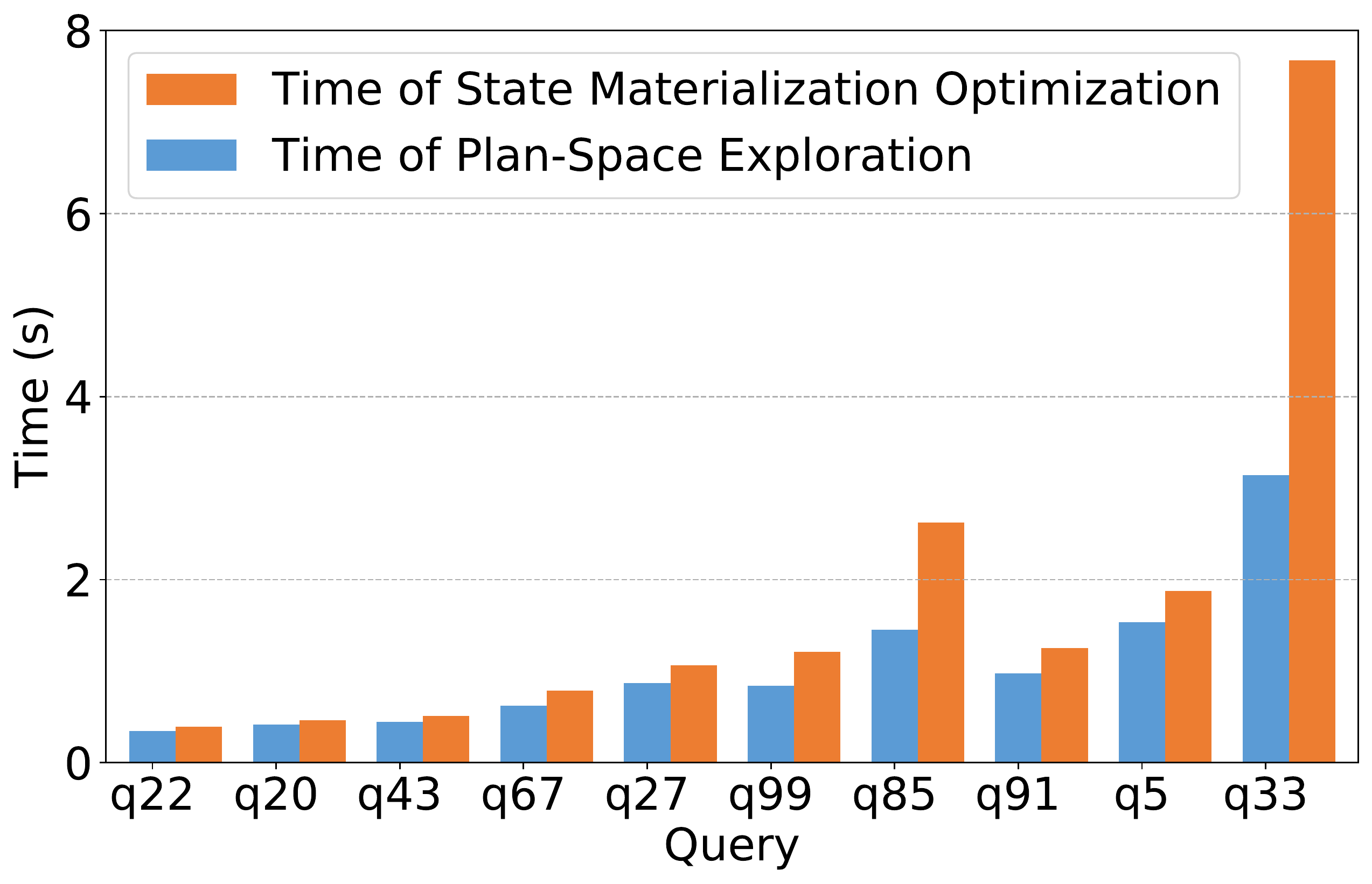}
  \label{fig:perf-query-complexity}
}
&
\subfigure[]{
  \includegraphics[height=0.18\textwidth]{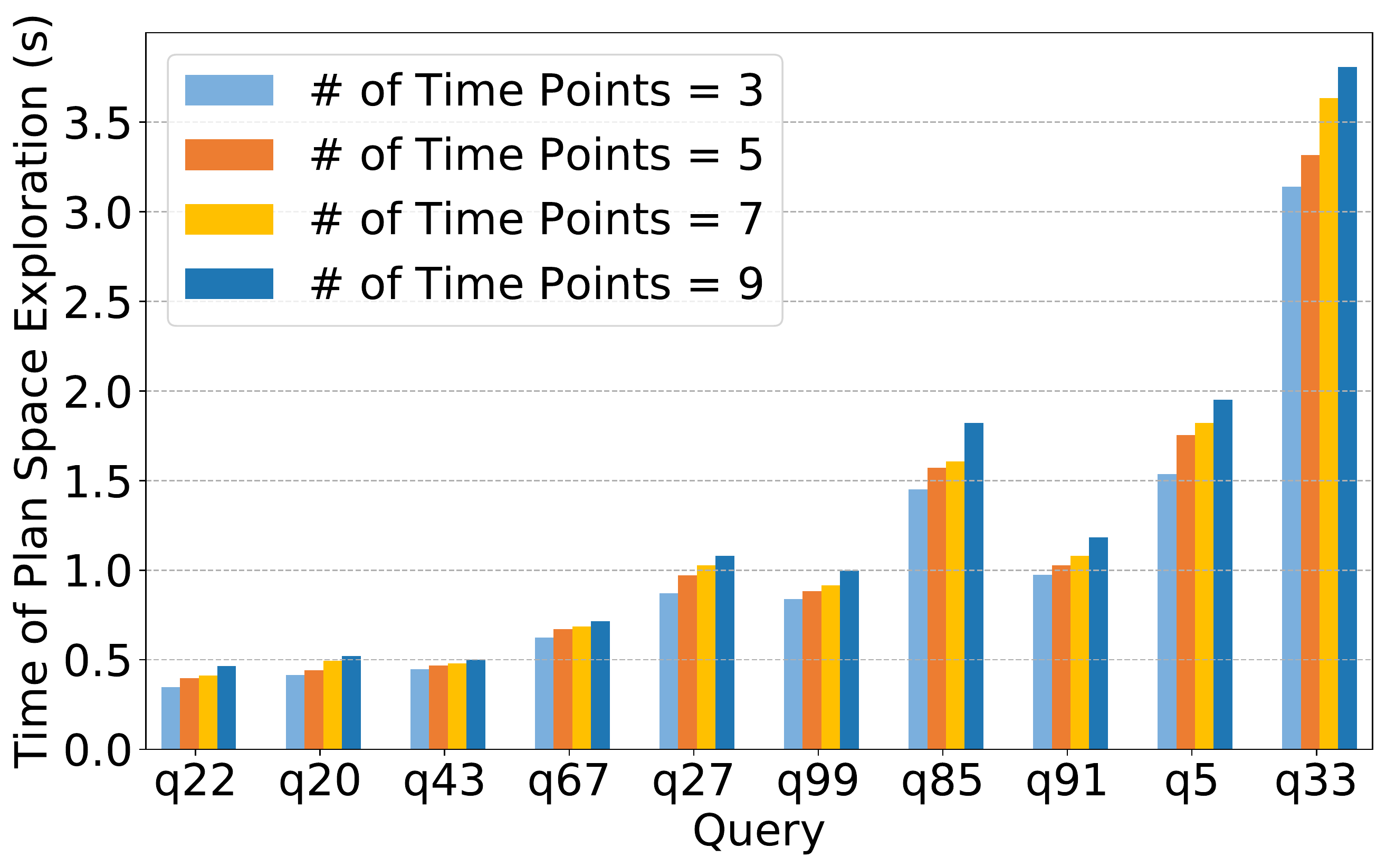}
  \label{fig:perf-size-pse}
}
&
\subfigure[]{
  \includegraphics[height=0.18\textwidth]{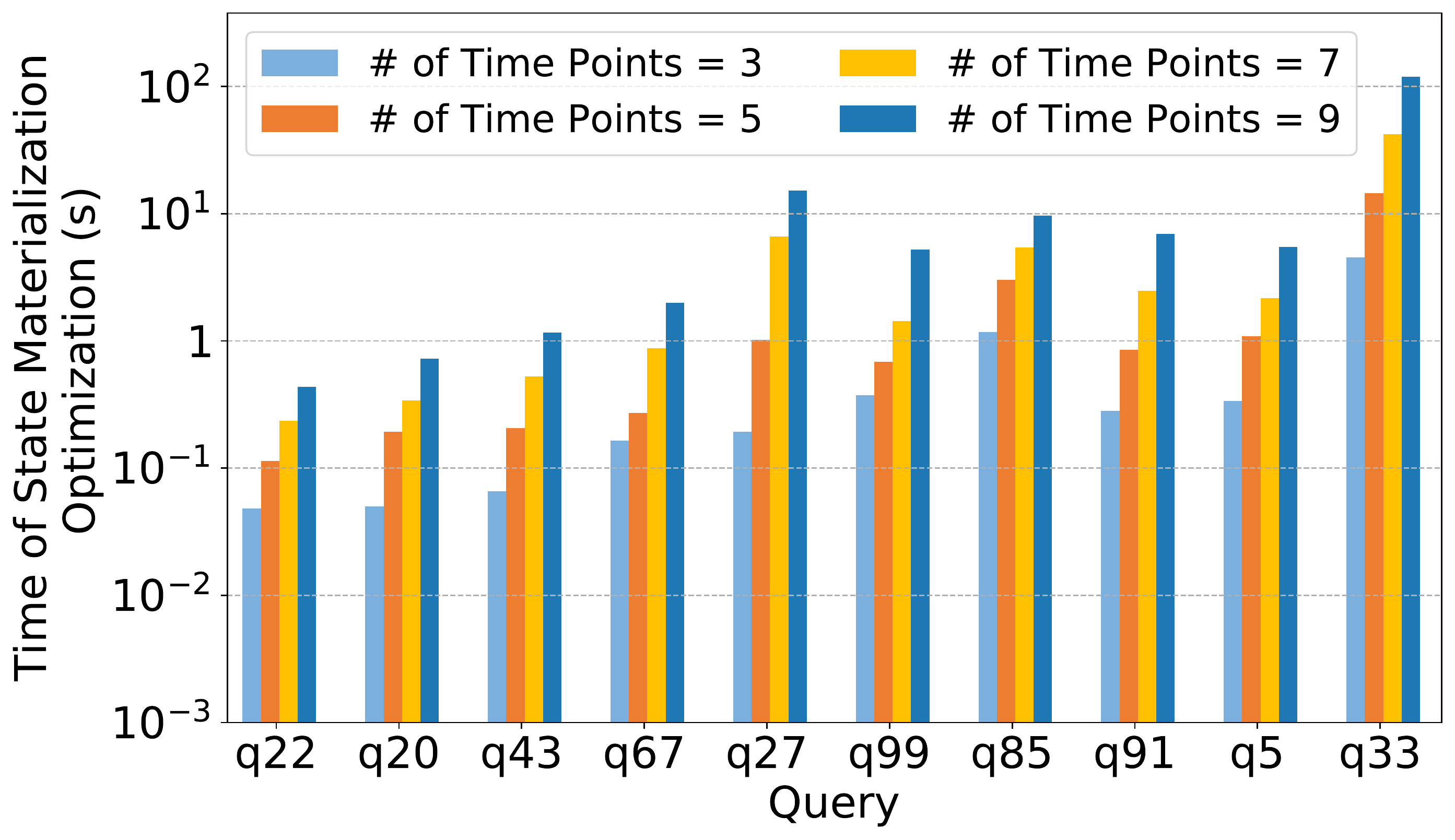}
  \label{fig:perf-size-sm}
}
\\
\subfigure[]{
  \includegraphics[height=0.18\textwidth]{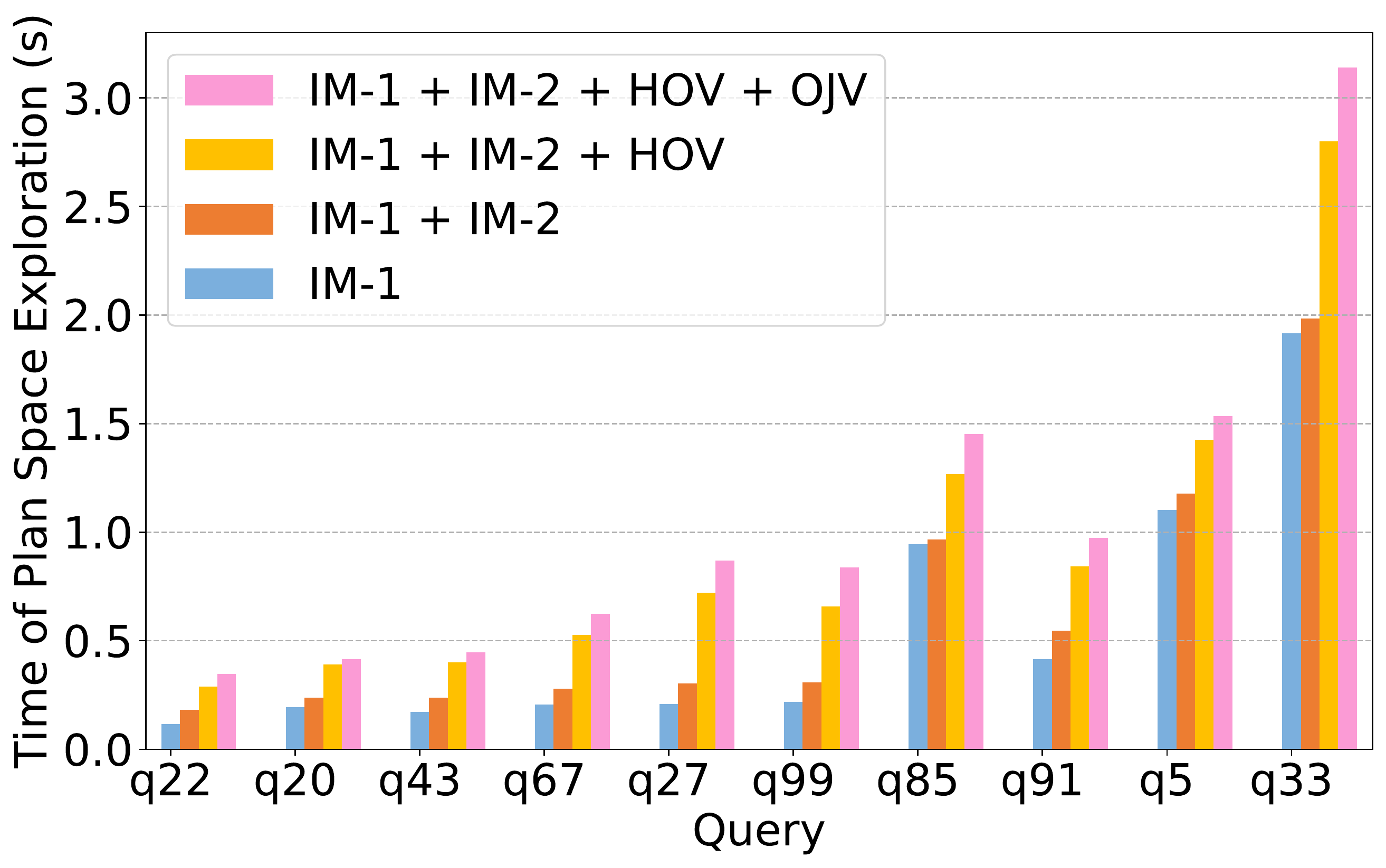}
  \label{fig:perf-pse-im}
}
&
\subfigure[]{
  \includegraphics[height=0.18\textwidth]{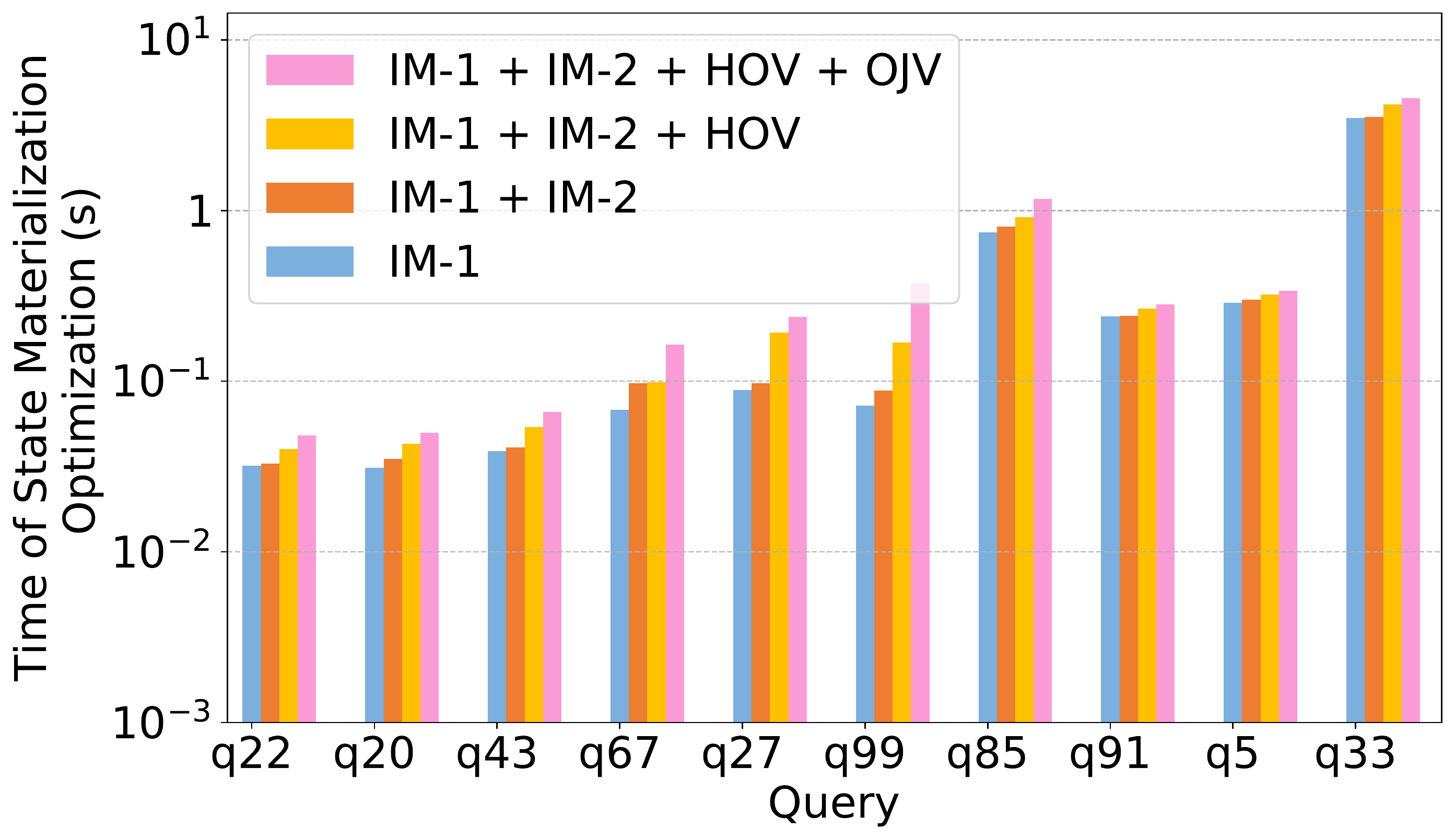}
  \label{fig:perf-sm-im}
}
&
\subfigure[]{
  \includegraphics[height=0.18\textwidth]{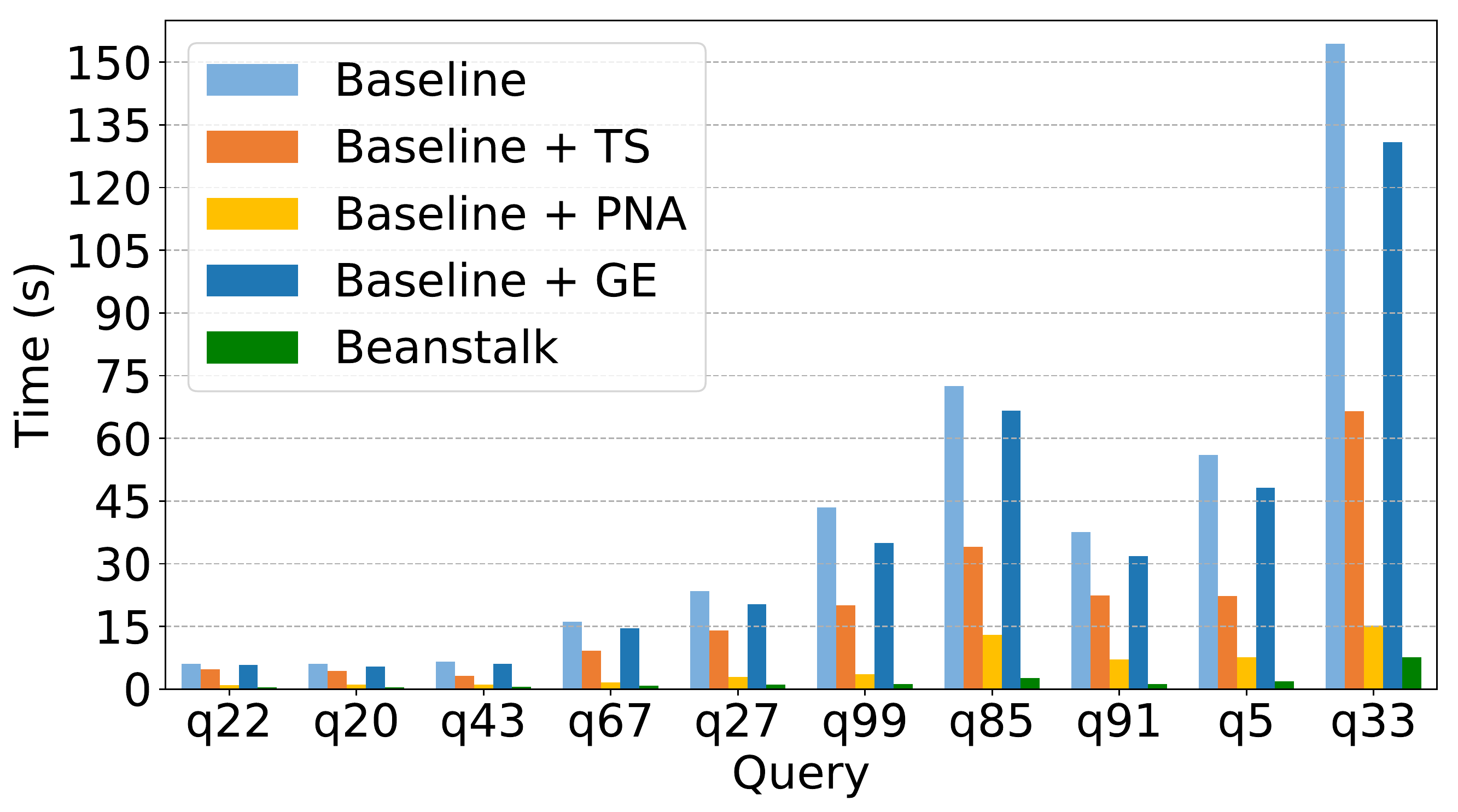}
  \label{fig:pse-opt-breakdown}
}
\end{tabular}
\caption{(a) Overall planning performance comparison on the TPC-DS benchmark between
traditional and incremental query planning. (b) Impact of the query complexity on the planning performance. (c) (d) Impact of the planning problem size on the planning performance.
(e)(f) Impact of the number of incremental methods on the planning performance.
(g) Effectiveness of the speed-up optimization techniques.
Note that the selected queries are ordered by their query complexity\iffull (as listed in Table~\ref{tbl:selected-queries})\fi.}
\label{fig:perf}
\end{figure*}

\noindent
\textbf{Overall Planning Performance}.
We first studied the overall query planning performance by comparing
IQP with traditional planning.
Fig.~\ref{fig:opt-perf} shows the end-to-end planning time on all TPC-DS queries.
As shown, although IQP planned
a much bigger plan space than traditional planning, 
\sys{} still delivered high planning performance:
IQP finished within $3$ seconds for $80\%$ queries, and for all queries finished
within $14$ seconds.
For over $80\%$ queries, the IQP optimization time was less than $24$X
of the traditional planning time.
Even though slower than traditional planning at optimization time on a single machine,
IQP generated much better incremental plans that
brought significant benefit in resource usage and query latency on a cluster.
\iffull
We can further reduce the planning time by adopting a parallel optimizer~\cite{orca}.
\fi
\iffull
\kai{
As a reference, we also reported the real CPU cost used by TDW,
the CPU costs saved by PDW compared to TDW, and the planning time
in Fig.~\ref{fig:tpcds-1t-run-time}.
We can see that for most queries, the CPU time on planning was
$2$-$3$ orders of magnitude smaller than the saved CPU costs.
This shows that the planning cost is negligible compared to the execution cost.
Thus the benefit of a better plan outweighs the extra time spent on planning.}
\else

\fi


\noindent
\textbf{Query Complexity.}
To study the impact of query complexity on performance,
we tested on the selected TPC-DS queries\iffull in Table~\ref{tbl:selected-queries}\fi,
and reported the broken-down optimization times in Fig.~\ref{fig:perf-query-complexity}.
As one can see, the planning time increased slowly when the query complexity increased,
because the plan space grew larger for complex queries.
The time spent on PSE
was less than that spent on SMO in general,
and also grew with a slower pace.
This result shows that query complexity has a smaller impact on PSE.

\noindent
\textbf{Size of IQP}.
To study the impact of the size of the planning problem on the planning time,
we gradually increased  the number of incremental runs planned from $3$ to $9$,
and reported the time on PSE and SMO
in Fig.~\ref{fig:perf-size-pse} and~\ref{fig:perf-size-sm}.
As depicted, the time on PSE stayed almost constant as
the size of IQP changed. E.g., when the number of incremental runs
grew $3$X, the time for q$33$ only slightly increased by $20\%$.
This was mainly due to the effective speed-up optimization techniques introduced
in \S\ref{sec:speed-up-opt}.
In comparison,
the SMO time increased superlinearly with increasing number of incremental runs,
due to the time complexity of the MQO algorithm we chose~\cite{mqo-pods}.



\iffull
\begin{figure*}[tb!]
\centering
\includegraphics[width=.96\textwidth]{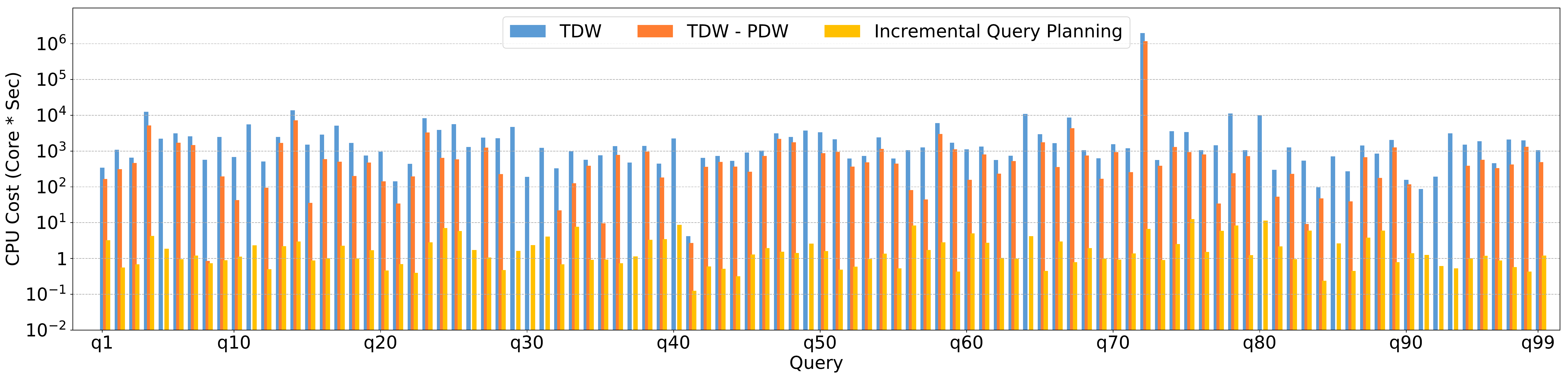}
\caption{Real resource consumption of \sys{}'s plan
as in Fig.~\ref{fig:opt-perf} on 1T TPC-DS benchmark.
}
\label{fig:tpcds-1t-run-time}
\end{figure*}
\fi

\noindent
\textbf{Number of Incremental Methods}.
Next, we evaluated how the performance scales when more incremental methods
are added into the optimizer by gradually adding methods \ivm{}, \scm{}, \hov{} and \ojv{}
into \sys{}. Fig.~\ref{fig:perf-sm-im} and~\ref{fig:pse-opt-breakdown} show
the time on PSE and SMO, respectively.
As illustrated, the time on both PSE and SMO increased when more incremental methods
were added, because more incremental methods increased the plan space.
There are two interesting findings.
\kai{(1) The PSE time did not grow linearly with the number of incremental methods,
but rather the size of the plan space that each method newly introduces.
For example, the difference of PSE time introduced by adding \hov{} was bigger than that introduced by adding \ojv{}. This was because \hov{} and \ojv{} use similar methods that update a single relation at a time, which are very different from \ivm{} and \scm{} that update all relations each time.}
(2) The number of incremental methods had less impact on the planning time than
the size of the IQP problem, which can be observed on the SMO time.
This is mainly because
the plan space explored by different incremental methods often have
overlaps, e.g., most incremental methods fundamentally share similar delta update rules
as \ivm{}, whereas the plan spaces of different incremental runs do not have overlaps.

\noindent
\textbf{Optimization Breakdown}.
In the end, we evaluated the effectiveness of the speed-up optimizations
by breaking down the optimization techniques discussed in \S\ref{sec:speed-up-opt},
i.e., translational symmetry (TS), pruning non-promising alternatives (PNA),
and guided exploration (GE).
Fig.~\ref{fig:pse-opt-breakdown} reports the PSE times of different combinations
of the speed-up optimizations.
We compared the implementations with no optimization (Baseline),
with each individual optimization (Baseline+TS, Baseline+PNA, Baseline+GE), and with all three optimizations (\sys{}).
As shown, the optimizations together brought an order of magnitude
performance improvements.
The most effective optimizations were PNA and TS.
PNA generally improved the PSE time by $5$-$12$X, while TS improved the PSE time by $1.5$-$2.5$X.

\section{Related Work}

\noindent
\textbf{Incremental Processing.}
There is a rich body of research
on various forms of incremental processing, ranging from incremental view maintenance,
stream computing, to approximate query answering and so on.
Incremental view maintenance has been intensively studied
before. It has been considered under both the set \cite{set-ivm-1, set-ivm-2} and bag \cite{bag-ivm-1, bag-ivm-2} semantics, for queries with outer joins~\cite{semi-outer-join,outerjoin-view}, and using higher-order maintenance methods~\cite{dbtoaster}.
Previous studies mainly focused on delta propagation rules for relational operators.
Stream computing~\cite{abadi2005design, ghanem2010supporting, motwani2003query, thakkar2011smm} adopts incremental processing and sublinear-space algorithms to process updates and deltas.
Many approximate query answering studies~\cite{acharya1999aqua, babcock2003dynamic, chaudhuri2007optimized} focused on constructing optimal samples to improve query accuracy.
Proactive or trigger-based incremental computation techniques~\cite{iolap,progressive-analysis} were used to achieve low query latency.
Zeng et al. \cite{iolap} proposed a probability-based pruning technique for incrementally computing complex
queries including those with nested queries.
These studies proposed incremental computation techniques in isolation, and do not have a general cost-based optimization framework, which is the focus of this paper. In addition, they can be integrated into \sys{}, and contribute to a unified plan space for searching the optimal incremental plan.

\noindent
\textbf{Query Planning for Incremental Processing.}
Previous work studied some optimization problems in incremental computation.
Viglas et al.~\cite{rate} proposed a rate-based cost model for stream processing.
The cost model is orthogonal to \sys{} and can be integrated.
DBToaster~\cite{dbtoaster} discussed a cost-based approach to deciding the views to materialize
under a higher-order view maintenance algorithm.
Tang et al.~\cite{intermittent} focused on selecting optimal
states to materialize for scenarios with intermittent data arrival.
They proposed a dynamic programming algorithm
for selecting intermediate states to materialize
given a fixed physical incremental plan and a memory budget,
by considering future data-arrival patterns.
These optimization techniques all focus on the optimal materialization problem
for a specific incremental plan or incremental method,
and thus are not general IQP solutions.

Flink~\cite{flink} uses Calcite~\cite{calcite} as the optimizer and support stream queries,
which only provides traditional optimizations on the logical plan generated
by a fixed incremental method,
but cannot explore the plan space of multiple incremental methods,
nor consider correlations between incremental runs.
On the contrary, \sys{} provides a general framework for users to integrate
various incremental methods, and searches the plan space
in a cost-based approach.


\noindent
\kai{\textbf{Semantic Models for Incremental Processing.}
CQL\cite{cql} exploited the relational model to provide strong
query semantics for stream processing.  Sax et al. \cite{two-sides} introduced the Dual Streaming Model
to reason about ordering in stream processing.
The key idea behind~\cite{cql,two-sides} is the duality of relations and streams,
i.e., time-varying relations can be modeled as a sequence of static relations,
or a sequence of change logs consisting of INSERT and DELETE operations.
The recent work~\cite{one-query} proposed to integrate streaming into the SQL standard,
and briefly mentioned that TVR's can serve as a unified basis of both relations and streams.
However, their models do not include a formal algebra and rewrite rules on TVR's, and thus cannot fully model incremental computation.
To the best of our knowledge, our \xmodel{} for the first time
formally defines an algebra on TVR's, especially it provides a principled
way to model different types of snapshot/deltas in multiplicity/attribute perspectives
and $+/-$ operators between them.
Furthermore, the \xmodel{} provides a trichotomy of all TVR rewrite rules
and shows that the three types of rewrite rules can subsume many existing incremental algorithms,
which provides a theoretical foundation for \sys{}, so that \sys{} can implement all incremental algorithms
that can be described using the three types of TVR rewrite rules.}




\section{Conclusion}
\label{sec:conclusion}

In this paper, we developed a novel, principled cost-based optimizer framework, called \sys{}, for optimizing incremental data processing.  We first proposed a theory called \xmodel{} as its foundation, which can formally model incremental processing in its most general form. We gave a full specification of \sys{}, which can not only unify various existing techniques to generate an optimal incremental plan, but also allow the developer to add their rewrite rules.  We studied how to explore the plan space and search for an optimal incremental plan.  We conducted thorough experimental evaluation of \sys{} in various incremental-query scenarios to show its effectiveness and efficiency.


\balance

\bibliographystyle{abbrv}
\bibliography{prog-exec}

\end{document}